\documentclass[%
 reprint,
 amsmath,amssymb,
 aps,
]{revtex4-2}

\usepackage{graphicx}% Include figure files
\usepackage{dcolumn}% Align table columns on decimal point
\usepackage{bm}% bold math
\usepackage{xcolor}
\usepackage{hyperref}
\usepackage{booktabs}
\usepackage{makecell}
\usepackage{array}
\usepackage{tikz}
\usetikzlibrary{arrows.meta, positioning, calc, fit}
\newcommand{\revA}[1]{#1}
\newcommand{\revB}[1]{#1}
\newcommand{\YS}[1]{#1}

\begin{document}
\title{A New Approach to the Calculation of Particle Creation from Analog Black Holes}

\author{Yang-Shuo Hsiung}
\email{physhsiung@gmail.com}
\affiliation{
Department of Physics and Center for Theoretical Physics,
Leung Center for Cosmology and Particle Astrophysics, 
Max Planck--IAS--NTU Center for \mbox{Particle Physics, Cosmology and Geometry},
National Taiwan University, Taipei 10617, Taiwan
}

\author{Pisin Chen}
\email{pisinchen@phys.ntu.edu.tw}
\affiliation{
Department of Physics and Center for Theoretical Physics,
Leung Center for Cosmology and Particle Astrophysics, 
Max Planck--IAS--NTU Center for \mbox{Particle Physics, Cosmology and Geometry},
National Taiwan University, Taipei 10617, Taiwan
}

\begin{abstract}
Accurate predictions of particle creation from accelerating mirrors are essential for interpreting forthcoming analog Hawking radiation experiments, yet the associated Bogoliubov integrals are generally analytically intractable for realistic trajectories. 
We introduce the Inertial Replacement Method (IRM), a hybrid analytic--numerical framework that reduces the infinite-domain Bogoliubov integral to a finite-domain problem by replacing the asymptotic tails of the trajectory with inertial extensions. 
This allows numerical evaluation to be restricted to the finite accelerating region, with controlled boundary contributions from the asymptotic segments. 
We derive perturbative bounds on the IRM-induced deviations, providing accuracy estimates and guidance for segmentation. 
The method is validated against trajectories with known analytic solutions and applied to a fully numerical \revB{Chen--Mourou plasma-mirror trajectory obtained from particle-in-cell (PIC) simulations} relevant to the planned AnaBHEL experiment.
Our results demonstrate that, within the IRM framework, the characteristic ray-tracing mapping in the finite accelerating region suffices to capture the overall shape of the particle number spectrum, while the result is insensitive to the further inclusion of the asymptotic tails. 
This suggests a more local perspective of the particle creation from the moving mirrors associated with its accelerating motion, rather than as a consequence of the detailed global structure of the spacetime.
Also, these results establish the IRM as a reliable and applicable computational framework for particle creation in realistic analog-gravity systems.
\end{abstract}

\maketitle
% \tableofcontents
% ====================================================================================================
\section{Introduction}
\label{sec:Introduction}
% ----------------------------------------------------------------------------------------------------
% Hawking radiation, information paradox, analog gravity system
Hawking radiation~\cite{1975_Hawking} is one of the most profound predictions of quantum field theory in curved spacetime. 
It reveals that black holes are not entirely black, but emit an approximately thermal spectrum of particles generated by quantum fluctuations near the event horizon. 
If this evaporation process were exactly thermal, an initially pure quantum state could evolve into a mixed state after complete evaporation, apparently conflicting with unitarity~\cite{1976_Hawking}.
This tension constitutes the black hole information paradox, which remains a central open problem in theoretical physics and motivates the search for experimental platforms in which Hawking-like radiation and information flow can be investigated.
However, direct observation of Hawking radiation from astrophysical black holes is infeasible.
For example, a solar-mass black hole radiates at only $T_{\rm BH}\sim10^{-8}\,\mathrm{K}$, far below the cosmic microwave background temperature and any realistic detection threshold.
This extreme weakness has therefore motivated the development of analog gravity systems, e.g.~\cite{1981_Unruh, 2005_Barcelo, 2016_Steinhauer, 2022_AnaBHEL}, in which horizon-like phenomena and quantum particle production can be explored under controlled laboratory conditions.

% ----------------------------------------------------------------------------------------------------
% Moving mirror
A powerful theoretical framework for studying Hawking-like radiation is the \emph{moving mirror model} first introduced by Moore~\cite{1970_Moore} and later investigated by Davies and Fulling~\cite{1976_FullingDavies,1977_DaviesFulling}.  
In this (1+1)D Minkowski spacetime, a prescribed mirror trajectory imposes a time-dependent boundary condition on a quantum field, leading to mixing of positive- and negative-frequency modes and consequent particle creation from the vacuum.  
For suitable trajectories, the model reproduces key features of black-hole evaporation, including effective horizon formation and thermal spectra.
Several analytically solvable examples have provided valuable insight into acceleration-induced radiation~\cite{2011_Good,2013_Good,2021_Lin,2024_Lin}.

% ----------------------------------------------------------------------------------------------------
% AnaBHEL
The Analog Black Hole Evaporation via Laser (AnaBHEL) experiment~\cite{2022_AnaBHEL}, originally proposed by Chen and Mourou~\cite{2017_Pisin}, aims to observe analog Hawking radiation using a state-of-the-art accelerating laser-plasma flying mirror.  
In this setup, an intense laser pulse propagating through plasma excites a relativistic wakefield whose compressed electron layer acts as a partially reflecting “flying mirror.”  
Its nonuniform acceleration induces a strong redshift of field modes, leading to quantum particle production from the vacuum and closely paralleling the mechanism of Hawking emission.
The AnaBHEL experiment therefore provides a promising platform for laboratory investigations of horizon-induced quantum radiation.

% ----------------------------------------------------------------------------------------------------
% Current challenges
Quantitative predictions for analog Hawking radiation require evaluation of the Bogoliubov coefficients, which quantify positive--negative frequency mixing. 
These coefficients are expressed as oscillatory integrals over the entire temporal domain, with the mirror's dynamics entering the integrand in a highly nontrivial manner. 
For generic mirror motions, the trajectory is not described by a simple analytic function but must instead be obtained from simulations or experiments. 
In such cases, closed-form evaluation of the Bogoliubov integrals is almost never possible, and analytic solutions exist only for a small class of specially designed trajectories.
Direct numerical integration is not a straightforward alternative. 
Because the integration domain extends over the entire real line, the integrand develops long, rapidly oscillatory tails. 
These oscillations generate delicate cancellations, substantial floating-point loss, and unstable numerical convergence. 
For realistic, experimentally motivated trajectories, naive numerical approaches therefore become slow or unreliable.
These difficulties motivate the development of a hybrid (or semi-analytic) method capable of treating non-ideal trajectories while preserving numerical stability and physical transparency.

% ----------------------------------------------------------------------------------------------------
% Introduction of IRM
To address this need, we introduce the \emph{Inertial Replacement Method} (IRM), a semi-analytic framework designed for mirror trajectories relevant to analog Hawking radiation. 
Such trajectories typically exhibit a characteristic structure: they are asymptotically inertial in the far past and approach a null worldline in the far future, ensuring the formation of an effective horizon analogous to the event horizon of a black hole. 
The IRM exploits this structure by replacing the asymptotic segments with analytically tractable inertial counterparts, while numerically resolving only the finite accelerating interval.
This decomposition reduces oscillatory contributions from asymptotic regions that would otherwise introduce analytic complexity or numerical instability and therefore provides a systematic and efficient procedure for computing Bogoliubov coefficients for both perfect and imperfect mirrors. 
The method is applicable to realistic, PIC-motivated trajectories such as the AnaBHEL profile~\cite{2020_Pisin}, which are expected to be relevant for future experimental implementations.

% ----------------------------------------------------------------------------------------------------
% Paper structure
Sec.~\ref{sec:Canonical_Approach} reviews the canonical moving-mirror formalism for perfectly and imperfectly reflecting mirrors.  
Sec.~\ref{sec:IRM_pf} introduces the IRM, applies it to perfectly reflecting mirror trajectories, and derives the corresponding bound on the induced deviation.
Sec.~\ref{sec:IRM_impf} extends the IRM framework to imperfectly reflecting mirrors and derives the associated bound.
Sec.~\ref{sec:IRM_impf_Chen_Trajectory} applies the IRM to a more realistic and experimentally motivated trajectory extracted from PIC simulations.  
Sec.~\ref{sec:Conclusion} summarizes the results and insights obtained from the IRM and discusses their implications for analog Hawking radiation and future directions.
Appx.~\ref{app:OPB_bound} and Appx.~\ref{app:DMB_bound} present the derivations of two distinct mathematical bounds on the integrals used in this work.  
Appx.~\ref{app:Bound_Summary} summarizes all IRM-induced bounds for perfectly and imperfectly reflecting mirrors.

Throughout this paper we work in natural units, $G=\hbar=c=1$, except in Sec.~\ref{sec:IRM_impf_Chen_Trajectory}, where physical units are retained for direct comparison with simulation data.

% ====================================================================================================
\section{Canonical Approach to Analog Hawking Radiation}
\label{sec:Canonical_Approach}
In this section we briefly review the standard formalism of the moving-mirror model in (1+1)D Minkowski spacetime and introduce the Bogoliubov coefficients that quantify particle creation induced by a prescribed mirror trajectory. 
The formalism for perfectly reflecting mirrors can be found in~\cite{1976_FullingDavies,1977_DaviesFulling}, Sec.~4.4 of~\cite{1982_BirrellDavies}, and Sec.~II of~\cite{2013_Good}. 
The extension to imperfectly reflecting mirrors is discussed in~\cite{2021_Lin,2024_Lin}.

% ----------------------------------------------------------------------------------------------------
\subsection{$(1+1)$D Perfectly Reflecting Moving Mirror}
% ----------------------------------------------------------------------------------------------------
% Moving mirror model
We begin with a real, free massless scalar field in $(1+1)$D Minkowski spacetime, whose Lagrangian is
\begin{equation}
\mathcal{L} = \frac{1}{2}\, \eta^{\mu\nu} \partial_\mu \Phi \partial_\nu \Phi.
\end{equation}
The field operator is expanded in frequency modes $\phi_\omega$ as
\begin{equation}\label{eq:pf_field_expansion}
\Phi(x)= \int_0^\infty d\omega \left[ a_{\omega}\, \phi_{\omega}(x) + a^{\dagger}_{\omega}\, \phi^{*}_{\omega}(x) \right],
\end{equation}
where each mode function satisfies the equation of motion
\begin{equation}\label{eq:pf_EOM}
\partial^\mu \partial_\mu \phi_{\omega}
= \left(\partial_t^2 - \partial_z^2\right)\phi_{\omega}
= 0.
\end{equation}
To model a perfectly reflecting mirror, we impose a Dirichlet boundary condition along its time-dependent trajectory $z(t)$,
\begin{equation}\label{eq:pf_BC}
\phi_{\omega}(t,z)\big|_{z=z(t)} = 0.
\end{equation}
This condition encodes the mirror's dynamics and ensures that no transmitted component can exist across a perfectly reflecting boundary.

\begin{figure}[!htbp]
    \includegraphics[width=1\linewidth]{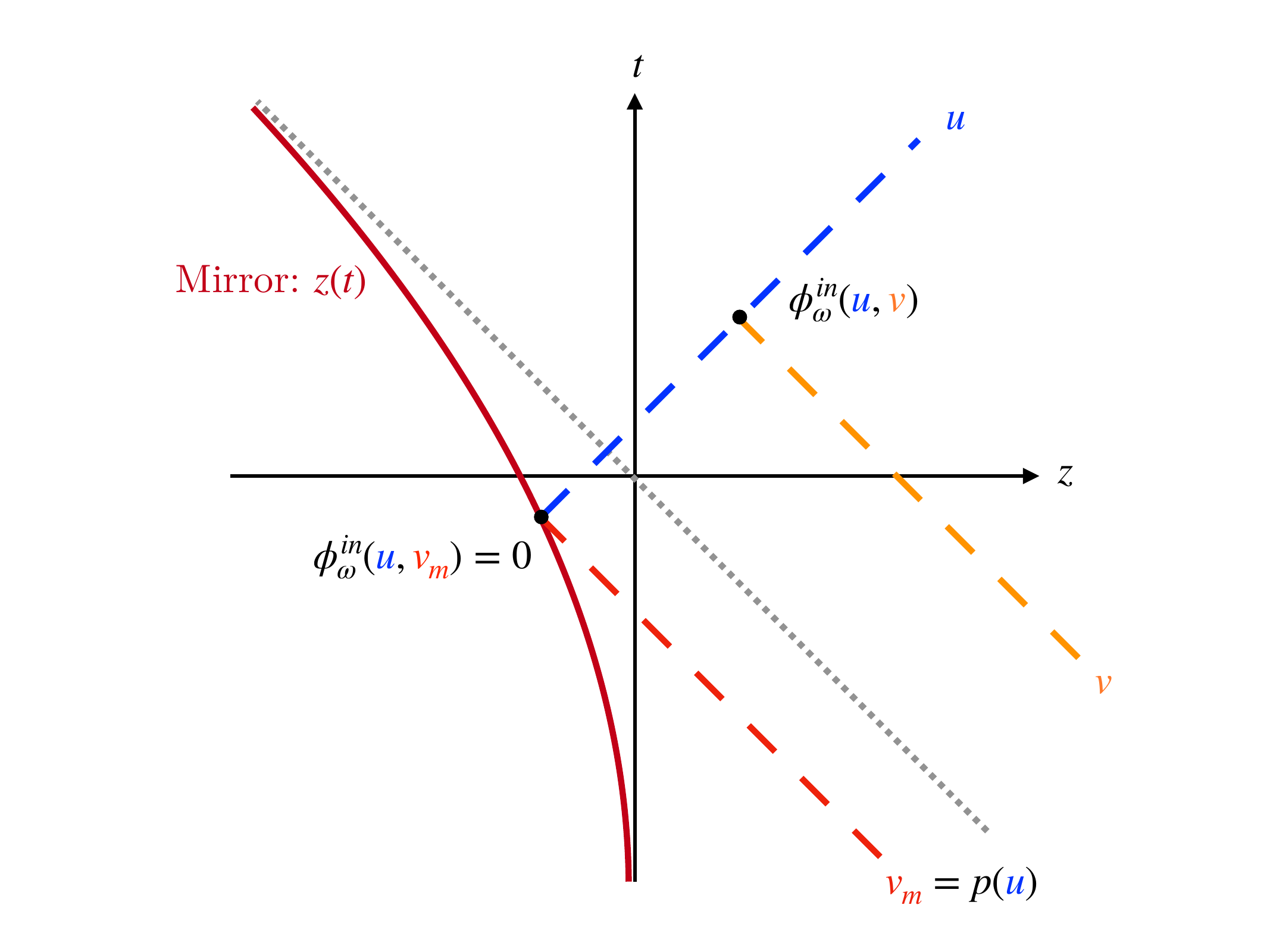}
    \caption{
        Schematic illustration of the ray-tracing functions \(v_m = p(u)\) and \(u_m = f(v)\).
        The field $\phi_\omega^{\mathrm{in}}(u,v)$ is causally connected with modes \(v\) and \(v_m=p(u)\) through reflection from the mirror.
    }
    \label{fig:pf_field_solution}
\end{figure}

% ----------------------------------------------------------------------------------------------------
% Mode solution
The solutions of Eq.~\eqref{eq:pf_EOM} satisfying the boundary condition Eq.~\eqref{eq:pf_BC} can be written in terms of null coordinates \(u = t - z\) and \(v = t + z\) as
\begin{subequations}
\begin{align}\label{eq:pf_mode_solution_in}
\phi_\omega^{\mathrm{in}}(u,v)
&=
\frac{i}{\sqrt{4\pi\omega}}
\left[
e^{-i\omega v}
-
e^{-i\omega p(u)}
\right],
\\
\label{eq:pf_mode_solution_out}
\phi_\omega^{\mathrm{out}}(u,v)
&=
\frac{i}{\sqrt{4\pi\omega}}
\left[
e^{-i\omega f(v)}
-
e^{-i\omega u}
\right].
\end{align}
\end{subequations}
The labels “in” and “out” reflect the fact that these modes behave like plane waves at past and future null infinity ($\mathcal{I}^-$ and $\mathcal{I}^+$), respectively.
The functions $p(u)$ and $f(v)$ are the ray-tracing functions that describe how null rays are reflected by the mirror trajectory, mapping incoming null coordinates to outgoing ones.
For example, for fixed $u$, the coordinate $v_m = p(u)$ is obtained by tracing the constant-$u$ null ray backward until it intersects the mirror trajectory. Explicitly, one finds
\begin{subequations}
\begin{align}
v_m = p(u) &= 2 t_m(u) - u,
\\
u_m = f(v) &= 2 t_m(v) - v,
\end{align}
\end{subequations}
where $t_m(u)$ and $t_m(v)$ denote the coordinate time at which the corresponding null ray intersects the mirror, as illustrated in Fig.~\ref{fig:pf_field_solution}. 

% ----------------------------------------------------------------------------------------------------
% Bogoliubov coefficient
The field expansion in Eq.~\eqref{eq:pf_field_expansion} may be written in either the in- or out-mode basis.
These two sets of modes are related by a Bogoliubov transformation,
\begin{equation}
\phi^{\mathrm{out}}_{\omega}
=
\int_{0}^{\infty} d\omega'
\left[
\alpha_{\omega\omega'}\,\phi^{\mathrm{in}}_{\omega'}
+
\beta_{\omega\omega'}\,\phi^{\mathrm{in}*}_{\omega'}
\right],
\end{equation}
where $\alpha_{\omega\omega'}$ and $\beta_{\omega\omega'}$ are the Bogoliubov coefficients that characterize the overlap between positive--positive and positive--negative frequency components, respectively.
\revB{
A non-inertial mirror trajectory induces mixing between positive- and negative-frequency modes, so that the in-vacuum is no longer the vacuum state with respect to the out modes.
The expectation value of the out-mode number operator in the in-vacuum is therefore
\begin{equation}
N_\omega=\int d\omega'\,|\beta_{\omega\omega'}|^{2},
\end{equation}
which gives the number spectrum of particles of frequency $\omega$ created by the non-inertial motion of the mirror.
In general, particle creation is not generated uniformly along the mirror trajectory.
Instead, different portions of the mirror motion may contribute differently to $\beta_{\omega\omega'}$, depending on the frequencies involved. 
For example, for an asymptotically null trajectory, the regime $\omega'\gg\omega$ is generally associated with the asymptotic-null portion of the trajectory, where extremely large redshifts are generated.
}

Evaluating the Klein--Gordon inner product between in- and out-modes yields
\begin{align}
\beta_{\omega\omega'}
&=
-\left( \phi^{\mathrm{out}}_{\omega},\, \phi^{\mathrm{in}*}_{\omega'} \right)
\notag \\
&=
\frac{1}{4\pi\sqrt{\omega\omega'}}
\int_{-\infty}^{\infty} dt\,
\big[ \omega_{+}\dot{z}(t) - \omega_{-} \big]\,
e^{\, i[\omega_{+} t - \omega_{-} z(t)]},
\label{eq:pf_beta}
\end{align}
where $\omega_{\pm} \equiv \omega \pm \omega'$.
Eq.~\eqref{eq:pf_beta} is one of two central integrals studied in this paper to investigate the particle number between two frequency modes of a perfectly reflecting mirror.

% ----------------------------------------------------------------------------------------------------
% Trivial case
As a consistency check, consider an eternally inertial mirror,
\[
z(t) = z_0 + \dot{z}_i (t - t_0).
\]
Substituting this trajectory into Eq.~\eqref{eq:pf_beta} gives
\begin{align}
\beta_{\omega\omega',i}
&=
\frac{1}{4\pi\sqrt{\omega\omega'}}
\big( \omega_+ \dot{z}_i - \omega_- \big)
e^{-i\omega_- (z_0 - \dot{z}_i t_0)} \notag \\
& \quad \times \, 2\pi\,
\delta_D\!\big[ \omega(1-\dot{z}_i) + \omega'(1+\dot{z}_i) \big]
= 0,
\end{align}
since the argument of the Dirac delta function cannot vanish for positive frequencies $\omega,\omega' > 0$. 
Thus, as expected, uniform (inertial) motion does not produce particles.
This result implies that particle creation arises only when the mirror undergoes non-uniform, time-dependent acceleration (redshifts) that mixes positive and negative frequency components of the field.

% ----------------------------------------------------------------------------------------------------
\subsection{$(1+1)$D Imperfectly Reflecting Moving Mirror}
% ----------------------------------------------------------------------------------------------------
% Moving mirror model
To describe a mirror with finite reflectivity, we follow the Barton--Calogeracos (BC) model~\cite{1995_BartonCalogeracos,2021_Lin}, in which a scalar field interacts with a delta-function potential localized on the mirror's worldline.
In $(1+3)$D Minkowski spacetime, the BC action for a massless scalar field is
\begin{align}
S_{\alpha}[\phi]
=
&-\frac{1}{2}\int d^{4}x\, \partial^{\mu}\phi\,\partial_{\mu}\phi \notag \\
&-\;
\frac{\alpha}{2}\int d^{4}x\,
\gamma^{-1}(t)\,\delta \big(x_{3}-z(t)\big)\,\phi^{2}(x),
\end{align}
where \(z(t)\) denotes the mirror trajectory, and \(\gamma(t)\equiv (1-\dot{z}^{\,2})^{-1/2}\) is the Lorentz factor.
The coupling constant \(\alpha\) (with dimensions of inverse length) characterizes the mirror’s reflectivity:
\begin{itemize}
    \item \(\alpha\to\infty\): perfectly reflecting mirror (Dirichlet limit),
    \item \(\alpha\) finite: imperfectly reflecting (semi-transparent) mirror,
    \item \(\alpha\to 0\): completely transparent mirror with no reflection.
\end{itemize}
Thus, the BC model interpolates smoothly between perfect reflection and complete transmission, allowing realistic modeling of plasma mirrors or dielectric interfaces.
Varying the action gives the equation of motion
\begin{equation}\label{eq:impf_EOM}
\partial^{\mu}\partial_{\mu}\phi(x)
=
\alpha\,\gamma^{-1}(t)\,
\delta\big(x_{3}-z(t)\big)\,\phi(x),
\end{equation}

Following the analysis in~\cite{2021_Lin}, the leading-order Bogoliubov coefficient in \((1+3)\)D can be obtained using the Born approximation.  
For our current purpose of developing a general method for evaluating Bogoliubov coefficients for generic trajectories, we focus on the $1+1$D limit by suppressing transverse momenta and considering a mirror trajectory \(z(t)\) in \(1+1\) dimensions.
Then the Bogoliubov coefficient becomes
\begin{equation}
\beta_{\omega\omega'}
=
\frac{-i\alpha}{4\pi\sqrt{\omega\,\omega'}}
\int_{-\infty}^{\infty} dt\;
\frac{1}{\gamma[z(t)]}\,
e^{i\left[\omega_{+}t - \omega_{-}z(t)\right]}
\label{eq:impf_beta}
\end{equation}
where \(\omega_{\pm}\equiv\omega\pm\omega'\). Eq.~\eqref{eq:impf_beta} is the second central integral on which this work focuses.
Its structure closely resembles the Bogoliubov coefficient for a perfectly reflecting mirror, with two key differences.
First, the integrand contains the Lorentz factor $\gamma^{-1}(t)$, which suppresses contributions when the mirror approaches relativistic speeds.
Second, the overall magnitude of mode mixing is controlled by the coupling constant $\alpha$: smaller values of $\alpha$ correspond to weaker reflectivity and therefore reduced particle production.
In the limit $\alpha \to \infty$, the delta-function potential enforces $\phi(z(t))=0$ in Eq.~\eqref{eq:impf_EOM}, recovering the Dirichlet boundary condition of a perfectly reflecting mirror.

% ====================================================================================================
\section{IRM for a Perfect Mirror}
\label{sec:IRM_pf}
In this section, we introduce the Inertial Replacement Method (IRM) in the context of a perfectly reflecting mirror. 
The goal is to present the core ideas and overall procedure of the IRM under its essential requirement: that the mirror trajectory is asymptotically inertial in the far past and asymptotically null in the far future.
The method, however, is not restricted to perfect reflectors \revA{or a particular mirror trajectory}; it can also be applied to imperfectly reflecting mirrors provided that this asymptotic condition is satisfied.

For a given trajectory \(z(t)\), the IRM consists of the following steps.
\begin{enumerate}
    \item Choose a threshold acceleration \(a_{\rm thres}\) and determine the splitting temporal boundaries \(t_A\) and \(t_B\) such that \( |\ddot z(t)| \le a_{\rm thres} \) holds for \( t \le t_A \) and \( t \ge t_B \)  
    (Sec.~\ref{subsec:Determine_the_Splitting_Boundary}).
    
    \item Divide the Bogoliubov integral into three segments:  
    Region I in the far past, Region II in the accelerating interval, and Region III in the far future  
    (Sec.~\ref{subsec:Determine_the_Splitting_Boundary}).
    
    \item Replace the asymptotic segments in Regions I and III with the inertial extensions and evaluate their contributions analytically  
    (Sec.~\ref{subsec:Inertial_Replacement_in_Region_I_and_III}).
    
    \item Derive perturbatively the bounds on the deviations induced by the replacements in Regions I and III
    (Sec.~\ref{subsec:Bound_on_IRM-Deviation_in_Region_III},  
    Sec.~\ref{subsec:Bound_on_IRM-Deviation_in_Region_I},  
    Sec.~\ref{subsec:Total_Bound_on_the_Particle-Number_Deviation}).
    
    \item Combine the analytic contributions from Regions I and III with the numerical evaluation in Region II  
    (Sec.~\ref{subsec:Approaching_the_Exact_Result}).
    
    \item Increase \(1/a_{\rm thres}\) so that \(t_A\) and \(t_B\) move deeper into the asymptotic region until convergence is reached
    (Sec.~\ref{subsec:Approaching_the_Exact_Result}).
\end{enumerate}

\begin{figure}[!htbp]
    \centering
    \includegraphics[width=0.8\linewidth]{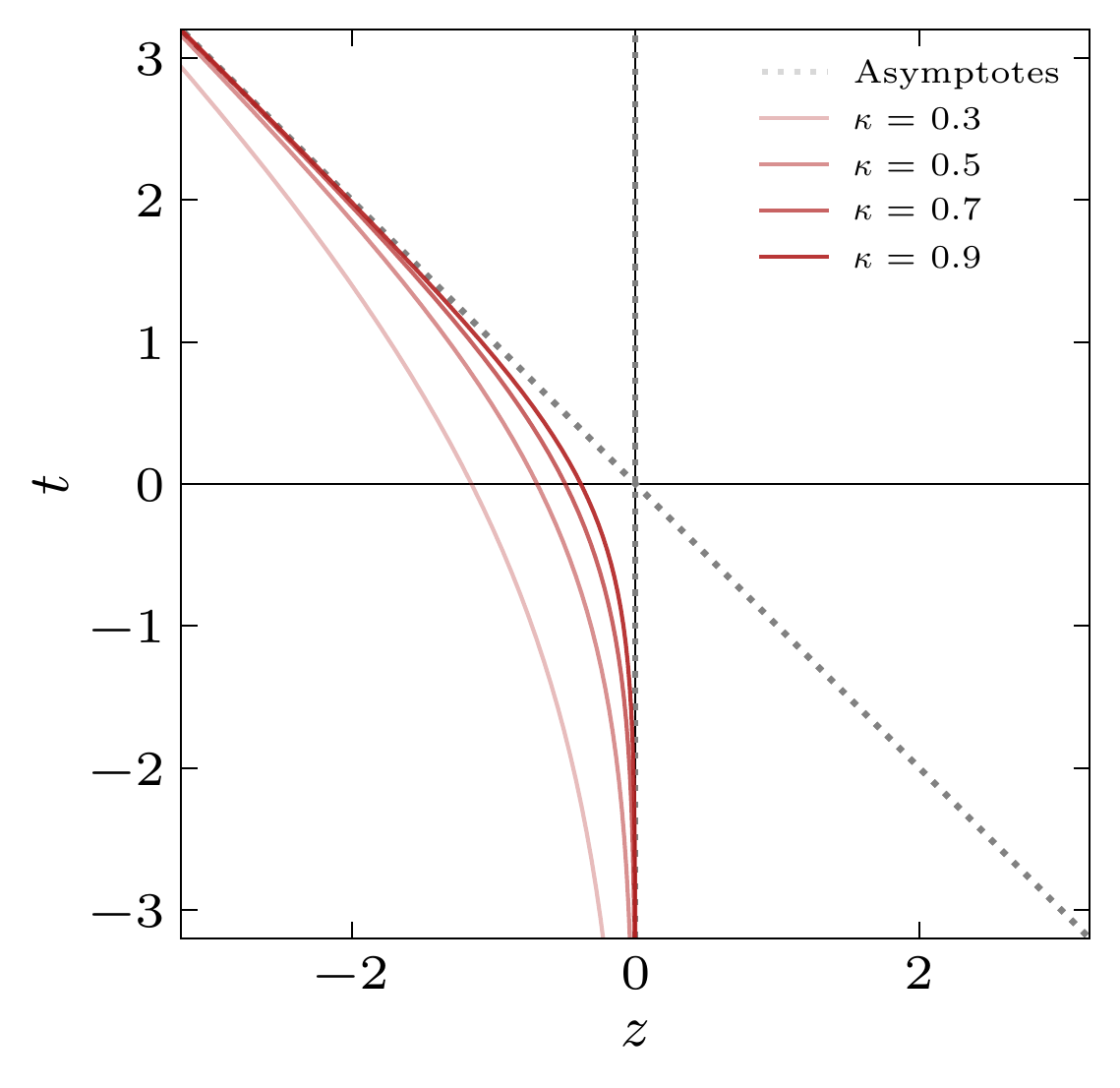}
    \caption{
    Logex trajectories for representative values of \(\kappa\). 
    The trajectory is asymptotically static in the far past and approaches a null worldline in the far future.
    }
    \label{fig:logex_traj}
\end{figure}

To illustrate the method concretely, we adopt the Logex trajectory as a working example throughout this section. 
The trajectory is defined by
\begin{equation}
z(t) = -\frac{1}{2\kappa}\,\ln\!\left(e^{2\kappa t}+1\right),
\label{eq:logex_traj}
\end{equation}
where the parameter \(\kappa\) sets the acceleration scale, playing a role analogous to the surface gravity in the black-hole case, and determines how rapidly the mirror approaches its asymptotic null trajectory.
Representative trajectories for several values of \(\kappa\) are shown in Fig.~\ref{fig:logex_traj}.

A key advantage of the Logex trajectory is that the associated Bogoliubov coefficient admits a closed-form analytic expression, derived in Sec.~III of~\cite{2011_Good}:
\begin{equation}
    \beta^{\mathrm{Ext}}_{\omega \omega'}
    =
    \frac{-\omega\,\omega'}
    {2\pi\sqrt{\omega\omega'}\,\kappa\,(\omega'-\omega)}
    \frac{
        \Gamma\!\left(-i\dfrac{\omega}{\kappa}\right)
        \Gamma\!\left(i\dfrac{\omega+\omega'}{2\kappa}\right)
    }{
        \Gamma\!\left(i\dfrac{\omega'-\omega}{2\kappa}\right)
    }.
    \label{eq:logex_beta_analytic}
\end{equation}
This analytic result of the exact (“Ext”) trajectory serves as a benchmark for validating the accuracy and convergence properties of the IRM.

% ----------------------------------------------------------------------------------------------------
\subsection{Determine the Splitting Boundary}
\label{subsec:Determine_the_Splitting_Boundary}
% ----------------------------------------------------------------------------------------------------
Since the mirror trajectory is asymptotically inertial in the far past and asymptotically null in the far future, its acceleration measured in the Lab frame vanishes in these limits,
\begin{equation}
    \lim_{t\to -\infty} \ddot{z}(t) = 0,
    \qquad
    \lim_{t\to +\infty} \ddot{z}(t) = 0 .
\end{equation}

To isolate the genuinely non-inertial portion of the motion, we introduce a threshold acceleration \(a_{\text{thres}}\).  
The boundaries \(t_A\) and \(t_B\) are defined implicitly by the condition
\begin{equation}
    |\ddot z(t)| > a_{\text{thres}}
    \quad \Longleftrightarrow \quad
    t \in (t_A,\, t_B),
    \label{eq:splitting_condition}
\end{equation}
so that the interval \((t_A,t_B)\) identifies the central accelerating region.

\begin{figure*}[!htbp]
    \centering
    \includegraphics[width=0.49\linewidth]{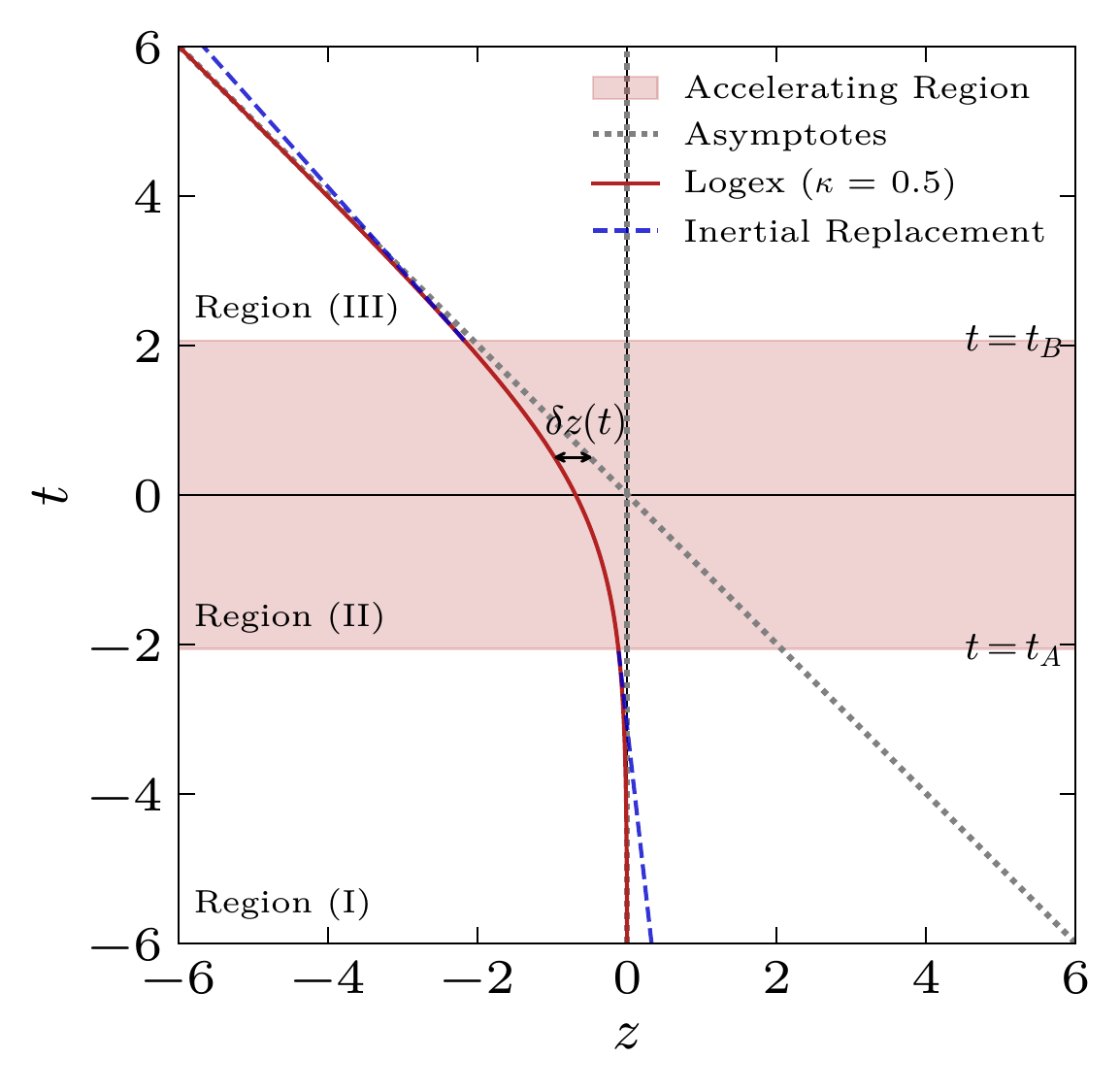}
    \includegraphics[width=0.49\linewidth]{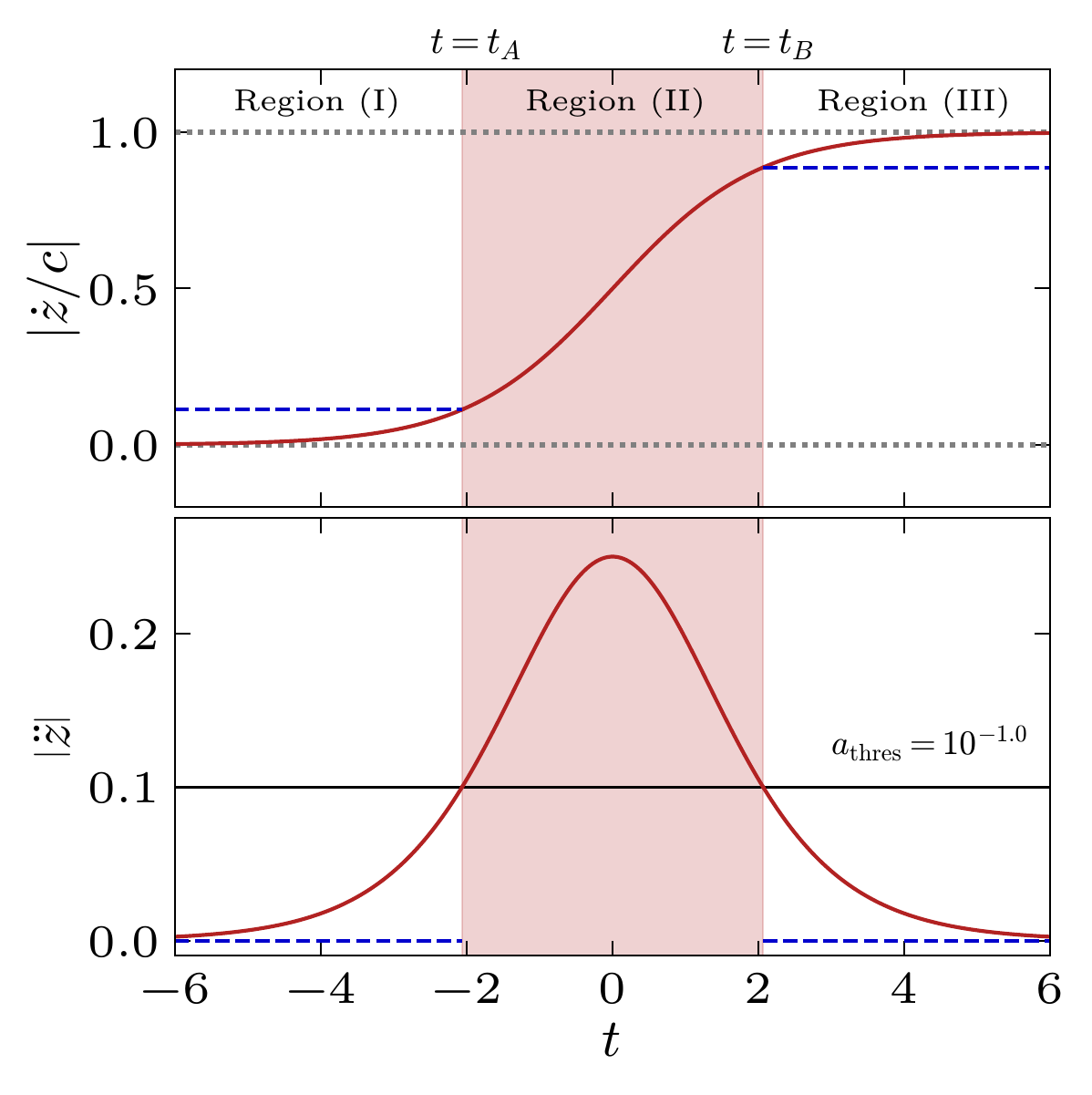}
    \caption{
    \textbf{Left}: splitting the \textbf{Logex} trajectory into asymptotic tails and a central accelerating region using the acceleration threshold \(a_{\text{thres}}\), together with the inertial replacement. 
    \textbf{Right}: velocity and acceleration profiles illustrating the threshold condition that defines \(t_A\) and \(t_B\). 
    }
    \label{fig:logex_dynamics_combined}
\end{figure*}

This divides the temporal domain into three segments,
\begin{align}
    & \text{Region I: } t < t_A, \notag \\
    & \text{Region II: } t_A < t < t_B, \notag  \\
    & \text{Region III: } t > t_B . 
\end{align}
\revA{The threshold acceleration $a_{\rm thres}$ determines the extent of the central accelerating region retained in Region II.
Smaller values of $a_{\rm thres}$ move the matching boundaries $t_A$ and $t_B$ deeper into the asymptotic regions and therefore enlarge Region II.}
An example of this splitting for the Logex trajectory is shown in Fig.~\ref{fig:logex_dynamics_combined}, where the accelerating region is determined by the chosen threshold.

Applying this decomposition to the Bogoliubov coefficient in Eq.~\eqref{eq:pf_beta}, we obtain
\begin{align}
    \beta^{\mathrm{Ext}}_{\omega\omega'}
    &= \int_{-\infty}^{\infty} I_{\omega \omega'}(t)\, dt \nonumber \\
    &= \int_{-\infty}^{t_A} I_{\omega \omega'}(t)\, dt
     + \int_{t_A}^{t_B} I_{\omega \omega'}(t)\, dt
     + \int_{t_B}^{\infty} I_{\omega \omega'}(t)\, dt \nonumber \\
    &\equiv
       \beta^{(\mathrm{I})}_{\omega\omega'}
     + \beta^{(\mathrm{II})}_{\omega\omega'}
     + \beta^{(\mathrm{III})}_{\omega\omega'} .
    \label{eq:beta_exact}
\end{align}
The contributions \(\beta^{(\mathrm{I})}_{\omega\omega'}\) and \(\beta^{(\mathrm{III})}_{\omega\omega'}\) arise from regions where the acceleration lies below the threshold and will later be replaced by inertial extensions within the IRM framework.  
The term \(\beta^{(\mathrm{II})}_{\omega\omega'}\) corresponds to the finite accelerating segment and will be evaluated numerically.

% ----------------------------------------------------------------------------------------------------
\subsection{Inertial Replacement in Regions I and III}
\label{subsec:Inertial_Replacement_in_Region_I_and_III}
% ----------------------------------------------------------------------------------------------------
The key observation underlying the IRM is that, in the far past and far future, the mirror trajectory becomes asymptotically inertial and null (see Fig.~\ref{fig:logex_dynamics_combined}).
In these regions the acceleration has already decayed to zero, so the exact trajectory may be approximated by an inertial line constructed at the splitting boundaries.
Since the integrand of the Bogoliubov coefficient depends only on \(z(t)\) and \(\dot{z}(t)\), replacing them by their inertial counterparts converts the integrand into a simple exponential with constant prefactors.
By construction, both \(z(t)\) and \(\dot{z}(t)\) match continuously at \(t_A\) and \(t_B\), \revA{so the inertial replacement introduces no singular contributions} and the substituted integrand remains well defined.
This eliminates the long-time oscillatory structure induced by residual acceleration and renders the infinite-domain integrals analytically tractable.

We define the IRM approximation to the Bogoliubov coefficient as
\begin{align}
    \beta^{\mathrm{IRM}}_{\omega\omega'}
    \equiv
    \beta^{(\mathrm{I})}_{\omega\omega',i}
    + \beta^{(\mathrm{II})}_{\omega\omega',N}
    + \beta^{(\mathrm{III})}_{\omega\omega',i}.
    \label{eq:beta_IRM}
\end{align}
Here \( \beta^{(\mathrm{I})}_{\omega\omega',i} \) and \( \beta^{(\mathrm{III})}_{\omega\omega',i} \) denote the analytic contributions obtained from inertial replacements in Regions~I and~III, while \( \beta^{(\mathrm{II})}_{\omega\omega',N} \) is the numerical contribution from the finite accelerating interval.
Compared with the exact decomposition in Eq.~\eqref{eq:beta_exact}, the contributions \( \beta^{(\mathrm{I})}_{\omega\omega'} \) and \( \beta^{(\mathrm{III})}_{\omega\omega'} \) are generally not available in closed form, whereas their inertial counterparts \( \beta^{(\mathrm{I})}_{\omega\omega',i} \) and \( \beta^{(\mathrm{III})}_{\omega\omega',i} \) admit explicit analytic expressions.
Eq.~\eqref{eq:beta_IRM} converges to the exact result in Eq.~\eqref{eq:beta_exact} as \( t_A \to -\infty \) and \( t_B \to +\infty \). 
In this limit the accelerating region covers the entire temporal domain, the inertial replacements become irrelevant, and the IRM-induced deviation vanishes. 
The IRM is therefore systematically improvable: enlarging the accelerating region recovers the exact result.

The next step is to explicitly evaluate $\beta^{(\mathrm{I})}_{\omega\omega',i}$ and $\beta^{(\mathrm{III})}_{\omega\omega',i}$.
First, in Region~III, the trajectory is replaced by
\begin{equation}
    z^{(\mathrm{III})}_i(t)
    = z_B + \dot{z}_B (t - t_B),
    \label{eq:traj_inertial_III}
\end{equation}
where \(z_B\) and \(\dot{z}_B\) are evaluated at \(t_B\). 
Substituting Eq.~\eqref{eq:traj_inertial_III} into Eq.~\eqref{eq:pf_beta} yields
\begin{align}
    \beta^{(\mathrm{III})}_{\omega \omega', i}
    &= \frac{1}{4\pi\sqrt{\omega \omega'}} 
    \int_{t_B}^{\infty} dt \, \left( \omega_+ \dot{z}_B - \omega_- \right)
    \, e^{i\big[\omega_+ t - \omega_- z^{(\mathrm{III})}_i(t)\big]} \nonumber \\
    &=
    \frac{i}{4\pi\sqrt{\omega \omega'}} \,
    \frac{\omega_+ \dot{z}_B - \omega_-}{\omega_+ - \omega_-\dot{z}_B} \,
    e^{i\left( \omega_+ t_B - \omega_- z_B\right)} 
    .
    \label{eq:pf_beta_inertial_III}
\end{align}
In the last step we used the standard \( i\epsilon \)-prescription to ensure the integral at \( t\to\infty \) vanishes.

Similarly, in Region~I, we have
\begin{equation}
    z^{(\mathrm{I})}_i(t)
    = z_A + \dot{z}_A (t - t_A),
    \label{eq:traj_inertial_1}
\end{equation}
and
\begin{align}
    \beta^{(\mathrm{I})}_{\omega \omega', i}
    &= 
    -\frac{i}{4\pi\sqrt{\omega \omega'}} \,
    \frac{\omega_+ \dot{z}_A - \omega_-}{\omega_+ - \omega_-\dot{z}_A} \,
    e^{i\left( \omega_+ t_A - \omega_- z_A\right)}
    .
    \label{eq:pf_beta_inertial_I}
\end{align}
Notice the minus sign arising from the reversed integration limits compared to Region~III.
These closed-form expressions demonstrate the essential advantage of the IRM: infinite-domain integrals reduce to simple analytic factors, while only the finite accelerating segment requires numerical evaluation.

We now quantify the deviation introduced by the IRM \revA{by defining the deviation of the Bogoliubov coefficient as the difference between the exact solution and the IRM result},
\small{
\begin{align}
\delta\beta_{\omega\omega'} 
&\equiv 
\beta^{\text{Ext}}_{\omega\omega'} - \beta^{\text{IRM}}_{\omega\omega'} 
\notag \\
&=
\left( \beta^{(\mathrm{I})}_{\omega\omega'} + \beta^{(\mathrm{II})}_{\omega\omega'} + \beta^{(\mathrm{III})}_{\omega\omega'} \right)
-
\left( \beta^{(\mathrm{I})}_{\omega\omega',i} + \beta^{(\mathrm{II})}_{\omega\omega',N} + \beta^{(\mathrm{III})}_{\omega\omega',i} \right)
\notag \\
&\equiv
\delta\beta^{(\mathrm{I})}_{\omega\omega'} + \delta\beta^{(\mathrm{III})}_{\omega\omega'}.
\label{eq:delta_beta_sum}
\end{align}
}
\revA{This deviation captures all effects introduced by the inertial replacement. 
In particular, if the replacement were to generate additional singular contributions, they would necessarily be contained in $\delta\beta_{\omega \omega'}$.}
Here we temporarily set \YS{\( \delta\beta^{(\mathrm{II})}_{\omega\omega'} \equiv \beta^{(\mathrm{II})}_{\omega\omega'}  - \beta^{(\mathrm{II})}_{\omega\omega',N} = 0 \),} as this term represents numerical integration error in Region~II rather than the intrinsic IRM-induced deviation. 
Such numerical errors depend on algorithmic precision and can be systematically reduced. 
Our goal in this section is to isolate the IRM-induced deviation itself; numerical implementation errors will be addressed later when analyzing explicit examples.

Although the exact quantities \( \beta^{(\mathrm{I})}_{\omega\omega'} \) and \( \beta^{(\mathrm{III})}_{\omega\omega'} \) are generally not available in closed form, the asymptotic structure of the trajectory in Regions~I and~III allows one to derive analytic bounds on their magnitudes. 
These bounds provide rigorous estimates for \( \Big|\delta\beta^{(\mathrm{I})}_{\omega\omega'}\Big| \) and \( \Big|\delta\beta^{(\mathrm{III})}_{\omega\omega'}\Big| \), establishing quantitative control over the IRM-induced deviation. 

% ----------------------------------------------------------------------------------------------------
\subsection{Bound on IRM Deviation in Region III}
\label{subsec:Bound_on_IRM-Deviation_in_Region_III}
% ----------------------------------------------------------------------------------------------------
The goal of this subsection is to derive an analytic bound on the IRM-induced deviation in Region~III,
\begin{align}
\Big|\delta \beta^{(\mathrm{III})}_{\omega\omega'}\Big|
\equiv 
\Big|\beta^{(\mathrm{III})}_{\omega\omega'} - \beta^{(\mathrm{III})}_{\omega\omega',i}\Big|,
\label{eq:delta_beta_III}
\end{align}
using only the asymptotic structure of the trajectory and the boundary data at \(t=t_B\).

We begin with the exact expression in Region~III,
\begin{align}
    \beta^{(\mathrm{III})}_{\omega\omega'}
    = \frac{1}{4\pi\sqrt{\omega \omega'}}
    \int_{t_B}^{\infty} dt\,
    \big[\omega_+ \dot z^{(\mathrm{III})}(t) - \omega_-\big]\,
    e^{\,i\omega_+ t - i\omega_- z^{(\mathrm{III})}(t)}.
    \label{eq:pf_beta_exact_3}
\end{align}
We introduce a small parameter that allows a perturbative expansion of the integrand.
To do so, we take advantage of the asymptote of the exact trajectory,
\begin{equation}
    z^{(\mathrm{III})}_{\mathrm{asym}}(t)
    = z'_B + \dot z^{(\mathrm{III})}_{\mathrm{asym}}(t - t_B),
    \label{eq:traj_asym_3}
\end{equation}
which the mirror approaches as $t\to\infty$, where $z'_B$ denotes the value of the asymptote evaluated at $t=t_B$.

We then decompose the exact trajectory into its asymptotic part and a deviation:
\begin{align}
    z^{(\mathrm{III})}(t)
    &\equiv z^{(\mathrm{III})}_{\mathrm{asym}}(t) - \delta z^{(\mathrm{III})}(t) \notag \\
    &= z'_B + \dot z^{(\mathrm{III})}_{\mathrm{asym}}(t - t_B) - \delta z^{(\mathrm{III})}(t) \notag \\
    &= C_B + \dot z^{(\mathrm{III})}_{\mathrm{asym}} t - \delta z^{(\mathrm{III})}(t),
    \label{eq:traj_exact_3}
\end{align}
where we have defined the constant
\begin{equation}
    C_B \equiv z'_B - \dot z^{(\mathrm{III})}_{\mathrm{asym}} t_B .
    \label{eq:constant_B}
\end{equation}
The function $\delta z^{(\mathrm{III})}(t)$ quantifies the deviation of the exact trajectory from its asymptote, and by definition, $\delta z^{(\mathrm{III})}(t)\to 0$ as $t\to\infty$.
The corresponding velocity is
\begin{equation}
    \dot z^{(\mathrm{III})}(t)
    = \dot z^{(\mathrm{III})}_{\mathrm{asym}}
    - \delta\dot z^{(\mathrm{III})}(t).
    \label{eq:vel_exact_3}
\end{equation}

By inserting Eqs.~\eqref{eq:traj_exact_3} and \eqref{eq:vel_exact_3} into \eqref{eq:pf_beta_exact_3} and subtracting $\beta^{(\mathrm{III})}_{\omega\omega',i}$, we obtain
\begin{widetext}
\begin{align}
    \delta \beta^{(\mathrm{III})}_{\omega\omega'} 
    &= \frac{1}{4\pi\sqrt{\omega \omega'}}
    \int_{t_B}^{\infty} dt\,
    \Big[\omega_+ \big( \dot z^{(\mathrm{III})}_{\mathrm{asym}} - \delta\dot z^{(\mathrm{III})}(t) \big)
    - \omega_- \Big]\,
    e^{\,i\omega_+ t - i\omega_- \big[C_B + \dot z^{(\mathrm{III})}_{\mathrm{asym}} t - \delta z^{(\mathrm{III})}(t)\big]}
    -\beta^{(\mathrm{III})}_{\omega\omega',i}
    \notag \\
    &=
    \frac{1}{4\pi\sqrt{\omega \omega'}}
    e^{-i\omega_- C_B}
    \int_{t_B}^{\infty} dt\,\Big[\big(\omega_+ \dot z^{(\mathrm{III})}_{\mathrm{asym}} - \omega_- \big)
    - \omega_+\,\delta\dot z^{(\mathrm{III})}(t)\Big]\,
    e^{\,i (\omega_+ - \omega_- \dot z^{(\mathrm{III})}_{\mathrm{asym}}) t}
    e^{\,i\omega_- \delta z^{(\mathrm{III})}(t)}
    -\beta^{(\mathrm{III})}_{\omega\omega',i}
    \notag \\
    % &\approx
    % \frac{1}{4\pi\sqrt{\omega \omega'}}
    % e^{-i\omega_- C_B}
    % \int_{t_B}^{\infty} dt\,
    % \Big[
    %     \Omega^{(\mathrm{III})}_{1}
    %     - \omega_+\,\delta\dot z^{(\mathrm{III})}(t)
    % \Big]
    % \big[1 + i\omega_- \delta z^{(\mathrm{III})}(t)\big]\,
    % e^{\,i\Omega^{(\mathrm{III})}_{2} t}
    % \notag \\
    & \approx
    \frac{1}{4\pi\sqrt{\omega \omega'}}
    \left[\,\underbrace{i\dfrac{ \Omega^{(\mathrm{III})}_{1}}{ \Omega^{(\mathrm{III})}_{2}} e^{-i \omega_- C_B} e^{i \Omega^{(\mathrm{III})}_{2} t_B}
    -4\pi\sqrt{\omega \omega'} \beta^{(\mathrm{III})}_{\omega\omega',i}}_{\text{zeroth-order}}
    \;+
    e^{-i \omega_- C_B}
    \int_{t_B}^{\infty} \!\!dt\left( \underbrace{i\omega_- \Omega^{(\mathrm{III})}_{1}\,\delta z^{(\mathrm{III})}(t)}_{\text{traj.\ correction}}
    \;-\;\underbrace{\omega_+\,\delta\dot z^{(\mathrm{III})}(t)}_{\text{vel.\ correction}}\,\right)\,
    e^{i\Omega^{(\mathrm{III})}_{2} t}
    \right]
    , 
    \label{eq:pf_delta_beta_III_1st_order}
\end{align}
\end{widetext}
where in the last equation we define
\begin{subequations}
\label{eq:Omega_12_III}
\begin{align}
\Omega^{(\mathrm{III})}_{1} \equiv \omega_+ \dot z^{(\mathrm{III})}_{\mathrm{asym}} - \omega_- , \\
\Omega^{(\mathrm{III})}_{2} \equiv \omega_+ - \omega_- \dot z^{(\mathrm{III})}_{\mathrm{asym}} .
\end{align}
\end{subequations}
and expand the exponential to the first order:
\begin{equation}
    e^{\,i\omega_- \delta z^{(\mathrm{III})}(t)}
    = 1 + i\omega_- \delta z^{(\mathrm{III})}(t)
    + O\!\left(\delta z^{(\mathrm{III})}(t)^2\right),
    \label{eq:exp_expansion}
\end{equation}
since the exact trajectory approaches its asymptote for $t>t_B$, the deviation $\delta z^{(\mathrm{III})}(t)$ can be suppressed to an arbitrarily small value by choosing $t_B$ sufficiently deep in the asymptotic region. 
For large values of $\omega_-$ (the high-frequency regime), maintaining the validity of the expansion requires either pushing $t_B$ further into the asymptotic region or retaining higher-order terms in Eq.~\eqref{eq:exp_expansion}. 
In this work, we adopt the former strategy and restrict attention to the controlled first-order regime.
Eq.~\eqref{eq:pf_delta_beta_III_1st_order} exhibits a transparent structure that the zeroth-order term represents the mismatch between the inertial replacement and the asymptote of the exact trajectory, while the remaining integral contains the first-order dynamical corrections arising from $\delta z^{(\mathrm{III})}(t)$ and $\delta \dot z^{(\mathrm{III})}(t)$, which are continuous and therefore integrable over $[t_B,\infty)$.

In principle, Eq.~\eqref{eq:pf_delta_beta_III_1st_order} admits alternative representations obtained by repeatedly performing integration by parts (IBP), which extract additional boundary terms and shift the integrand to higher-order derivatives of the trajectory. 
Each additional IBP step, however, requires the corresponding boundary data to be well defined.
Within the IRM construction, continuity is imposed only on the trajectory and its first derivative at the splitting boundary \( t=t_B \). 
Higher derivatives are not guaranteed to match across the inertial replacement. 
Consequently, IBP can be carried out consistently only up to integrands involving \( \delta\ddot z^{(\mathrm{III})}(t) \); further iterations would require additional smoothness assumptions.
Moreover, when the contributions from all regions are combined, these extracted complex boundary terms can interfere destructively, leading to partial cancellations that can reduce the magnitude of the total deviation and are therefore essential for obtaining a tight bound.

% ----------------------------------------------------------------------------------------------------
To bound the integrals appearing in Eq.~\eqref{eq:pf_delta_beta_III_1st_order}, we employ two standard estimates for oscillatory integrals of the form
\begin{equation}
    I = \int_{t_0}^{\infty} f(t)\, e^{i\Omega_0 t}\, dt ,
    \label{eq:oscillatory_integral}
\end{equation}
where \(f(t)\) is assumed to be real, smooth, monotonic, and decaying as \(t\to\infty\).
Under these conditions, the following bounds hold.

% ----------------------------------------------------------------------------------------------------
\noindent\textbf{Oscillatory-Phase Bound (OPB)}  
(Dirichlet/van der Corput-type~\cite{1993_Stein}):
\begin{equation}
\left| \int_{t_0}^{\infty} f(t)\, e^{i\Omega_0 t}\, dt \right|
\;\le\;
\frac{8\, |f(t_0)|}{|\Omega_0|} ,
\label{eq:OPB}
\end{equation}
which is effective when the phase oscillates rapidly relative to the variation scale of \(f(t)\).
This estimate captures the universal \(1/|\Omega_0|\) suppression characteristic of highly oscillatory integrals.
The derivation is provided in Appendix~\ref{app:OPB_bound}.
In the present setting, OPB applies to \( f(t) = \delta z^{(\mathrm{III})}(t) \) and \( f(t) = \delta\dot z^{(\mathrm{III})}(t) \), since both are continuous at \(t_B\).
However, \( \delta\ddot z^{(\mathrm{III})}(t) \) need not be continuous at the boundary, so the assumptions required for OPB are not guaranteed in that case.

% ----------------------------------------------------------------------------------------------------
\noindent\textbf{Direct-Magnitude Bound (DMB):}
\begin{equation}
\left| \int_{t_0}^{\infty} f(t)\, e^{i\Omega_0 t}\, dt \right|
\;\le\;
\left| F(t_0) \right| ,
\qquad
F(t_0) \equiv \int_{t_0}^{\infty} f(t)\, dt ,
\label{eq:DMB}
\end{equation}
which provides a non-oscillatory estimate that remains valid even when \(|\Omega_0|\) is small, where OPB becomes ineffective.
Its derivation is given in Appendix~\ref{app:DMB_bound}.
For DMB to be useful, the tail integral \(F(t_0)\) must be explicitly computable.  
This holds for \( f(t) = \delta\dot z^{(\mathrm{III})}(t) \) and \(f(t) = \delta\ddot z^{(\mathrm{III})}(t)\), but not for \( f(t) = \delta z^{(\mathrm{III})}(t) \), whose primitive does not admit a closed form.
Accordingly, DMB is applied only in the former two cases.

For example, the simplest bound of Eq.~\eqref{eq:pf_delta_beta_III_1st_order} is obtained by applying the OPB Eq.~\eqref{eq:OPB} directly to both integrals.
We find
\begin{widetext}
\begin{align}
    & \textit{Strategy 1} \text{ (0 IBP, OPB}\to \delta z^{(\mathrm{III})}, \text{ OPB}\to \delta\dot z^{(\mathrm{III})}\text{)}: \notag \\
    &\left| \delta \beta^{(\mathrm{III})}_{\omega\omega'} \right|
    \leq
    \frac{1}{4\pi\sqrt{\omega \omega'}}
    \left(
    \left|
    \,i\dfrac{ \Omega^{(\mathrm{III})}_{1}}{ \Omega^{(\mathrm{III})}_{2}} e^{-i \omega_- C_B}e^{i \Omega^{(\mathrm{III})}_{2} t_B}
    \;
    -{4\pi\sqrt{\omega \omega'}} \beta^{(\mathrm{III})}_{\omega \omega',i}
    \right|
    +\left|
    i\omega_- \Omega^{(\mathrm{III})}_{1} 
    \int_{t_B}^{\infty} dt\, \delta z^{(\mathrm{III})}(t) e^{i\Omega^{(\mathrm{III})}_{2} t}\right|
    % \right. \notag \\
    % & \qquad \qquad \qquad \qquad \quad \left.
    +
    \left|
    \omega_+ 
    \int_{t_B}^{\infty} dt\,
    \delta\dot z^{(\mathrm{III})}(t)  e^{i\Omega^{(\mathrm{III})}_{2} t}
    \right | \right) \notag \\
    & \ \qquad \quad \leq \frac{1}{4\pi\sqrt{\omega \omega'}} \left(
    \left|
    \,i\dfrac{ \Omega^{(\mathrm{III})}_{1}}{ \Omega^{(\mathrm{III})}_{2}} e^{-i \omega_- C_B}e^{i \Omega^{(\mathrm{III})}_{2} t_B}
    \;
    -{4\pi\sqrt{\omega \omega'}}\beta^{(\mathrm{III})}_{\omega \omega',i}
    \right|
    + \left|
    \dfrac{8\omega_- \Omega^{(\mathrm{III})}_{1} \delta z_B}
    {\Omega^{(\mathrm{III})}_{2}}
    \right|
    +
    \left|
    \dfrac{8\omega_+ \delta \dot{z}_B}
    {\Omega^{(\mathrm{III})}_{2}}
    \right|
    \right)
    .
    \label{eq:pf_delta_beta_III_bound_m1}
\end{align}
\end{widetext}
which corresponds to the first row of Table~\ref{tb:pf_delta_beta_III_bounds} in Appendix~\ref{app:Bound_Summary}.

% ----------------------------------------------------------------------------------------------------
In total, there are eight admissible combinations, corresponding to zero, one, or two applications of IBP together with either OPB or DMB applied to the remaining integrals.  
Since the algebra of the remaining cases is lengthy and does not introduce additional conceptual insight, we summarize all eight results in Table~\ref{tb:pf_delta_beta_III_bounds}.  
Throughout this procedure, we postpone absolute-value estimates and preserve the full complex structure of Eq.~\eqref{eq:pf_delta_beta_III_1st_order} until contributions from all regions are assembled, allowing phase interference to be taken into account before bounding magnitudes.
In practice, given \(\omega\), for each pair \((\omega', a_{\mathrm{thres}})\), we evaluate all admissible combinations of IBP orders and applicable bounds, and select the combination that yields the minimal bound.  
This adaptive optimization ensures that the IRM-induced deviation remains conservative, robust, and tight across all relevant frequency regimes.

It is illuminating to check that the second and third terms in Eq.~\eqref{eq:pf_delta_beta_III_bound_m1} vanish as \(t_B\to\infty\), since both \(\delta z_B \to 0\) and \(\delta\dot z_B \to 0\) in the asymptotic limit.
As for the first term in Eq.~\eqref{eq:pf_delta_beta_III_bound_m1}, we express it explicitly in terms of the boundary data \((z_B,\dot z_B)\), obtaining
\begin{align}
    & i\frac{\Omega^{(\mathrm{III})}_{1}}{\Omega^{(\mathrm{III})}_{2}}
      e^{-i\omega_- C_B} e^{i\Omega^{(\mathrm{III})}_{2} t_B}
      - 4\pi\sqrt{\omega\omega'}\,\beta^{(\mathrm{III})}_{\omega\omega',i}
      \notag\\
    &=
    i\frac{\omega_+(\dot z_B + \delta\dot z_B)-\omega_-}
            {\omega_+ - \omega_-\left(\dot z_B + \delta\dot z_B\right)}
      e^{i(\omega_+ t_B - \omega_- z_B)} e^{-i\omega_- \delta z_B}
    \notag \\
    &\quad
    - i\frac{\omega_+\dot z_B - \omega_-}
            {\omega_+ - \omega_-\dot z_B}
      e^{i(\omega_+ t_B - \omega_- z_B)} ,
\end{align}
which likewise vanishes as \(t_B\to\infty\).

Therefore,
\begin{equation}
\lim_{t_B\to\infty}
\left| \delta\beta^{(\mathrm{III})}_{\omega\omega'} \right|
= 0 .
\end{equation}
This demonstrates that the IRM-induced deviation in Region~III can be made arbitrarily small by pushing the splitting boundary \(t=t_B\) deeper into the asymptotic regime, i.e. the IRM result converges smoothly to the exact solution.

% ----------------------------------------------------------------------------------------------------
\subsection{Bound on IRM Deviation in Region I}
\label{subsec:Bound_on_IRM-Deviation_in_Region_I}
% ----------------------------------------------------------------------------------------------------
In Region~I, we consider the bound
\begin{align}
\Big|\delta \beta^{(\mathrm{I})}_{\omega\omega'}\Big|
\equiv 
\Big|\beta^{(\mathrm{I})}_{\omega\omega'} - \beta^{(\mathrm{I})}_{\omega\omega',i}\Big|,
\label{eq:delta_beta_I}
\end{align}
where
\begin{align}
    \beta^{(\mathrm{I})}_{\omega\omega'} 
    = \frac{1}{4\pi\sqrt{\omega \omega'}} 
      \int_{-\infty}^{t_A} dt \,
      \big[\omega_+ \dot{z}^{(\mathrm{I})}(t) - \omega_- \big] \,
      e^{i\omega_+ t - i\omega_- z^{(\mathrm{I})}(t)} .
    \label{eq:pf_beta_exact_1}
\end{align}
The analysis proceeds in complete analogy with Region~III, under the substitutions
\begin{equation}
(\mathrm{III})\!\to\!(\mathrm{I}),\qquad
B\!\to\!A,\qquad
[t_B,\infty)\to(-\infty,t_A].
\label{eq:region_III_to_I}
\end{equation}
Because the lower integration limit extends to $t=-\infty$, the $i\epsilon$ prescription suppresses the boundary contribution at $t=-\infty$ 
and introduces an overall minus sign in the terms generated by integration by parts.
Applying the first-order expansion and following the same algebra as in Region~III gives
% % \begin{widetext}
% \begin{align}
%     &\delta \beta^{(\mathrm{I})}_{\omega\omega'} 
%     \approx
%     \frac{1}{4\pi\sqrt{\omega \omega'}}
%     \left[\,\underbrace{-i\dfrac{ \Omega^{(\mathrm{I})}_{1}}{ \Omega^{(\mathrm{I})}_{2}} e^{-i \omega_- C_A} e^{i \Omega^{(\mathrm{I})}_{2} t_A}
%     -4\pi\sqrt{\omega \omega'} \beta^{(\mathrm{I})}_{\omega\omega',i}}_{\text{zeroth-order}}
%     \;
%     \right. \notag \\
%     &+
%     \left.
%     e^{-i \omega_- C_A}
%     \int_{-\infty}^{t_A} \!\!dt\left( \underbrace{i\omega_- \Omega^{(\mathrm{I})}_{1}\,\delta z^{(\mathrm{I})}(t)}_{\text{traj.\ correction}}
%     \;-\;\underbrace{\omega_+\,\delta\dot z^{(\mathrm{I})}(t)}_{\text{vel.\ correction}}\,\right)\,
%     e^{i\Omega^{(\mathrm{I})}_{2} t}
%     \right]
%     .
%     \label{eq:pf_delta_beta_I_1st_order}
% \end{align}
% % \end{widetext}
\begin{widetext}
\begin{align}
    &\delta \beta^{(\mathrm{I})}_{\omega\omega'} 
    \approx
    \frac{1}{4\pi\sqrt{\omega \omega'}}
    \left[\,\underbrace{-i\dfrac{ \Omega^{(\mathrm{I})}_{1}}{ \Omega^{(\mathrm{I})}_{2}} e^{-i \omega_- C_A} e^{i \Omega^{(\mathrm{I})}_{2} t_A}
    -4\pi\sqrt{\omega \omega'} \beta^{(\mathrm{I})}_{\omega\omega',i}}_{\text{zeroth-order}}
    \;+
    e^{-i \omega_- C_A}
    \int_{-\infty}^{t_A} \!\!dt\left( \underbrace{i\omega_- \Omega^{(\mathrm{I})}_{1}\,\delta z^{(\mathrm{I})}(t)}_{\text{traj.\ correction}}
    \;-\;\underbrace{\omega_+\,\delta\dot z^{(\mathrm{I})}(t)}_{\text{vel.\ correction}}\,\right)\,
    e^{i\Omega^{(\mathrm{I})}_{2} t}
    \right]
    .
    \label{eq:pf_delta_beta_I_1st_order}
\end{align}
\end{widetext}
As in Region~III, there are eight admissible combinations of IBP order and bounding strategy.  
These are summarized in Table~\ref{tb:pf_delta_beta_I_bounds}.  
Aside from relabeling and the overall sign difference in the integral contributions, the structure of the bounds is identical to that of Region~III.

Similarly, the bound on Eq.~\eqref{eq:pf_delta_beta_I_1st_order} vanishes as \(t_A\to -\infty\), since both \(\delta z_A \to 0\) and \(\delta\dot z_A \to 0\).  
Therefore,
\begin{equation}
\lim_{t_A\to -\infty}
\left| \delta \beta^{(\mathrm{I})}_{\omega\omega'} \right|
= 0 .
\end{equation}
This establishes that the IRM-induced deviation in Region~I can be made arbitrarily small by extending the splitting boundary sufficiently deep into the asymptotic inertial regime. 
Together with the result for Region~III, this confirms the asymptotic exactness of the IRM.

% ----------------------------------------------------------------------------------------------------
\subsection{Total Bound on $\left|\beta_{\omega\omega'}\right|^2$ Residual}
\label{subsec:Total_Bound_on_the_Particle-Number_Deviation}
% ----------------------------------------------------------------------------------------------------
%  Squared-Magnitude Deviation
Now that the IRM has been applied to compute the asymptotic contributions $\beta^{(\mathrm{III})}_{\omega\omega',i}$ and $\beta^{(\mathrm{I})}_{\omega\omega',i}$ in Eqs.~\eqref{eq:pf_beta_inertial_III} and \eqref{eq:pf_beta_inertial_I}, we regard them as analytic approximations to the exact asymptotic integrals.  
The associated IRM-induced deviations are quantified by $|\delta\beta^{(\mathrm{III})}_{\omega\omega'}|$ and $|\delta\beta^{(\mathrm{I})}_{\omega\omega'}|$, summarized in Tables~\ref{tb:pf_delta_beta_III_bounds} and~\ref{tb:pf_delta_beta_I_bounds}.
Since the physical observable is the squared magnitude, \( \left|\beta_{\omega\omega'}\right|^2\), whose integration over frequency mode $\omega'$ gives the particle number spectrum, we must combine the contributions from all three regions and translate the deviation into an estimate of the squared-magnitude bound.

As discussed above, complex boundary terms are retained without taking absolute values in order to preserve phase interference.  
We therefore decompose the deviation into complex and absolute components:
\begin{subequations}
\begin{align}
\delta\beta^{\text{com}}_{\omega\omega'} 
&\equiv 
\delta\beta^{(\mathrm{I}),\text{com}}_{\omega\omega'} 
+ \delta\beta^{(\mathrm{III}),\text{com}}_{\omega\omega'},
\label{eq:delta_beta_com}
\\
\Big|\delta\beta^{\text{abs}}_{\omega\omega'}\Big|
&\equiv 
\Big|\delta\beta^{(\mathrm{I}),\text{abs}}_{\omega\omega'}\Big|
+
\Big|\delta\beta^{(\mathrm{III}),\text{abs}}_{\omega\omega'}\Big|.
\label{eq:delta_beta_abs}
\end{align}
\end{subequations}
Here, \( \delta\beta^{\text{com}}_{\omega\omega'} \) collects the complex contributions from Regions~I and~III prior to taking magnitudes, whereas \( \Big|\delta\beta^{\text{abs}}_{\omega\omega'}\Big| \) consists of the remaining terms that have already been bounded in absolute value via OPB or DMB.
All quantities can be looked up from Tables~\ref{tb:pf_delta_beta_III_bounds} and~\ref{tb:pf_delta_beta_I_bounds}.

Then the squared-magnitude residual can be estimated as:
\begin{widetext}
\begin{align}
\Delta \left|\beta_{\omega\omega'}\right|^2
&\equiv 
\left| 
\left|\beta^{\text{Ext}}_{\omega\omega'}\right|^2
-
\left|\beta^{\text{IRM}}_{\omega\omega'}\right|^2
\right|
=
\left| 
\left| \beta^{\mathrm{IRM}}_{\omega\omega'}+\delta\beta_{\omega\omega'}\right|^2
- \left|\beta^{\mathrm{IRM}}_{\omega\omega'}\right|^2
\right|
=
\left| 
2\,\mathrm{Re}\!\left[(\beta^{\text{IRM}}_{\omega\omega'})^*\delta\beta_{\omega\omega'}\right]
+ \left|\delta\beta_{\omega\omega'}\right|^2
\right|
\notag \\
&\le 
2\,\left|(\beta^{\text{IRM}}_{\omega\omega'})^*\,\delta\beta_{\omega\omega'}\right|
+ \left|\delta\beta_{\omega\omega'}\right|^2
\le
2\,\left|(\beta^{\text{IRM}}_{\omega\omega'})^*\,\delta\beta^{\text{com}}_{\omega\omega'}\right|
+2\,\left|\beta^{\text{IRM}}_{\omega\omega'}\right|\,\left|\delta\beta^{\text{abs}}_{\omega\omega'}\right| 
+\left( \left|\delta\beta^{\text{com}}_{\omega\omega'}\right| + \left|\delta\beta^{\text{abs}}_{\omega\omega'}\right| \right)^2
\equiv 
\Delta \left|\beta^{\text{IRM}}_{\omega\omega'}\right|^2.
\label{eq:Delta_beta}
\end{align}
\end{widetext}
The quantity $\Delta |\beta^{\text{IRM}}_{\omega\omega'}|^2$ therefore serves as a conservative bound of the squared-magnitude residual between the IRM result and the exact solution, $\Delta |\beta_{\omega\omega'}|^2$.

% ----------------------------------------------------------------------------------------------------
% Squared-Magnitude Deviation on Logex
To examine how \(\Delta \left|\beta^{\text{IRM}}_{\omega\omega'}\right|^2\) depends on the choice of bounding prescriptions, we apply the IRM to the Logex trajectory in Eq.~\eqref{eq:logex_traj} with $\kappa = 0.5$ and explore the parameter space.
Throughout this example we fix $\omega = 1$ and vary $\omega'$, allowing us to probe the full frequency range.

% ----------------------------------------------------------------------------------------------------
\begin{figure*}[!htbp]
    \centering
    \includegraphics[width=0.9\linewidth]{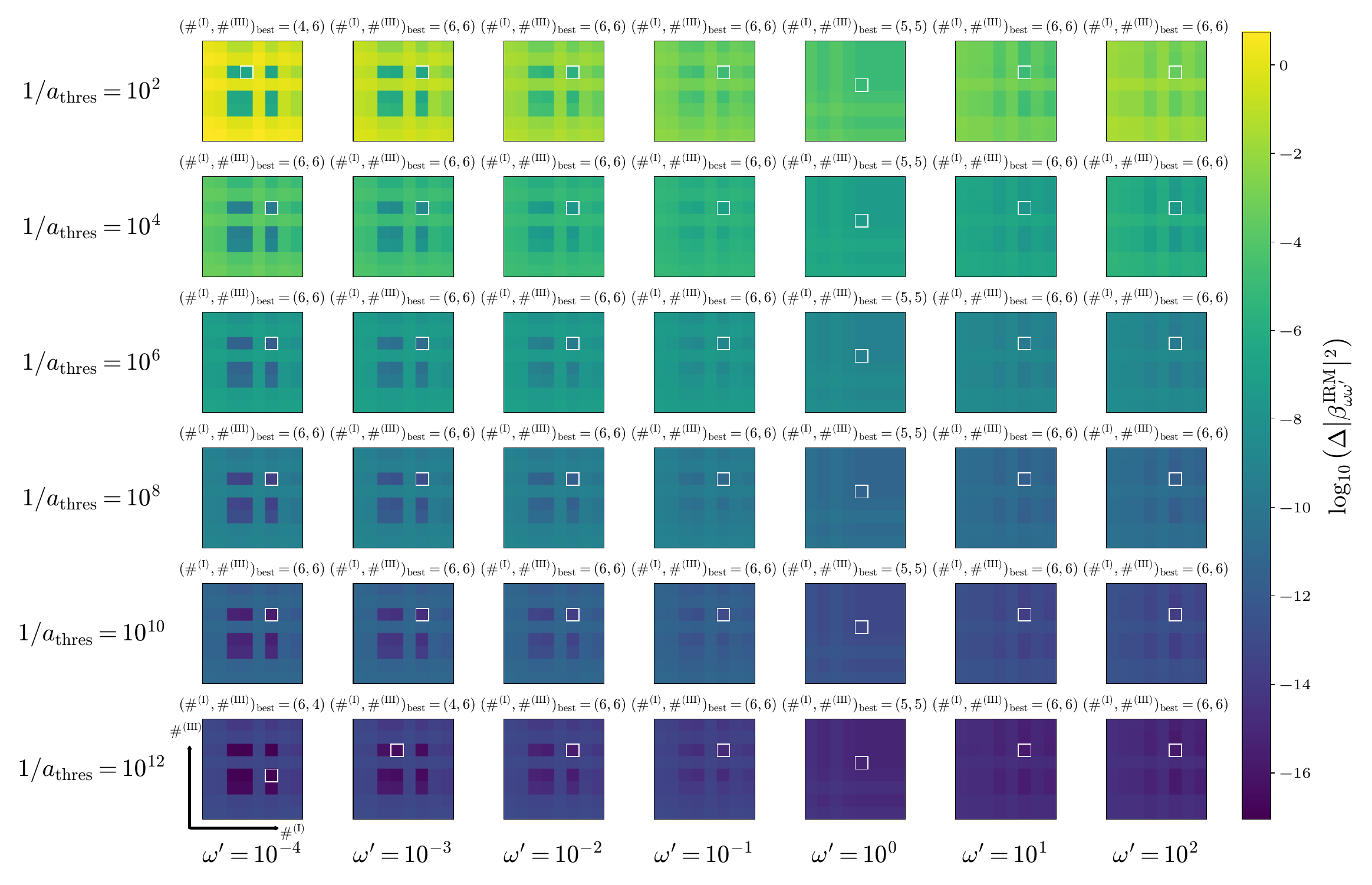}
    \caption{
    Dependence of $\Delta\big|\beta^{\mathrm{IRM}}_{\omega\omega'}\big|^2$ over all $8\times8$ Region~I/III bounding combinations for different values of $a_{\mathrm{thres}}$ and $\omega'$.
    The white squares mark the optimal pair $(\#^{(\mathrm{I})},\#^{(\mathrm{III})})_{\mathrm{best}}$.
    }
    \label{fig:logex_Dbeta2_wprime_athres_ij}
\end{figure*}
Fig.~\ref{fig:logex_Dbeta2_wprime_athres_ij} shows how $\Delta\big|\beta^{\mathrm{IRM}}_{\omega\omega'}\big|^2$ varies across all $8\times8$ combinations of Region~I and Region~III bounding strategies.
For each $(\omega',a_{\mathrm{thres}})$, the optimal strategy pair $(\#^{(\mathrm{I})},\#^{(\mathrm{III})})_{\mathrm{best}}$ marked by the white squares differs, indicating that no single bounding prescription is uniformly optimal throughout the parameter space.
This dependence motivates the automated selection of the tightest admissible bound over all strategies in our numerical IRM implementation.
\revA{
The detailed frequency dependence of the individual and optimized bounds, as well as their scaling with $1/a_{\rm thres}$, is trajectory-dependent and is therefore presented in Appendix~\ref{app:supplementary_analyses}.
}

% ----------------------------------------------------------------------------------------------------
\subsection{Approaching the Exact Solution}
\label{subsec:Approaching_the_Exact_Result}
% ----------------------------------------------------------------------------------------------------
To perform an explicit calculation, we restore the numerical integration error previously omitted in Eq.~\eqref{eq:delta_beta_sum}.
In our implementation, the Region~II integral is evaluated using \texttt{scipy.integrate.quad\_vec}, whose internal estimator provides a complex floating-point uncertainty, denoted $\delta\beta^{(\mathrm{II}),\mathrm{com}}_{\omega\omega'}$.
The total complex deviation therefore becomes
\begin{equation}
\delta\beta^{\mathrm{com}}_{\omega\omega'}
=
\delta\beta^{(\mathrm{I}),\mathrm{com}}_{\omega\omega'}
+
\delta\beta^{(\mathrm{II}),\mathrm{com}}_{\omega\omega'}
+
\delta\beta^{(\mathrm{III}),\mathrm{com}}_{\omega\omega'},
\label{eq:delta_beta_com_include_2}
\end{equation}
which replaces Eq.~\eqref{eq:delta_beta_com}.

% ----------------------------------------------------------------------------------------------------
\begin{figure*}[!htbp]
\centering
\begin{tikzpicture}[
    >=Latex,
    node distance=8mm and 10mm,
    every node/.style={font=\footnotesize},
    narrowbox/.style={
        draw,
        rounded corners,
        align=center,
        minimum height=1cm,
        text width=1.5cm
    },
    box/.style={
        draw,
        rounded corners,
        align=center,
        minimum height=1.2cm,
        text width=3cm
    },
    widebox/.style={
        draw,
        rounded corners,
        align=center,
        minimum height=1.2cm,
        text width=4cm
    },
    line/.style={thick, ->},
    dashline/.style={thick, ->, dashed},
    eqlabel/.style={font=\footnotesize, fill=white, inner sep=2pt}
]

% -------------------------------------------------
% Input and boundary
\node[widebox] (input) {Choose trajectory and parameters $\{z(t), \omega,\omega',a_{\rm thres}\}$};
\node[widebox, below=7mm of input] (boundary) {Splitting boundaries $\{t_A,t_B\}$ determined by $a_{\rm thres}$};

% -------------------------------------------------
% Regions
\node[box, fill=red!12, below=7mm of boundary] (r2) {Region II:\\$t_A<t<t_B$\\Accelerating region};
\node[box, fill=blue!8, below=7mm of boundary, right=30mm of r2] (r3) {Region III:\\$t>t_B$\\Asymptotic tail};
\node[box, fill=blue!8, below=7mm of boundary, left=30mm of r2] (r1) {Region I:\\$t<t_A$\\Asymptotic tail};

% -------------------------------------------------
% Region I branch
\node[narrowbox, fill=blue!8, below left=10mm and -5mm of r1] (b11) {$\beta^{(\mathrm {I})}_{\omega\omega',i}$};
\node[narrowbox, fill=blue!8, below =10mm of r1] (b12) {$\delta \beta^{(\mathrm {I), com}}_{\omega\omega'}$};
\node[narrowbox, fill=blue!8, below right=10mm and -5mm of r1] (b13) {$\left| \delta \beta^{(\mathrm {I), abs}}_{\omega\omega'}\right|$};

% -------------------------------------------------
% Region II branch
\node[narrowbox, fill=red!12, below left=10mm and -5mm of r2] (b21) {$\beta^{(\mathrm {II})}_{\omega\omega',N}$};
\node[narrowbox, fill=red!12, below =10mm of r2] (b22) {$\delta \beta^{(\mathrm {II), com}}_{\omega\omega'}$};
\node[narrowbox, fill=red!12, below right=10mm and -5mm of r2] (b23) {$0$};

% -------------------------------------------------
% Region III branch
\node[narrowbox, fill=blue!8, below left=10mm and -5mm of r3] (b31) {$\beta^{(\mathrm {III})}_{\omega\omega',i}$};
\node[narrowbox, fill=blue!8, below =10mm of r3] (b32) {$\delta \beta^{(\mathrm {III), com}}_{\omega\omega'}$};
\node[narrowbox, fill=blue!8, below right=10mm and -5mm of r3] (b33) {$\left| \delta \beta^{(\mathrm {III), abs}}_{\omega\omega'}\right|$};

% -------------------------------------------------
% Result and Bound
\node[box, below right=15mm and 25mm of b21] (com) {$\delta \beta^{\rm{com}}_{\omega\omega'}$ via Eq.~\eqref{eq:delta_beta_com_include_2}};
\node[box, right=5mm of com] (abs) {$\left| \delta \beta^{\rm{abs}}_{\omega\omega'}\right|$ via Eq.~\eqref{eq:delta_beta_abs}};
\node[box, fill=gray!8, below=35mm of b13] (result) {$\beta^{\rm {IRM}}_{\omega\omega'}$};
\node[box, fill=gray!8, below=35mm of b31] (bound) {$\Delta \left| \beta^{\rm {IRM}}_{\omega\omega'}\right|^2$ via Eq.~\eqref{eq:Delta_beta}};

% -------------------------------------------------
% Input -> boundary -> regions
\draw[line] (input) -- node[eqlabel, midway, right, xshift=2mm, yshift=0.5mm] {Eq.~\eqref{eq:splitting_condition}} (boundary);
\draw[line] (boundary) -- (r2);
\draw[line] (boundary.south west) -- (r1.north);
\draw[line] (boundary.south east) -- (r3.north);

% -------------------------------------------------
% Region -> beta contributions
\draw[line] (r1) -- node[eqlabel, midway, left, xshift=-2mm, yshift=0.5mm] {Eq.~\eqref{eq:pf_beta_inertial_I}} (b11);
\draw[dashline] (r1) -- (b12);
\draw[dashline] (r1) -- node[eqlabel, midway, xshift=-5mm, yshift=0.5mm] {Table~\ref{tb:pf_delta_beta_I_bounds}} (b13);

\draw[line] (r2) -- (b21);
\draw[dashline] (r2) -- node[eqlabel, midway, xshift=-5mm, yshift=0.5mm] {numerical} (b22);
\draw[dashline] (r2) -- (b23);

\draw[line] (r3) -- node[eqlabel, midway, left, xshift=-2mm, yshift=0.5mm] {Eq.~\eqref{eq:pf_beta_inertial_III}} (b31);
\draw[dashline] (r3) -- (b32);
\draw[dashline] (r3) -- node[eqlabel, midway, xshift=-5mm, yshift=0.5mm] {Table~\ref{tb:pf_delta_beta_III_bounds}} (b33);

% -------------------------------------------------
% beta contributions -> IRM result
\draw[line] (b11.south) -- (result.north);
\draw[line] (b21.south) -- (result.north);
\draw[line] (b31.south) -- (result.north);

% -------------------------------------------------
% delta beta^com contributions -> com
\draw[dashline] (b12.south) -- (com.north);
\draw[dashline] (b22.south) -- (com.north);
\draw[dashline] (b32.south) -- (com.north);

% -------------------------------------------------
% abs contributions -> abs
\draw[dashline] (b13.south) -- (abs.north);
\draw[dashline] (b33.south) -- (abs.north);

% -------------------------------------------------
% Final bound construction
\draw[line] (result) -- (bound);
\draw[dashline] (com.south) -- (bound.north);
\draw[dashline] (abs.south) -- (bound.north);

% -------------------------------------------------
\end{tikzpicture}
\caption{
Schematic workflow of the IRM calculation for perfectly reflecting mirrors. 
The case of imperfectly reflecting mirrors follows identically, with the corresponding quantities appropriately replaced.
}
\label{fig:IRM_calculation_process}
\end{figure*}

\begin{figure*}[!htbp]
    \includegraphics[width=0.49\linewidth]{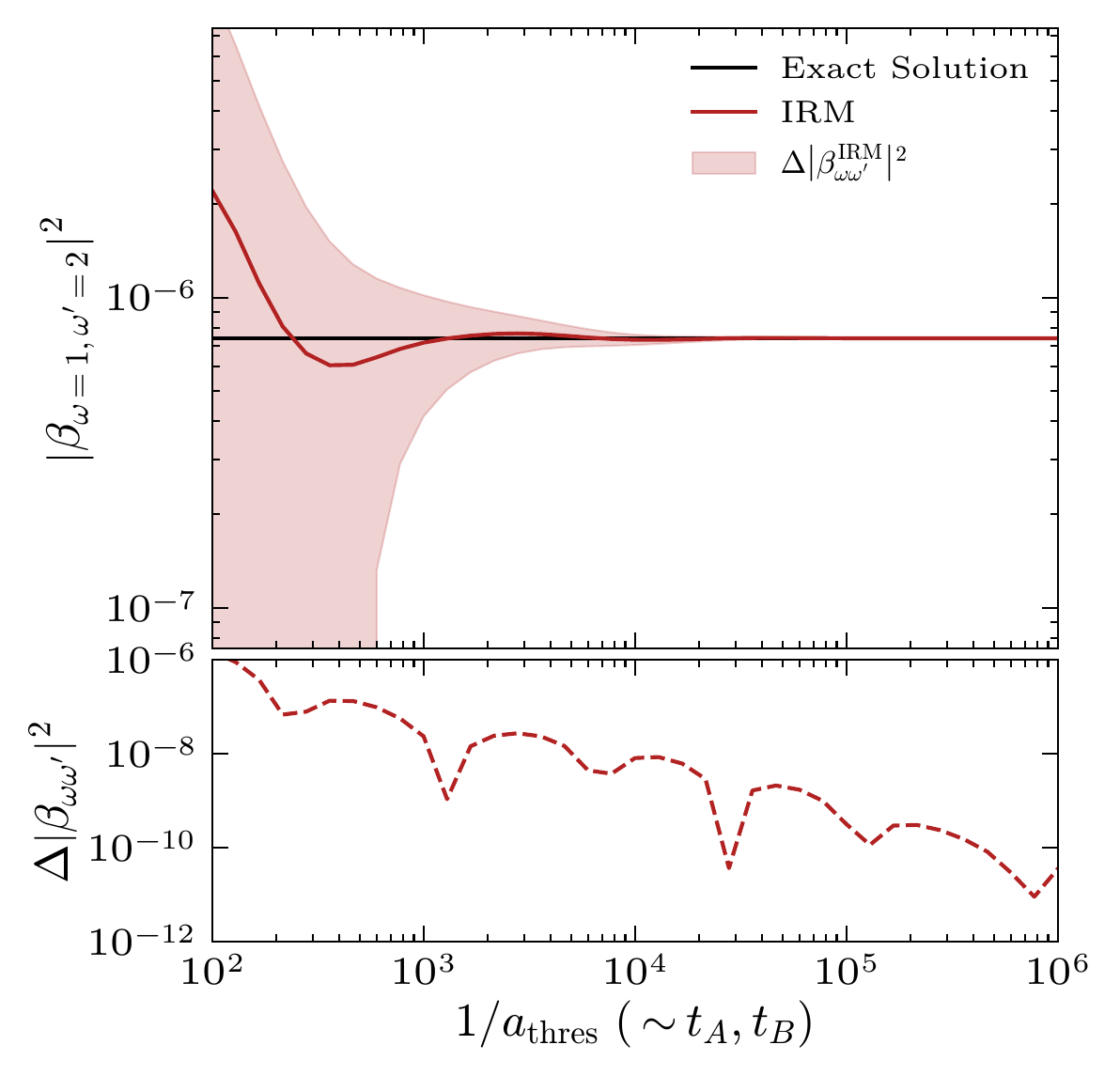}
    \includegraphics[width=0.49\linewidth]{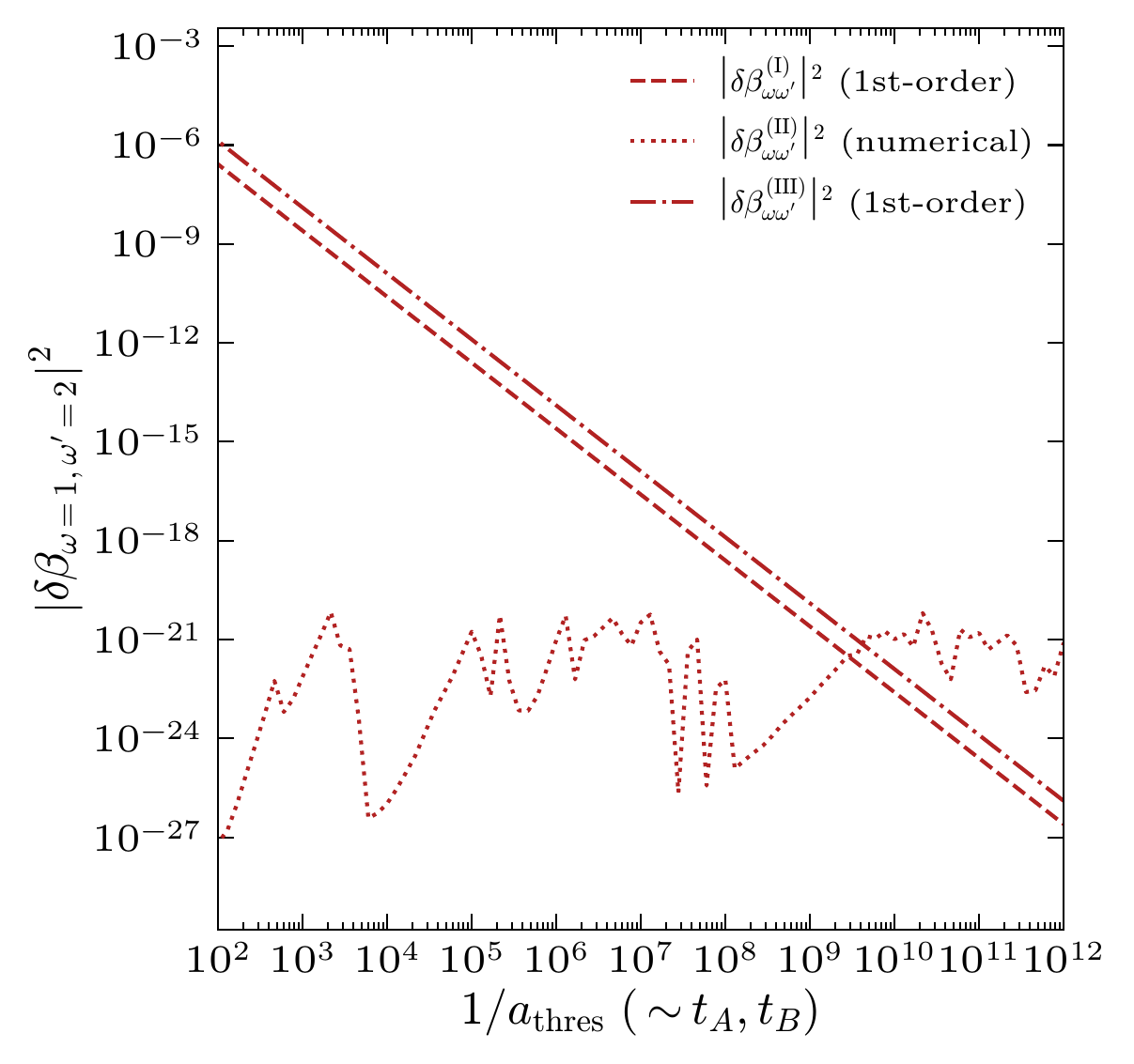}
    \caption{
    \textbf{Left:} $|\beta^{\mathrm{IRM}}_{\omega\omega'}|^{2}$ of \textbf{Logex} trajectory with $\kappa=0.5$ is compared with the exact solution $|\beta^{\mathrm{Ext}}_{\omega\omega'}|^{2}$ as a function of $1/a_{\mathrm{thres}}$, with the residual \YS{$\Delta \left| \beta_{\omega\omega'}\right|^{2}$} shown below.
    \textbf{Right:} Comparison of the IRM-induced deviation and the estimated numerical calculation error.
    }
    \label{fig:logex_convergence_combined}
\end{figure*}

The procedure for performing calculations within the IRM framework is summarized in Fig.~\ref{fig:IRM_calculation_process}. 
In practice, one first specifies a trajectory and the parameters $\{\omega, \omega', a_{\text{thres}}\}$, from which the matching points $\{t_A, t_B\}$ are implicitly determined. 
The inertial contributions $\beta^{(\mathrm{I})}_{\omega\omega',i}$ and $\beta^{(\mathrm{III})}_{\omega\omega',i}$ are then computed using Eq.~\eqref{eq:pf_beta_inertial_I} and Eq.~\eqref{eq:pf_beta_inertial_III}, while the non-inertial contribution $\beta^{(\mathrm{II})}_{\omega\omega',N}$ is obtained via numerical integration. 
These contributions combine to yield the IRM result $\beta^{\mathrm{IRM}}_{\omega\omega'}$ through Eq.~\eqref{eq:beta_IRM}, shown as the red solid curve in Fig.~\ref{fig:logex_convergence_combined}. 
To estimate the IRM-induced deviation, a bounding strategy is selected independently for Regions~I and III from Tables~\ref{tb:pf_delta_beta_I_bounds} and \ref{tb:pf_delta_beta_III_bounds}. 
The complex contributions, together with the numerical integration error from Region~II, are substituted into Eq.~\eqref{eq:delta_beta_com_include_2}, while the corresponding absolute terms are inserted into Eq.~\eqref{eq:delta_beta_abs}. 
Finally, these results are combined in Eq.~\eqref{eq:Delta_beta} to obtain a conservative bound on the squared-magnitude residual, shown as the red shaded region in Fig.~\ref{fig:logex_convergence_combined}.

% ----------------------------------------------------------------------------------------------------
We now apply the IRM to the Logex trajectory with $\kappa=0.5$, fixing $(\omega,\omega')=(1,2)$ and test convergence by pushing $1/a_{\mathrm{thres}}$ deep into the asymptotic regime.
The results are shown in Fig.~\ref{fig:logex_convergence_combined}.
The left panel compares $|\beta^{\mathrm{IRM}}_{\omega\omega'}|^{2}$ with the exact solution from Eq.~\eqref{eq:logex_beta_analytic}.
The shaded region represents the total squared-magnitude bound, $\Delta \left|\beta^{\text{IRM}}_{\omega\omega'}\right|^2$.
\revA{The quantity $1/a_{\rm thres}$ may be viewed as a convergence-control parameter.
As $1/a_{\mathrm{thres}}$ increases, Region II is enlarged, the IRM result converges to the exact solution with the residual $\Delta \left|\beta_{\omega\omega'}\right|^2$ decreasing accordingly.}
This behavior reflects the vanishing of the boundary data $\{\delta z_A,\delta z_B,\delta\dot z_A,\delta\dot z_B\}$ as $t_A,t_B\to\infty$ and that the IRM serves as a good approximation to the exact solution.
The right panel separates the estimated deviations from Regions~I, II, and III.
The IRM-induced deviations in Regions~I and III decrease with $1/a_{\mathrm{thres}}$, while the numerical calculation error from Region~II grows as the integration interval enlarges since the effective accuracy of the quadrature is reduced and the integrand becomes increasingly oscillatory.
This illustrates a fundamental tradeoff: reducing $a_{\mathrm{thres}}$ suppresses the analytic IRM-induced deviation but amplifies the numerical error.
Their intersection determines the optimal balance between these competing effects; in this example, the minimum occurs near $1/a_{\mathrm{thres}}\approx 10^{10}$.
However, this limitation is not intrinsic to the IRM.
Improving numerical precision or employing more specialized oscillatory integration techniques can further suppress the Region~II uncertainty, allowing convergence deeper into the asymptotic regime.

% ----------------------------------------------------------------------------------------------------
\begin{figure}[!htbp]
    \includegraphics[width=1\linewidth]{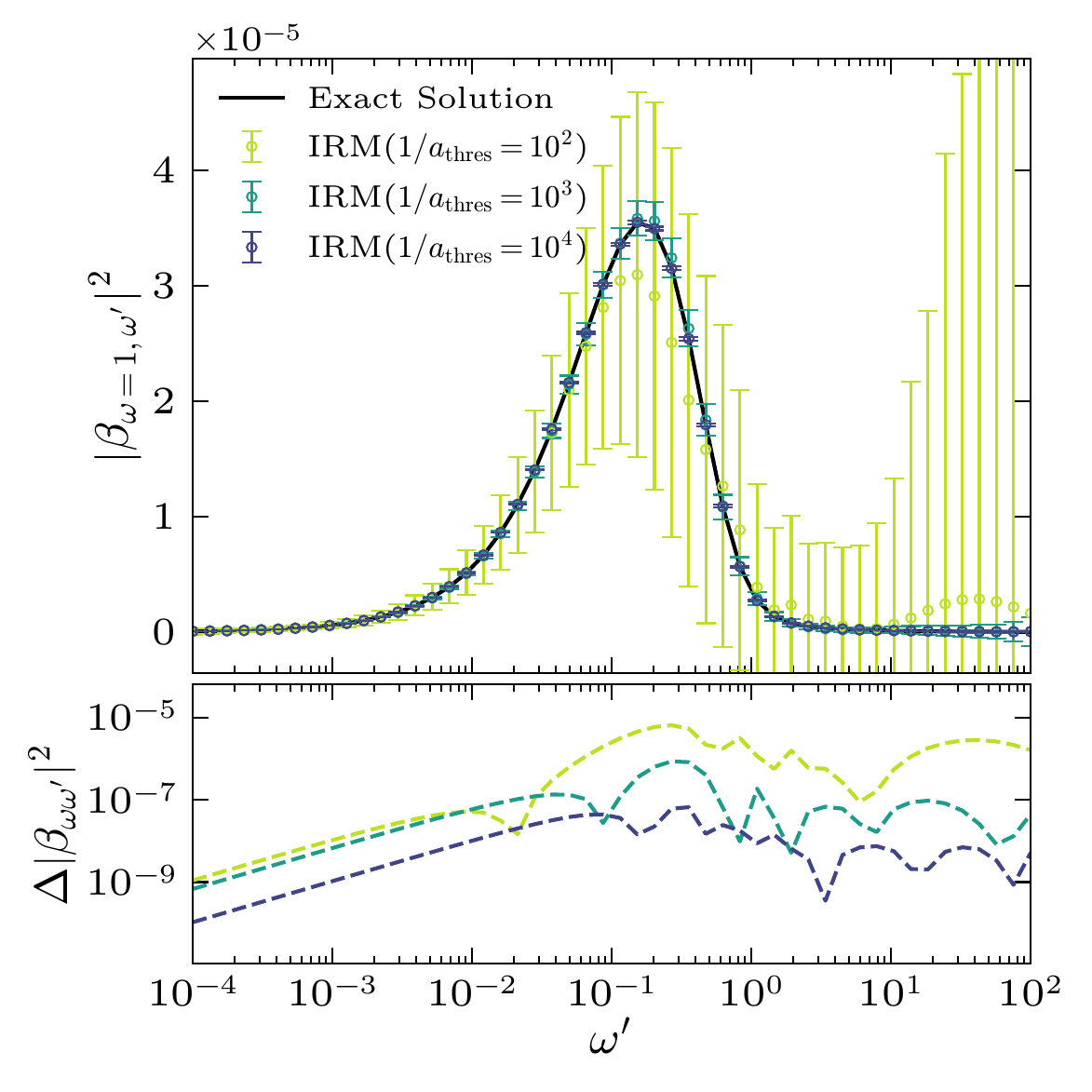}
    \caption{
    $|\beta^{\mathrm{IRM}}_{\omega\omega'}|^{2}$ of \textbf{Logex} trajectory with $\kappa=0.5$ is compared with the exact solution $|\beta^{\mathrm{Ext}}_{\omega\omega'}|^{2}$ as a function of $\omega'$, with the residual \YS{$\Delta \left| \beta_{\omega\omega'}\right|^{2}$} shown below.
    }
    \label{fig:logex_beta2_wprime}
\end{figure}
Finally, sweeping over $\omega'$ at fixed $\omega$ yields the full squared-magnitude spectrum $|\beta^{\mathrm{IRM}}_{\omega\omega'}|^{2}$, shown in Fig.~\ref{fig:logex_beta2_wprime}.
The IRM results (scattered points) agree closely with the analytic spectrum (black line) across a broad frequency range.
At low and intermediate frequencies, both the amplitude and spectral shape are accurately reproduced, indicating that the asymptotic deviations are small and the numerical integration remains stable.
At higher frequencies, mild deviations appear in the tail, where the integrand becomes highly oscillatory and the first-order expansion in Eq.~\eqref{eq:exp_expansion} gradually loses accuracy.
Nevertheless, increasing $1/a_{\mathrm{thres}}$ from $10^{2}$ to $10^{4}$ significantly suppresses the deviation, even though this regime remains well below the numerical saturation scale $1/a_{\mathrm{thres}}\sim10^{10}$.

Overall, the spectral comparison confirms that, once convergence is achieved, the IRM reproduces the full particle-production spectrum with high accuracy.
Combined with the convergence analysis above, this demonstrates that the IRM is a robust and systematically improvable method that successfully bridges analytic control and numerical implementation for asymptotic mirror trajectories.

% ====================================================================================================
\section{IRM for an Imperfect Mirror}
\label{sec:IRM_impf}
In the previous section, we introduced the IRM and demonstrated its performance for a perfectly reflecting mirror following the Logex trajectory.
The underlying idea of the IRM, however, is not restricted to perfect reflectors.
The method extends directly to imperfect mirrors, provided the asymptotic behaviors at early and late times.
Since the overall procedure remains unchanged, we emphasize here only the additional features specific to the imperfect case.

As a concrete example, we consider the gravitational collapse (Grav) trajectory \cite{2008_Haro, 2010_Haro, 2011_Haro}, defined by
\begin{equation}
z(t)=
\begin{cases}
0, & t \le 0, \\[8pt]
-t + \dfrac{1 - W\!\left(e^{1 - 2\kappa t}\right)}{\kappa}, & t>0 ,
\end{cases}
\label{eq:grav_traj}
\end{equation}
where $W$ denotes the Lambert function.
This trajectory reproduces the late-time behavior of a collapsing object forming a horizon.

For this trajectory, the Bogoliubov coefficient defined in Eq.~\eqref{eq:impf_beta} can still be obtained analytically.
The exact closed-form solution is given in \cite{2021_Lin}:
% \begin{widetext}
\begin{align}
\beta_{\omega\omega'}^{\text{Ext}}
=&
-\frac{\alpha}{4\pi\sqrt{\omega\,\omega'}(\omega+\omega')}
-\frac{\alpha\, e^{i \omega'/\kappa} e^{i\pi/4} e^{-\pi\omega/(2\kappa)}}
       {4\pi\kappa\sqrt{\omega\,\omega'}}
\notag \\
&\times
\left(\frac{\kappa}{\omega'}\right)^{\frac12-\frac{i\omega}{\kappa}}
\!\left[
\Gamma\!\left(\frac12 - \frac{i\omega}{\kappa}\right)
-
\Gamma\!\left(\frac12 - \frac{i\omega}{\kappa}, \frac{i\omega'}{\kappa}\right)
\right].
\label{eq:grav_beta_analytic}
\end{align}
% \end{widetext}

% ----------------------------------------------------------------------------------------------------
\begin{figure*}[!htbp]
    \centering
    \includegraphics[width=0.49\linewidth]{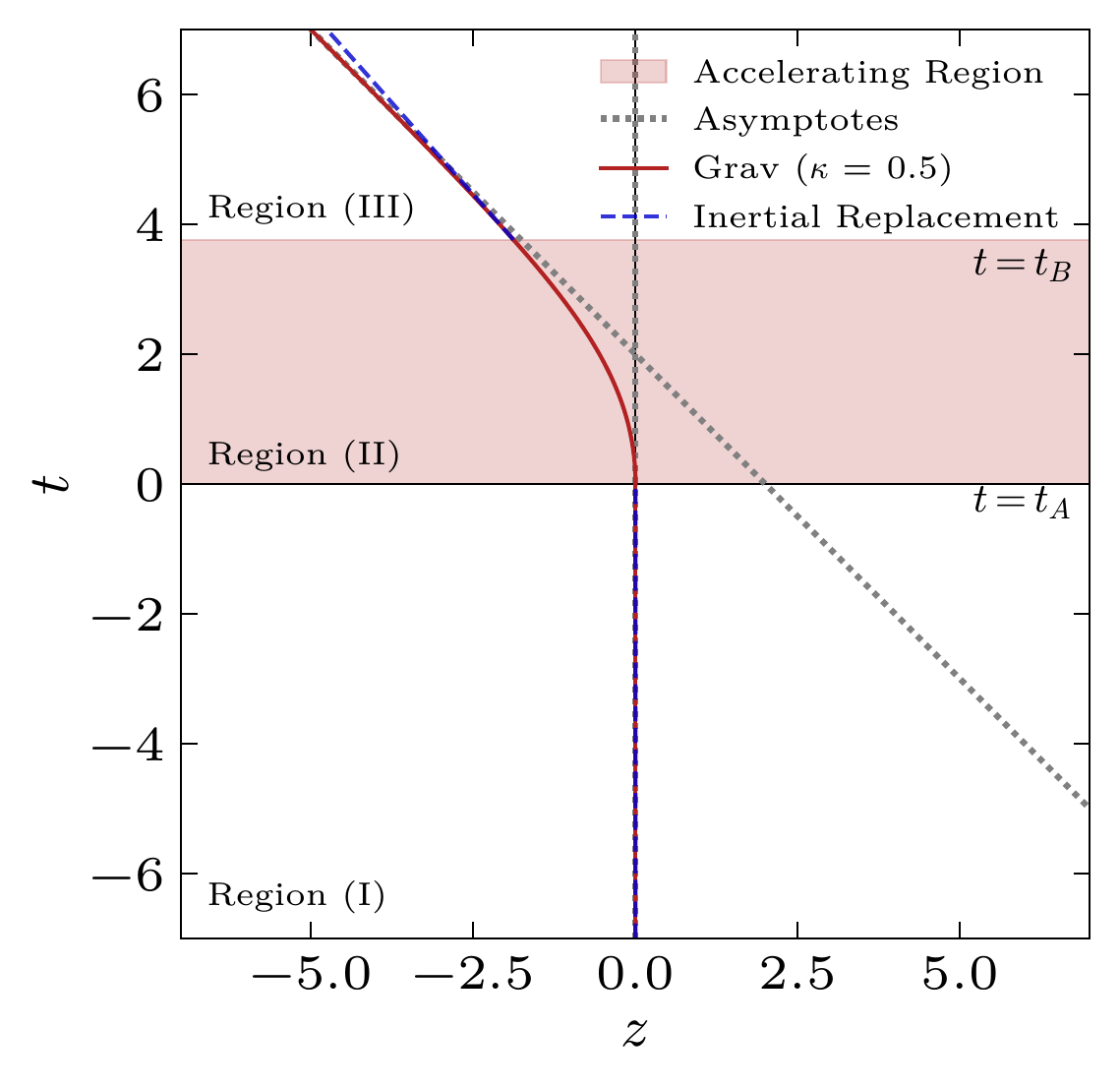}
    \includegraphics[width=0.49\linewidth]{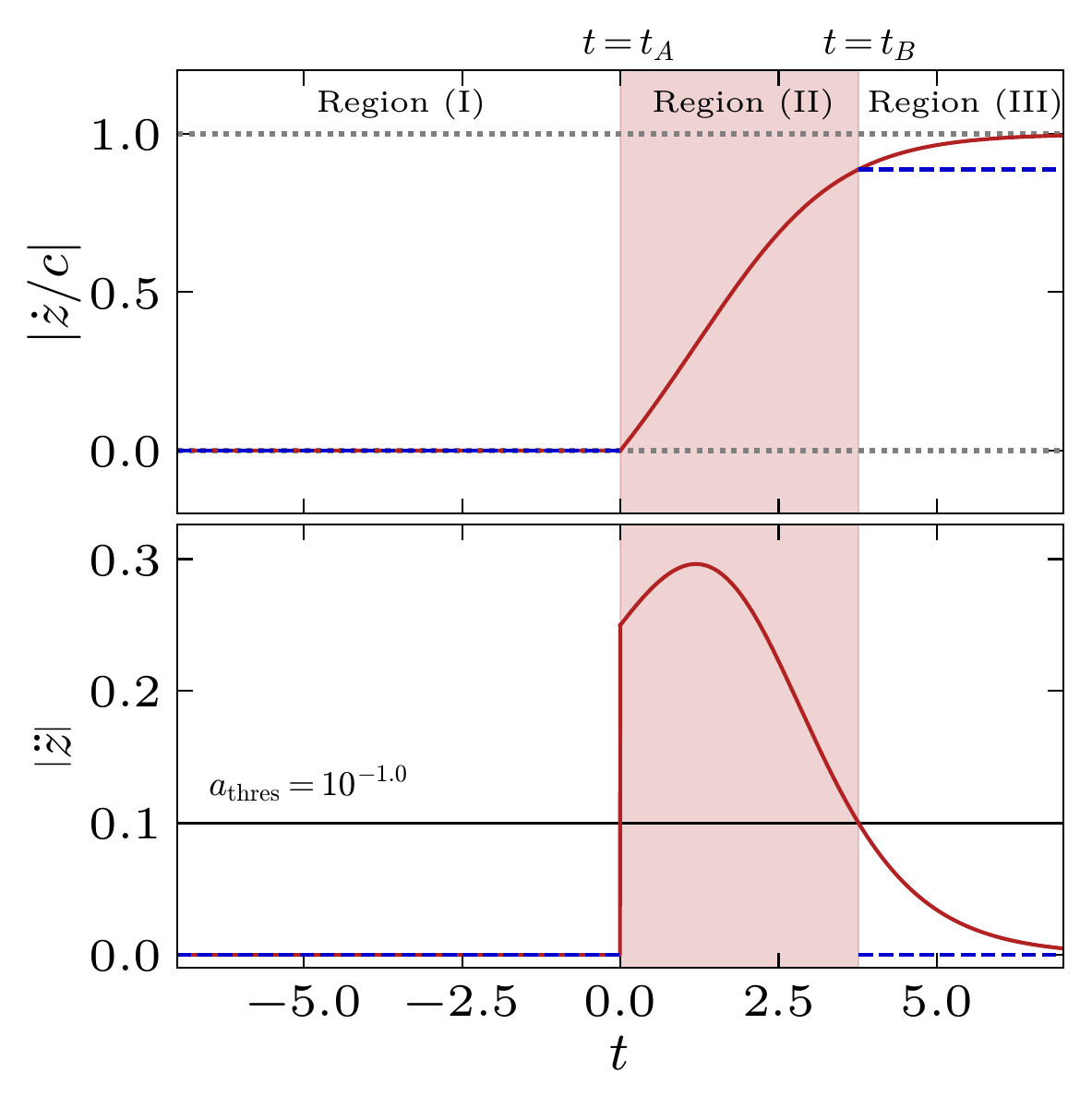}
    \caption{
    \textbf{Left}: splitting the \textbf{Grav} trajectory into asymptotic tails and a central accelerating region using the acceleration threshold \(a_{\text{thres}}\), together with the inertial replacement. 
    \textbf{Right}: velocity and acceleration profiles illustrating the threshold condition that defines \(t_A\) and \(t_B\). 
    }
    \label{fig:grav_dynamics_combined}
\end{figure*}
Using the splitting condition in Eq.~\eqref{eq:splitting_condition}, the trajectory is decomposed into three segments, as illustrated in Fig.~\ref{fig:grav_dynamics_combined}.
Applying the inertial replacements Eqs.~\eqref{eq:traj_inertial_1} and \eqref{eq:traj_inertial_III} in Regions~I and~III, the corresponding inertial contributions are
\begin{subequations}
\begin{align}
\beta^{(\mathrm{I})}_{\omega\omega',i}
=
\frac{-\alpha}{4\pi\sqrt{\omega\omega'}}
\frac{\sqrt{1-\dot z_A^{\,2}}}{\omega_+ - \omega_- \dot z_A}\,
e^{i(\omega_+ t_A - \omega_- z_A)},
\label{eq:impf_beta_inertial_I}
\\
\beta^{(\mathrm{III})}_{\omega\omega',i}
=
\frac{\alpha}{4\pi\sqrt{\omega\omega'}}
\frac{\sqrt{1-\dot z_B^{\,2}}}{\omega_+ - \omega_- \dot z_B}\,
e^{i(\omega_+ t_B - \omega_- z_B)} .
\label{eq:impf_beta_inertial_III}
\end{align}
\end{subequations}
which are directly analogous to the corresponding expressions for the perfect mirror, Eqs.~\eqref{eq:pf_beta_inertial_I} and \eqref{eq:pf_beta_inertial_III}.

% ----------------------------------------------------------------------------------------------------
Now we derive the bound on the IRM-induced deviation in Region~III $\bigl|\delta\beta^{(\mathrm{III})}_{\omega\omega'}\bigr|$ for the imperfect mirror. 
The steps parallel the perfect-mirror analysis, but the integrand differs because of the factor $\sqrt{1-\dot z^2}$.
Substituting Eqs.~\eqref{eq:traj_exact_3} and \eqref{eq:vel_exact_3} into the Region~III integral gives
\begin{align}
\beta^{(\mathrm{III})}_{\omega\omega'}
& = -\frac{i\alpha}{4\pi\sqrt{\omega\omega'}}
\int_{t_B}^{\infty} dt \,
\sqrt{1-\!\left(\dot z^{(\mathrm{III})}(t)\right)^{2}}\,
e^{\,i\omega_{+} t - i\omega_{-} z^{(\mathrm{III})}(t)} \notag \\
&= -\frac{i\alpha}{4\pi\sqrt{\omega\omega'}}
e^{-i\omega_- C_B} 
\int_{t_B}^{\infty} dt \,
e^{i\Omega^{(\mathrm{III})}_{2} t} e^{\,i\omega_- \delta z^{(\mathrm{III})}(t)} 
\, \notag \\
& \quad \times \sqrt{1-(\dot z^{(\mathrm{III})}_{\rm asym})^{2}
+2\dot z^{(\mathrm{III})}_{\rm asym}\,\delta\dot z^{(\mathrm{III})}(t)
-\left(\delta\dot z^{(\mathrm{III})}(t)\right)^{2}},
\label{eq:impf_beta_exact_III_expand}
\end{align}
where $C_B$ and $\Omega^{(\mathrm{III})}_{2}$ are defined in Eqs.~\eqref{eq:constant_B} and~\eqref{eq:Omega_12_III}.
For a horizon-forming trajectory, we have $\dot z^{(\mathrm{III})}_{\rm asym}=-1$.
Expanding the square root up to the order $\left(-\delta\dot z^{(\mathrm{III})}(t)\right)^{3/2}$ and the exponential phase factor via Eq.~\eqref{eq:exp_expansion} to the leading order yields
\begin{widetext}
\begin{align}
    \delta \beta^{(\mathrm{III})}_{\omega\omega'} 
    & \approx
    \frac{\alpha}{4\pi\sqrt{\omega \omega'}} \left[
    - \frac{4\pi\sqrt{\omega \omega'}}{\alpha} \beta^{(\mathrm{III})}_{\omega \omega' ,i}
    - i e^{-i \omega_- C_B} 
    \right. \notag \\
    & \left. \quad 
    \times \int_{t_B}^{\infty} \!\!dt
    \left(
    \underbrace{\sqrt{2} \left( - \delta\dot{z}^{(\mathrm{III})}(t) \right)^{1/2}
    -\dfrac{\sqrt{2}}{4} \left( - \delta\dot{z}^{(\mathrm{III})}(t) \right)^{3/2}}_{\text{vel.\ correction}}
    + \underbrace{\sqrt{2} i\omega_- \delta z^{(\mathrm{III})}(t) \left( - \delta\dot{z}^{(\mathrm{III})}(t) \right)^{1/2}}_{\text{coup.\ correction}}
    \right)
    e^{i \Omega^{(\mathrm{III})}_{2} t}
    \right ]
    .
    \label{eq:impf_delta_beta_III_1st_order}
\end{align}
\end{widetext}
Note that $\left( -\delta\dot z^{(\mathrm{III})}(t) \right )\ge 0$.

As in the perfect-mirror case, the integrals in Eq.~\eqref{eq:impf_delta_beta_III_1st_order} can be bounded using IBP together with OPB, Eq.~\eqref{eq:OPB}, or DMB, Eq.~\eqref{eq:DMB}.
However, integration by parts is only useful for the $(-\delta\dot z^{(\mathrm{III})})^{3/2}$ term.
Applying IBP to the other two terms generates factors of the form $\delta\ddot z^{(\mathrm{III})}/(-\delta\dot z^{(\mathrm{III})})^{1/2}$, for which the monotonically decreasing assumption required by OPB/DMB is not guaranteed.
Therefore, we either apply OPB directly to each term in Eq.~\eqref{eq:impf_delta_beta_III_1st_order} or perform IBP once on the $(-\delta\dot z)^{3/2}$ term and bound the associated integral by DMB.
All resulting bounds are summarized in Table~\ref{tb:impf_delta_beta_III_bounds_null}.

% ----------------------------------------------------------------------------------------------------
For Region~I, the analysis is slightly more involved, since the expansion depends on the early-time asymptotic behavior of the trajectory.
Via the substitutions in Eq.~\eqref{eq:region_III_to_I}, Eq.~\eqref{eq:impf_beta_exact_III_expand} becomes
\begin{align}
    \beta^{(\mathrm{I})}_{\omega\omega'} 
    &= 
    -\frac{i\alpha}{4\pi\sqrt{\omega \omega'}} 
    e^{-i\omega_- C_A} 
    \int_{-\infty}^{t_A} dt\,
    e^{i \Omega^{(\mathrm{I})}_2 t}\,
    e^{ i\omega_- \delta z^{(\mathrm{I})}(t)}
    \notag \\
    & \times\sqrt{1-\left( \dot{z}^{(\mathrm{I})}_{\text{asym}}\right)^2
           + 2 \dot{z}^{(\mathrm{I})}_{\text{asym}}\,\delta\dot{z}^{(\mathrm{I})}(t)
           - \left( \delta\dot{z}^{(\mathrm{I})}(t)\right)^2}.
    \label{eq:impf_beta_exact_I_expand}
\end{align}
The appropriate expansion of the square root depends on the early-time asymptotic regime.
In the following, we consider two representative cases: \textit{asymptotic-null} and \textit{asymptotic-static}.

% ----------------------------------------------------------------------------------------------------
For trajectories that are \textit{asymptotically null} in the far past, one has $\dot z^{(\mathrm{I})}_{\mathrm{asym}}=-1$.
In this case the Region~I expansion and bounding procedure is identical to the Region~III analysis.
Accordingly, $\delta\beta^{(\mathrm{I})}_{\omega\omega'}$ is obtained from Eq.~\eqref{eq:impf_delta_beta_III_1st_order} via the substitutions in Eq.~\eqref{eq:region_III_to_I} together with the sign change of the IBP boundary term due to the reversed integration limits.
The resulting bounds are listed in Table~\ref{tb:impf_delta_beta_I_bounds_null}.

% ----------------------------------------------------------------------------------------------------
We now consider the \textit{asymptotically static} case in the far past, $\dot z^{(\mathrm{I})}_{\rm asym}=0$, as a simple representative case.
The more general case for the asymptotically inertial one can be treated analogously, at the cost of additional algebra.
Starting from Eq.~\eqref{eq:impf_beta_exact_I_expand}, the Region~I deviation reads
\begin{widetext}
\begin{align}
&\delta \beta^{(\mathrm{I})}_{\omega\omega'}
=
\frac{-i\alpha}{4\pi\sqrt{\omega \omega'}} 
e^{-i\omega_- C_A} 
\int_{-\infty}^{t_A}\! dt\,
\sqrt{1-\bigl(\delta\dot z^{(\mathrm{I})}(t)\bigr)^2}\;
e^{i\Omega^{(\mathrm{I})}_2 t}\,
e^{i\omega_- \delta z^{(\mathrm{I})}(t)}
-\beta^{(\mathrm{I})}_{\omega\omega',i}
\notag\\
& \qquad \ \approx
\frac{\alpha}{4\pi\sqrt{\omega \omega'}}
\Bigg[
\frac{-1}{\Omega^{(\mathrm{I})}_2}\,e^{-i\omega_- C_A}e^{i\Omega^{(\mathrm{I})}_2 t_A}
-\frac{4\pi\sqrt{\omega \omega'}}{\alpha}\,\beta^{(\mathrm{I})}_{\omega\omega',i}
% \notag\\
% &\hspace{3.2cm}
-i\,e^{-i\omega_- C_A}\!\int_{-\infty}^{t_A}\! dt\,
\Bigg(
i\omega_-\,\delta z^{(\mathrm{I})}(t)
-\frac{\omega_-^2}{2}\bigl(\delta z^{(\mathrm{I})}(t)\bigr)^2
-\frac{1}{2}\bigl(\delta\dot z^{(\mathrm{I})}(t)\bigr)^2
\Bigg)
e^{i\Omega^{(\mathrm{I})}_2 t}
\Bigg],
\label{eq:impf_delta_beta_I_1st_order_static}
\end{align}
\end{widetext}
where we expand $\sqrt{1-\delta\dot z^2} \approx 1-\delta\dot z^2/2$ and $\exp\left[{i\omega_- \delta z} \right] \approx 1+i\omega_- \delta z- \omega_-^2\delta z^2/2$, retaining terms up to quadratic order in the deviation parameters.
One may further apply integration by parts for multiple times to extract boundary contributions and then bound the remaining oscillatory integrals using OPB \eqref{eq:OPB} or DMB \eqref{eq:DMB}.
We summarize all \(24\) admissible strategies and the resulting bounds in Table~\ref{tb:impf_delta_beta_I_bounds_static}.

Finally, in the asymptotic limit $t_A\to -\infty$ and $t_B\to\infty$, one finds
\begin{equation}
\lim_{t_A\to -\infty}\left|\delta\beta^{(\mathrm{I})}_{\omega\omega'}\right|=0,
\qquad
\lim_{t_B\to\infty}\left|\delta\beta^{(\mathrm{III})}_{\omega\omega'}\right|=0 .
\end{equation}
This shows again that the IRM-induced deviation can be made arbitrarily small by pushing the matching points sufficiently far into the asymptotic regime.

% ----------------------------------------------------------------------------------------------------
Restoring $\delta\beta^{(\mathrm{II}),\mathrm{com}}_{\omega\omega'}$ and using Eq.~\eqref{eq:Delta_beta} to estimate the total bound on the squared magnitude, we apply the IRM to the Grav trajectory Eq.~\eqref{eq:grav_traj} with \YS{$\kappa=0.5 \ \rm{eV}$.}
\begin{figure*}[!htbp]
    \centering
    \includegraphics[width=0.49\linewidth]{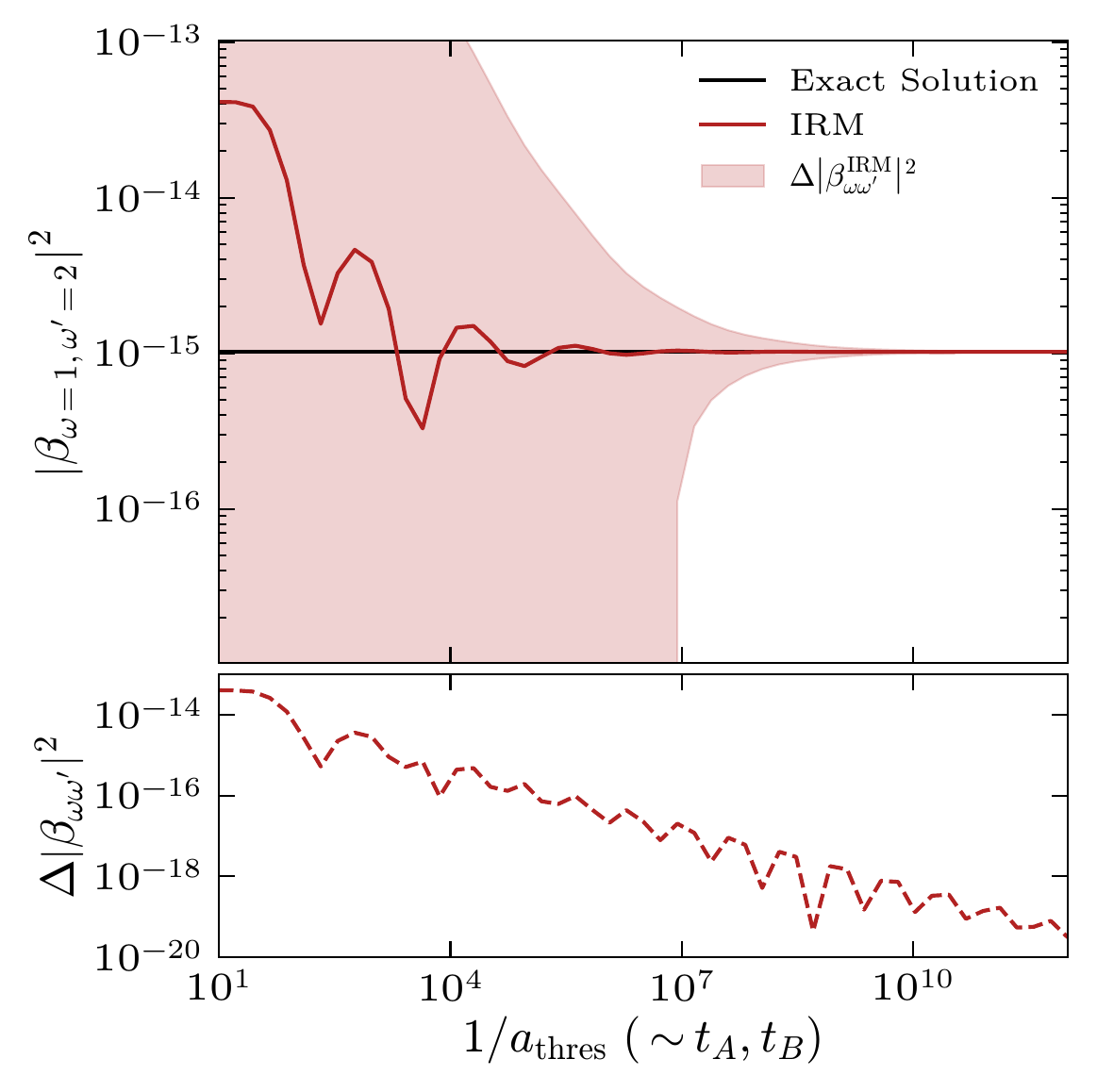}
    \includegraphics[width=0.49\linewidth]{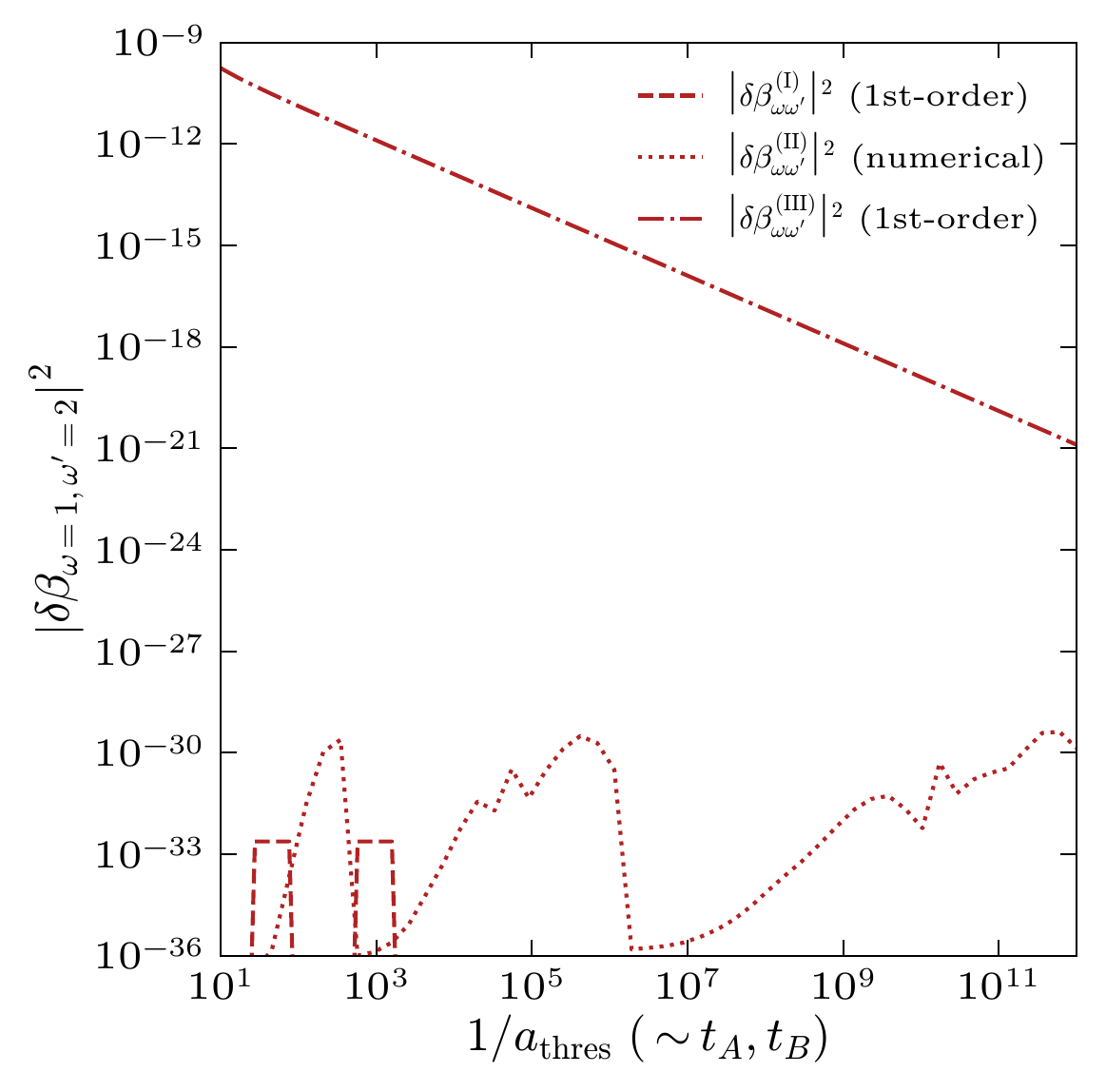}
    \caption{
    \textbf{Left:} $|\beta^{\mathrm{IRM}}_{\omega\omega'}|^{2}$ of \textbf{Grav} trajectory with \YS{$\kappa=0.5\ \rm{eV}$ and $\alpha=10^{-4}\ \rm{eV}$} is compared with the exact solution $|\beta^{\mathrm{Ext}}_{\omega\omega'}|^{2}$ as a function of $1/a_{\mathrm{thres}}$, with the residual \YS{$\Delta \left| \beta_{\omega\omega'}\right|^{2}$} shown below.
    \textbf{Right:} Comparison of the IRM-induced deviation and the estimated numerical calculation error.
    }
    \label{fig:grav_convergence_combined}
\end{figure*}
The convergence behavior of the IRM for the Grav trajectory is shown in Fig.~\ref{fig:grav_convergence_combined}.
In the left panel, as $1/a_{\mathrm{thres}}$ increases, the IRM prediction rapidly converges toward the exact solution, with the residual decreasing smoothly over many orders of magnitude.
The right panel decomposes the total deviation into contributions from Regions~I, II (numerical), and III.
The Region~I IRM-induced deviation is extremely small, reflecting its intrinsic static nature of the early-time Grav trajectory.
The small bumps arise from the discrete procedure used to determine $t_A$ and $t_B$: for a given $a_{\rm thres}$ the time axis is discretized and the boundaries are located using the criterion in Eq.~\eqref{eq:splitting_condition}.
These fluctuations are therefore numerical artifacts that diminish as the time grid is refined.
The oscillatory structure of the bound in Region~II originates from finite numerical precision and would be further reduced with higher floating-point accuracy.
Even at $1/a_{\mathrm{thres}}\sim10^{11}$ the numerical uncertainty remains well below the first-order asymptotic deviations, indicating that the crossover point where numerical error dominates has not yet been reached, which confirms that the IRM remains effective for imperfect mirrors.

% ----------------------------------------------------------------------------------------------------
\begin{figure}[!htbp]
    \includegraphics[width=1\linewidth]{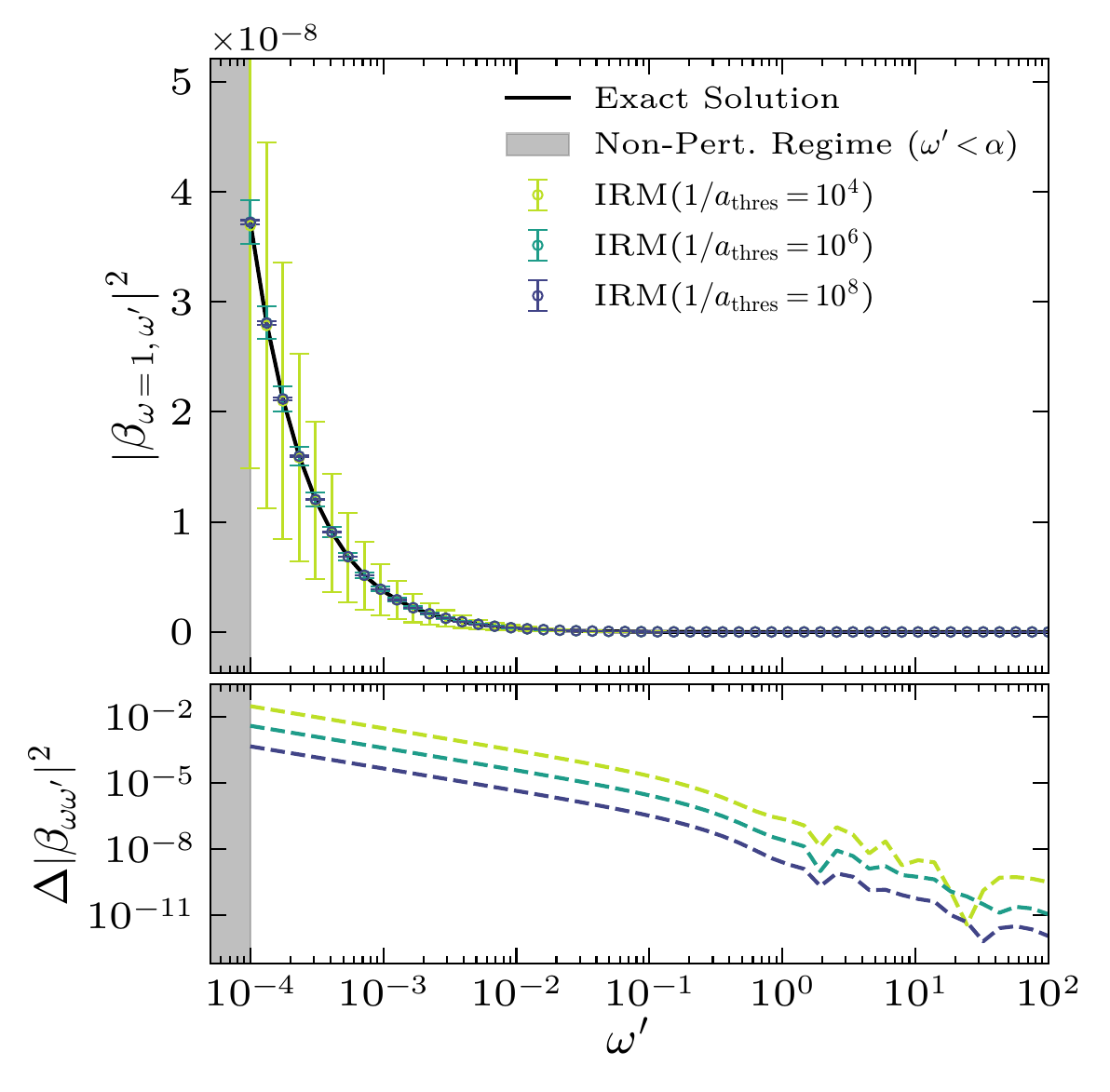}
    \caption{
    $|\beta^{\mathrm{IRM}}_{\omega\omega'}|^{2}$ of \textbf{Grav} trajectory with \YS{$\kappa=0.5\ \rm{eV}$ and $\alpha=10^{-4}\ \rm{eV}$} is compared with the exact solution $|\beta^{\mathrm{Ext}}_{\omega\omega'}|^{2}$ as a function of $\omega'$, with the residual \YS{$\Delta \left| \beta_{\omega\omega'}\right|^{2}$} shown below.
    }
    \label{fig:grav_beta2_wprime}
\end{figure}
\YS{
Sweeping over $\omega'$ with $\omega=1$ gives the frequency dependence shown in Fig.~\ref{fig:grav_beta2_wprime}, which displays a monotonically decreasing spectrum as $\omega'$ increases unlike the Logex trajectory, which exhibits a pronounced peak in Fig.~\ref{fig:logex_beta2_wprime}.
As we can see, the IRM results (colored points) closely track the analytic curve (solid black) across the entire frequency range.  
Even for a moderate threshold such as $1/a_{\mathrm{thres}}=10^{4}$ the IRM already reproduces the analytic spectrum with good accuracy, while larger thresholds ($1/a_{\mathrm{thres}}=10^{6},10^{8}$) suppress the deviation by several additional orders of magnitude.
This demonstrates that the IRM remains reliable even for trajectories without a spectral peak, accurately reproducing the Grav spectrum.
}

\YS{
A twofold clarification of the apparent infrared divergence is provided.
First, this behavior is already present in the analytic expression and does not arise from the IRM approximation. 
Second, the Bogoliubov coefficient in Eq.~\eqref{eq:impf_beta} is derived within a semitransparent mirror model using a perturbative expansion in the coupling $\alpha$, which is valid when $\alpha \ll \omega, \omega'$. 
In this regime, only the leading-order contribution is retained. 
In the low-frequency, or non-perturbative, regime $\omega' \lesssim \alpha$, shown as the gray shaded regions in Fig.~\ref{fig:grav_beta2_wprime} and Fig.~\ref{fig:anabhel_beta2_wprime}, this perturbative expansion breaks down, as higher-order terms in $\alpha/\omega'$ are no longer suppressed and become equally important. 
Consequently, taking the limit $\omega' \to 0$ within the truncated expression amounts to an extrapolation beyond its domain of validity. 
The resulting inverse-frequency growth should therefore not be interpreted as a physical infrared divergence.
This should be clearly distinguished from the case of a perfectly reflecting mirror, where prolonged acceleration can give rise to a genuine infrared divergence, which is intrinsically non-perturbative and cannot be captured by a finite-order expansion.
}

% ====================================================================================================
\section{The Chen--Mourou Trajectory}
\label{sec:IRM_impf_Chen_Trajectory}
% ----------------------------------------------------------------------------------------------------
Our main interest in this section is the experimentally motivated trajectory of a laser-plasma flying mirror currently being explored by the AnaBHEL collaboration \cite{2022_AnaBHEL}. 
We refer to this family as the Chen--Mourou trajectory, proposed by P.~Chen and G.~Mourou in \cite{2017_Pisin,2020_Pisin}. 
We focus in particular on the imperfectly reflecting case, for which the particle-number spectrum $\left|\beta_{\omega\omega'}\right|^{2}$ is analytically intractable. 
In this situation, the IRM provides a practical framework for obtaining the particle-number spectrum when no closed-form solution is available.
Note that in this section we restore the physical units.

% ----------------------------------------------------------------------------------------------------
The flying mirror is generated by a decreasing plasma-density profile designed to produce relativistic acceleration through laser-plasma wakefield dynamics.  
We perform a $(1+1)$D \revB{PIC} simulation using the \revB{Extendable PIC Open Collaboration (EPOCH) code~\cite{2015_Arber}} to model the mirror dynamics produced by a constant-plus-exponential plasma density profile,
\begin{equation}
n_p(z) =
\begin{cases}
4n_{p0}, & z \le z_0, \\[6pt]
n_{p0}\,\big[1 + e^{-(z-z_0)/D}\big]^{2}, & z > z_0 ,
\end{cases}
\label{eq:anabhel_np}
\end{equation}
which yields a late-time behavior resembling a Davies--Fulling trajectory and therefore provides a physically motivated analog of horizon formation in the laboratory.
The relevant parameters are as follows: 
$\lambda_0 = 800\,\mathrm{nm}$ is the wavelength of the driver laser,
\YS{$n_{p0} = 0.0005\,n_c$} is the asymptotic plasma density,
$n_c = 1.72\times10^{27}\,\mathrm{m^{-3}}$ is the critical density corresponding to the laser wavelength $\lambda_0$,
\YS{$z_0 = 600\,\mu\mathrm{m}$} marks the location where the density begins to fall,
and $D = 100\,\mu\mathrm{m}$ is the characteristic length of the exponential down-ramp, which controls the acceleration scale of the mirror.

\begin{figure*}[!htbp]
    \includegraphics[width=0.46\linewidth]{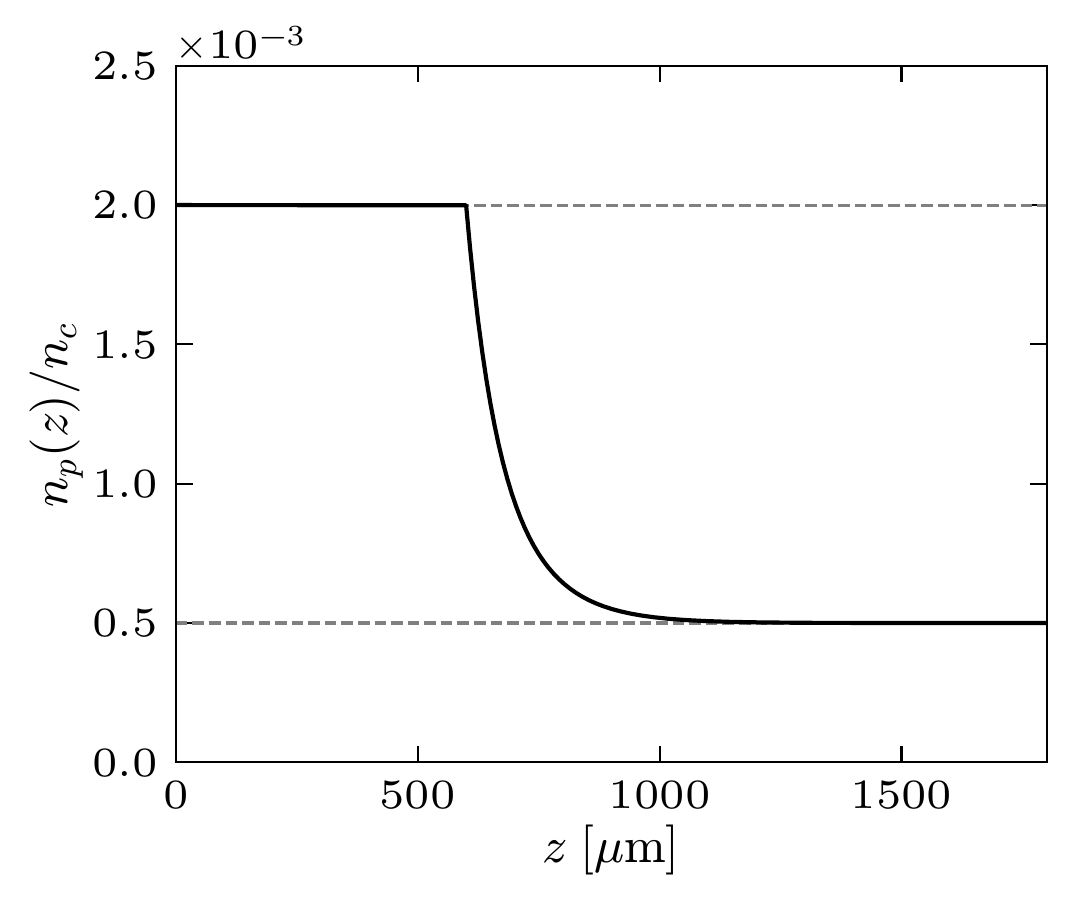}
    \includegraphics[width=0.53\linewidth]{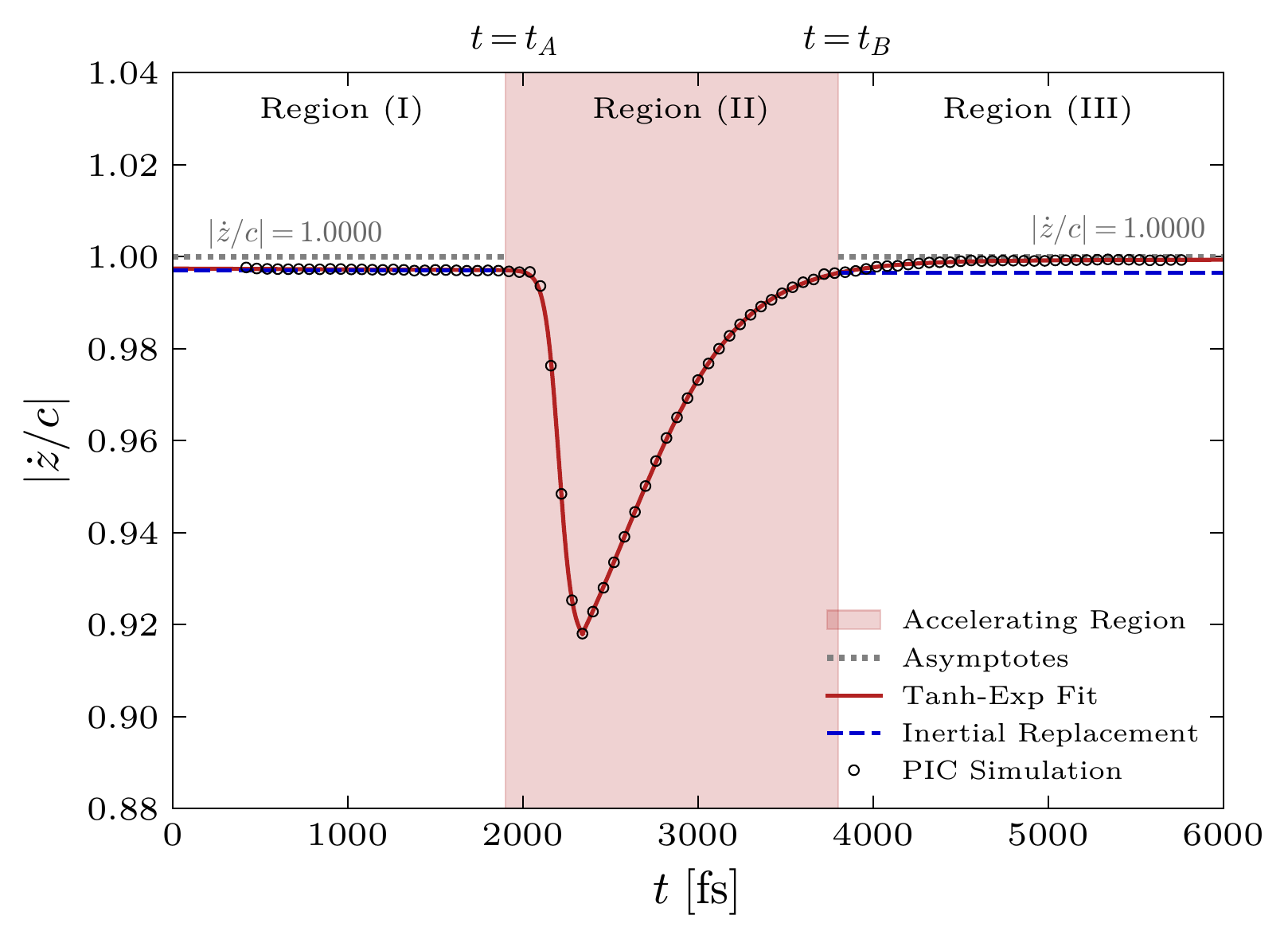}
    \caption{
    \textbf{Left:} Constant-plus-exponential plasma density profile used in the PIC simulation.  
    \textbf{Right:} Mirror velocity of the Chen--Mourou trajectory extracted from the PIC simulation and fitted by the Tanh--Exp model \YS{and the inertial extensions used in Regions I and III.}
    The asymptotic-null limits at early and late times are preserved.
    }
    \label{fig:anabhel_np_v}
\end{figure*}
The plasma density profile is shown in the left panel of Fig.~\ref{fig:anabhel_np_v}.
The gradual density down-ramp causes the local plasma frequency to decrease, which increases the wakefield phase velocity.
The right panel of Fig.~\ref{fig:anabhel_np_v} shows the mirror velocity extracted from the PIC simulation \revA{(black circles)}.
The mirror propagates near the speed of light on the high-density plateau, briefly decelerates upon entering the density gradient, and \revA{subsequently approaches the null asymptote} again as it propagates into the lower-density region.

To construct a smooth analytic trajectory suitable for the IRM analysis, we fit the PIC-extracted velocity $|\dot z|/c$ using a phenomenological \emph{tanh--exponential} (Tanh--Exp) model, which naturally provides the required asymptotic behavior.
A single global fit cannot simultaneously reproduce the full \revA{set of PIC-extracted velocity data} while preserving the asymptotic-null limits at early and late times.
We therefore divide the PIC data into two segments and fit each branch separately.
This procedure is fully consistent with the IRM, since the splitting point can be chosen within Region~II, where only continuity and integrability of the fitting function are required.

In practice, we identify the time $t=t_{\rm crit}$ at which the mirror reaches its minimum velocity $\dot z=\dot z_{\rm crit}$, close to the extreme value of the acceleration.
The fitted velocity profiles are then given by
\begin{widetext}
\begin{subequations}
\begin{align}
\frac{\dot z^{\rm L}(t)}{c}
&=
\frac{1}{2}\!\left[
-1
+ A_{\rm tanh}^{\rm L}
\bigg(
1 - \tanh\!\left(-\frac{t - t_0^{\rm L}}{\tau_{\tanh}^{\rm L}}\right)
\bigg)
\right]
+
\frac{1}{2}\!\left[-1 + A_{\exp}^{\rm L} e^{t / \tau_{\exp}^{\rm L}}\right],
\qquad t \le t_{\rm crit}, 
\label{eq:anabhel_vL_fit}
\\
\frac{\dot z^{\rm R}(t)}{c}
&=
\frac{1}{2}\!\left[
-1
+ A_{\rm tanh}^{\rm R}
\bigg(
1 - \tanh\!\left(\frac{t - t_0^{\rm R}}{\tau_{\tanh}^{\rm R}}\right)
\bigg)
\right]
+
\frac{1}{2}\!\left[-1 + A_{\exp}^{\rm R} e^{-t / \tau_{\exp}^{\rm R}}\right],
\qquad t \ge t_{\rm crit}.
\label{eq:anabhel_vR_fit}
\end{align}
\end{subequations}
\end{widetext}
The free fitting parameters are
\[
\{t_0^{\rm L},\, \tau_{\tanh}^{\rm L},\, A_{\exp}^{\rm L}>0,\, \tau_{\exp}^{\rm L}\},
\qquad
\{t_0^{\rm R},\, \tau_{\tanh}^{\rm R},\, A_{\exp}^{\rm R}>0,\, \tau_{\exp}^{\rm R}\},
\]
while $A_{\tanh}^{\rm L}>0$ and $A_{\tanh}^{\rm R}>0$ are fixed by enforcing continuity at the critical point,
\begin{equation}
\dot z^{\rm L}(t_{\rm crit})=\dot z^{\rm R}(t_{\rm crit})=\dot z_{\rm crit},
\end{equation}
where $\dot z_{\rm crit}$ is the minimum mirror velocity at the critical point.
The best-fit parameters, obtained by minimizing the least-squares residuals, are summarized in Table~\ref{tb:anabhel_best_params}.
\begin{table}
    \renewcommand{\arraystretch}{1.5}
    \caption{\label{tb:anabhel_best_params} Best-fit parameters for the Tanh--Exp velocity model.  
    Left side (L) corresponds to the pre-minimum branch, and Right side (R) corresponds to the post-minimum branch.}
    \begin{ruledtabular}
    \begin{tabular}{l | r | r | r | r | r}
        Branch &
        $t_{0}$ [fs] &
        $A_{\rm tanh}$ &
        $\tau_{\rm tanh}$ [fs] &
        $A_{\rm exp}$ &
        $\tau_{\rm exp}$ [fs] \\
        \hline
        Left (L)  & 2203.63 & 0.0812 & 76.06  & 0.0052 & 17318.00 \\
        \hline
        Right (R) & 2585.37 & 0.1189 & 646.00 & 0.0026 & 9219.35 \\
    \end{tabular}
    \end{ruledtabular}
\end{table}
The resulting piecewise Tanh--Exp fit, shown as the red curve in \revA{the right panel of} Fig.~\ref{fig:anabhel_np_v}, reproduces the PIC velocity profile with high accuracy across the entire trajectory.
This smooth matching follows from the favorable asymptotic behavior of the hyperbolic-tangent and exponential functions, which provide a controlled and physically consistent transition between the accelerating region and the far-past and far-future inertial motion.
Such well-behaved asymptotics are essential for applying the IRM.
Convergence requires pushing the boundaries $t_A$ and $t_B$ deep into regions where the trajectory approaches inertial motion with sufficiently small deviations.
The Tanh--Exp model ensures that these deviations decay smoothly and monotonically, thereby providing a reliable and convergent foundation for the IRM computation.

% ----------------------------------------------------------------------------------------------------
In addition, the fitted velocity is integrable on each branch, allowing the reconstructed trajectory $z(t)$ to be defined smoothly and used directly in the evaluation of the Bogoliubov-coefficient integrals.
Integrating Eq.~\eqref{eq:anabhel_vL_fit} and Eq.~\eqref{eq:anabhel_vR_fit} yields
\begin{widetext}
\begin{subequations}
\begin{align}
\frac{z^{\rm L}(t)}{c}
&=
\frac{1}{2}\!\left[
- t
+ A_{\rm tanh}^{\rm L}
\!\left(
t + \tau_{\rm tanh}^{\rm L}
\ln\!\left[
\cosh\!\left(
-\frac{t - t_{0}^{\rm L}}{\tau_{\rm tanh}^{\rm L}}
\right)
\right]
\right)
\right]
+\frac{1}{2}\!\left(
- t + A_{\exp}^{\rm L}\tau_{\exp}^{\rm L}
e^{t/\tau_{\exp}^{\rm L}}
\right)
+ C_{\rm L},
\label{eq:anabhel_xL_fit} \\
\frac{z^{\rm R}(t)}{c}
&=
\frac{1}{2}\!\left[
- t
+ A_{\rm tanh}^{\rm R}
\!\left(
t - \tau_{\rm tanh}^{\rm R}
\ln\!\left[
\cosh\!\left(
\frac{t - t_{0}^{\rm R}}{\tau_{\rm tanh}^{\rm R}}
\right)
\right]
\right)
\right]
+\frac{1}{2}\!\left(
- t - A_{\exp}^{\rm R}\tau_{\exp}^{\rm R}
e^{-t/\tau_{\exp}^{\rm R}}
\right)
+ C_{\rm R}.
\label{eq:anabhel_xR_fit}
\end{align}
\end{subequations}
\end{widetext}
The integration constants $C_{\rm L}$ and $C_{\rm R}$ are fixed by imposing continuity of the trajectory at the minimum-velocity point,
\begin{align}
z^{\rm L}(t_{\rm crit}) = z^{\rm R}(t_{\rm crit}) = z_{\rm crit}.
\end{align}

% ----------------------------------------------------------------------------------------------------
We now apply the IRM to compute the Bogoliubov coefficient for the Chen--Mourou trajectory using Eq.~\eqref{eq:beta_IRM} and estimate the corresponding bound on the squared magnitude using Eq.~\eqref{eq:Delta_beta}.
Since the Chen--Mourou trajectory is asymptotically null in both the far past and far future, the relevant bounds on the IRM-induced deviations are those in Table~\ref{tb:impf_delta_beta_I_bounds_null} for Region~I and Table~\ref{tb:impf_delta_beta_III_bounds_null} for Region~III.

Convergence requires pushing the asymptotic boundaries $t_A$ and $t_B$ deep into the inertial regimes by increasing $1/a_{\rm thres}$ so that the IRM-induced deviation terms become sufficiently small.
In this regime the Region~II integral becomes highly oscillatory, especially for large $\omega'$, where the Bogoliubov coefficient is intrinsically small and delicate cancellations occur.
To ensure numerical reliability, we employ the arbitrary-precision package \texttt{mpmath} together with \texttt{mpmath\_quad} for the numerical integration.
Each IRM evaluation is performed twice, with the second run using twice the working precision.
If the floating-point uncertainty, i.e. the difference between two results, is larger than $0.1\%\times|\beta^{\rm IRM}_{\omega\omega'}|^{2}$, the working precision is doubled and the comparison repeated.
This adaptive procedure guarantees that floating-point uncertainty in the asymptotic tails at large $t_A$ and $t_B$ remains well below the $0.1\%$ target.

A second numerical issue arises when the Region~II integration domain becomes large.  
Even with high working precision, the oscillatory integrand may be insufficiently resolved if sampled too coarsely.  
To address this issue, the integration interval is divided into many subintervals (e.g., 512), and the partial integrals are summed.
The calculation is then repeated with 1024 and 2048 subintervals (corresponding to $\times2$ and $\times4$ refinements) to verify stability.

\revA{
The detailed convergence analysis, together with the numerical precision and interval-refinement tests, is presented in Appendix~\ref{app:supplementary_analyses}.
Based on this analysis, we adopt the conservative choices:
\begin{subequations}
\begin{align}
& 1/a_{\rm thres} = 10^{18} \quad (\omega' < 1~\mathrm{eV}), \\
& 1/a_{\rm thres} = 10^{32} \quad (1~\mathrm{eV} \le \omega' \le 100~\mathrm{eV}),
\end{align}
\end{subequations}
for all subsequent calculations. 
These choices ensure that the estimated IRM-induced deviations remain well below the magnitude of the computed spectrum throughout the frequency range considered.
}

% ----------------------------------------------------------------------------------------------------
\begin{figure}[!htbp]
    \includegraphics[width=1\linewidth]{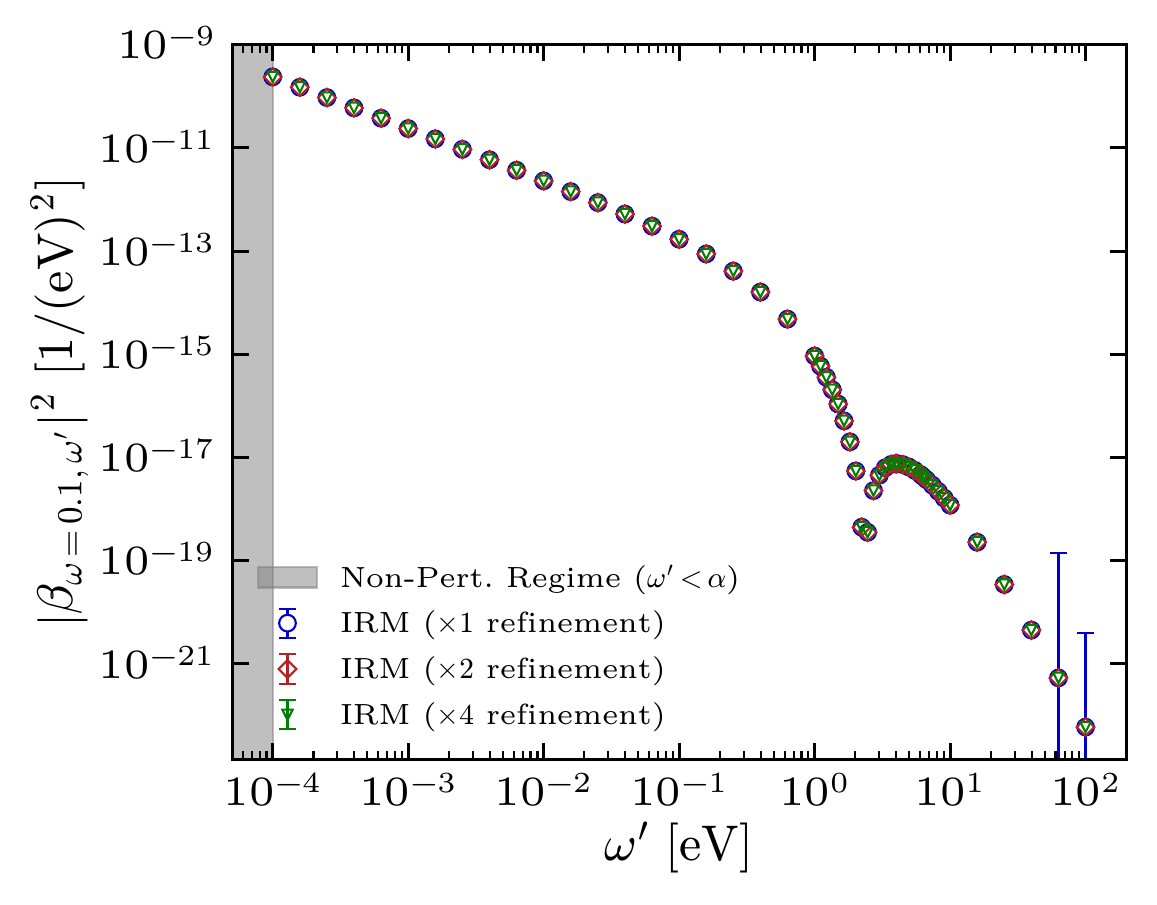}
    \caption{
    Frequency dependence of $|\beta^{\mathrm{IRM}}_{\omega\omega'}|^{2}$ with \YS{$\alpha=10^{-4}\ \rm{eV}$} for the \textbf{Chen--Mourou} trajectory.
    The IRM results with multiple refinement levels (colored markers) are shown together with the estimated uncertainty $\Delta|\beta^{\mathrm{IRM}}_{\omega\omega'}|^{2}$.  
    }
    \label{fig:anabhel_beta2_wprime}
\end{figure}
The resulting IRM spectrum for the Chen--Mourou trajectory is shown in Fig.\ref{fig:anabhel_beta2_wprime}, where we fix $\omega = 0.1\,\mathrm{eV}$ and scan over $10^{-4}\,\mathrm{eV} \le \omega' \le 100\,\mathrm{eV}$.  
The three sets of points correspond to different refinement levels of the Region~II integration domain ($\times1$, $\times2$, and $\times4$ subdivisions).
The error bar represents the bound of the total squared-magnitude uncertainty $\Delta|\beta^{\mathrm{IRM}}_{\omega\omega'}|^{2}$, which combines the IRM-induced deviation from Regions~I and ~III with the floating-point uncertainty in Region~II.
They do not include the interval-refinement error, which is instead assessed independently by comparing the results obtained with different refinement levels.

Across the entire frequency range the IRM spectrum is smooth and numerically stable.
At low and intermediate frequencies the error bars are extremely small, reflecting the strong suppression of both the analytic IRM-induced deviations and the numerical uncertainty.
At high frequencies ($\omega' \gtrsim 50\,\mathrm{eV}$), the $\times1$ refinement exhibits visible error bars.
This behavior does not indicate floating-point failure, but arises from insufficient resolution of the highly oscillatory Region~II integrand on the coarse grid, particularly in the far asymptotic tail where the integrand becomes extremely small.
In this regime a few outer sub-intervals are not fully resolved; although their contributions to the total integral are negligible, they can produce large local discrepancies when the working precision is doubled in the floating-point consistency test, thereby inflating the precision-based error estimate.

After grid refinement ($\times2$ and $\times4$), the oscillatory structure is properly resolved and the refined results coincide, demonstrating that the numerical integration is stable in the high-frequency regime.
Quantitatively, the maximal difference between the $\times1$ and $\times2$ results is \YS{$\mathcal{O}(10^{-67})$}, whereas the smallest values of $|\beta^{\mathrm{IRM}}_{\omega\omega'}|^{2}$ are \YS{$\mathcal{O}(10^{-23})$}.
This confirms that the refinement error is negligible compared with the physical magnitude of the spectrum.

% ====================================================================================================
\section{Conclusion}
\label{sec:Conclusion}
% ----------------------------------------------------------------------------------------------------
% IRM results review
In this work, we develop a new method, the Inertial Replacement Method (IRM), which provides a more generic and robust framework for computing Bogoliubov coefficients for moving-mirror trajectories associated with analog black holes. 
The IRM has been validated for several trajectories with known analytic solutions (Logex, Grav, Arcx, Uniform, and Hyperlog, whose analytic expressions were derived in~\cite{2011_Good}) and applied to a fully numerical trajectory extracted from PIC simulations and motivated by prospective experimental realizations.  
Across all cases studied, the IRM shows excellent numerical stability, rapid convergence under refinement, and quantitative agreement with analytic benchmarks.

% IRM insight
A central implication of the success of the IRM is that particle production from a moving mirror, as an analog black hole, can be accurately approximated by replacing the asymptotic tail with its inertial counterpart. 
While conventional treatments of particle creation, such as the Unruh effect~\cite{1981_Unruh} and the \YS{Davies--Fulling mirror~\cite{1976_FullingDavies, 1977_DaviesFulling}}, are often formulated in terms of the global structure of spacetime, \YS{where a characteristic logarithmic ray-tracing mapping defined globally between in- and out-going null coordinates leads to mode mixing and the emergence of particles, our results indicate that the particle number spectrum is less sensitive to this mapping in the far asymptotic region than in the finite accelerating region, as evidenced by the rapid convergence of $|\beta^{\rm IRM}_{\omega\omega'}|^2$ to the exact result $|\beta^{\rm Ext}_{\omega\omega'}|^2$ even at relatively small values of $1/a_{\rm thres}$ in Fig.~\ref{fig:logex_beta2_wprime} and Fig.~\ref{fig:grav_beta2_wprime}.}
The underlying reason for this behavior is not yet fully understood and requires further investigation.

% Entanglement issue
This observation also provides a perspective on the role of partner modes in moving-mirror radiation. 
In the standard picture of an accelerating mirror, each emitted Hawking particle is entangled with a partner mode that continues to follow the mirror trajectory, and such discussions are often framed in terms of an eternally accelerating mirror. 
Our results suggest that even if the mirror becomes inertial at a finite time, so that the partner mode gets reflected by the mirror earlier than in the eternally accelerating case, the resulting radiation spectrum remains essentially unchanged. 
This perspective may be related to recent studies of partner modes and late-time bursts in moving-mirror models~\cite{2015_Hotta,2024_Osawa}, as well as to investigations of the entanglement structure between Hawking and partner modes.

% Future study
Besides the extensions mentioned above, this work opens several promising directions for further study.  
First, a more quantitative analysis of the relative contributions from different trajectory segments, particularly through two-point correlation functions for piecewise trajectories, could provide further insight into the causal structure underlying particle creation. 
Varying the junction locations may also help assess the sensitivity of the spectrum to the precise boundaries of the accelerating region.  
Second, including higher-order terms in the exponential expansion of Eq.~\eqref{eq:exp_expansion} would clarify the behavior of the IRM in the high-frequency regime, where the first-order estimation becomes less accurate.
Finally, extending the IRM to (1+3)D \YS{Chen--Mourou trajectories is an important step toward realistic modeling of the AnaBHEL experiment and enables a more direct comparison with physical measurements.}

% ====================================================================================================
\begin{acknowledgments}
We would like to thank Robert M. Wald of the University of Chicago, Kuan-Nan Lin of APCTP, Yung-Kun Liu of NTU, Harry Goodhew of NTU, and Justin Feng of CEICO for useful discussions. 
Y.-S. H. acknowledges financial support from LeCosPA and support from NTU for participation in the UCTS program, during which part of this work was completed. 
P. C. acknowledges support from Taiwan's National Science and Technology Council (NSTC) under Project No.~112-2112-M-002-013 and from LeCosPA at NTU.
\end{acknowledgments}

% ====================================================================================================
\appendix
% ----------------------------------------------------------------------------------------------------
\section{Oscillatory-Phase Bound for Oscillatory Integrals}
\label{app:OPB_bound}
% ----------------------------------------------------------------------------------------------------
In this appendix, we derive an oscillatory-phase bound using standard tools from harmonic analysis (see, e.g. \cite{1993_Stein}).  
Consider the oscillatory integral
\begin{align}
I &\equiv \int_{t_0}^{\infty} f(t)\, e^{i\Omega_0 t}\, dt \notag \\
&= \int_{t_0}^{\infty} f(t) \cos(\Omega_0 t)\, dt
  + i\int_{t_0}^{\infty} f(t)\sin(\Omega_0 t)\, dt ,
\label{eq:oscillatory_integral_split}
\end{align}
where \(f(t)\) is real, monotonically decreasing, and sufficiently smooth, with \(f(t)\to0\) as \(t\to\infty\).
We bound the real and imaginary parts separately:
\begin{equation}
|I|
\le \left|\int_{t_0}^{\infty} f(t)\cos(\Omega_0 t)\, dt\right|
   +\left|\int_{t_0}^{\infty} f(t)\sin(\Omega_0 t)\, dt\right|.
\end{equation}
Let us begin with the cosine term. Define
\begin{equation}
C(t) = \int_{t_0}^{t} \cos(\Omega_0\tau)\, d\tau ,
\quad\Rightarrow\quad
C'(t)=\cos(\Omega_0 t).
\end{equation}
A direct evaluation gives
\begin{align}
|C(t)|
&= \left|\frac{1}{\Omega_0}\big[\sin(\Omega_0 t)-\sin(\Omega_0 t_0)\big]\right| \notag \\
& \le \frac{|\sin(\Omega_0 t)|+|\sin(\Omega_0 t_0)|}{|\Omega_0|} \notag \\
& \le \frac{2}{|\Omega_0|}.
\end{align}
On the other hand, using integration by parts,
\begin{align}
\int_{t_0}^{\infty} f(t)\cos(\Omega_0 t)\, dt
&= \left[f(t)C(t)\right]_{t_0}^{\infty}
  -\int_{t_0}^{\infty} f'(t)C(t)\, dt \notag \\
&= - f(t_0)C(t_0)
   -\int_{t_0}^{\infty} f'(t)C(t)\, dt ,
\end{align}
and therefore
\begin{align}
\left|\int_{t_0}^{\infty} f(t)\cos(\Omega_0 t)\, dt\right|
&\le |f(t_0)|\,|C(t_0)|
   + \int_{t_0}^{\infty} |f'(t)|\, |C(t)|\, dt \notag \\
&\le \frac{2|f(t_0)|}{|\Omega_0|}
   + \frac{2}{|\Omega_0|}\!\int_{t_0}^{\infty}\! |f'(t)|\, dt.
\end{align}
Since \(f(t)\) is assumed monotonically decreasing,  
\begin{align}
\int_{t_0}^{\infty} |f'(t)|\, dt
= -\int_{t_0}^{\infty} f'(t)\, dt
= \big|f(t_0)-\lim_{t\to\infty}f(t)\big|
= |f(t_0)|,
\end{align}
so we obtain
\begin{equation}
\left|\int_{t_0}^{\infty} f(t)\cos(\Omega_0 t)\, dt\right|
\le \frac{4|f(t_0)|}{|\Omega_0|}.
\end{equation}
For the sine term, an identical calculation using
\begin{align}
S(t)=\int_{t_0}^{t}\sin(\Omega_0\tau)\, d\tau,
\quad |S(t)|\le \frac{2}{|\Omega_0|}
\end{align}
gives the same bound:
\begin{equation}
\left|\int_{t_0}^{\infty} f(t)\sin(\Omega_0 t)\, dt\right|
\le \frac{4|f(t_0)|}{|\Omega_0|}.
\end{equation}
Adding the cosine and sine contributions yields
\begin{equation}
\left| I \right| \equiv
\left| \int_{t_0}^{\infty} f(t)\, e^{i\Omega_0 t}\, dt \right|
\le \frac{8\, |f(t_0)|}{|\Omega_0|}.
\label{eq:oscillatory_bound}
\end{equation}

This is a Dirichlet/van der Corput-type estimate.
It exhibits the characteristic \(1/|\Omega_0|\) suppression associated with rapidly oscillating phases and provides a sharp and practical upper bound for the high-frequency regime relevant to the IRM error analysis.

% ====================================================================================================
\section{Direct-Magnitude Bound for Oscillatory Integrals}
\label{app:DMB_bound}
% ----------------------------------------------------------------------------------------------------
In this appendix, we derive the direct-magnitude bound used in the main text.  
Consider the oscillatory integral in Eq.~\eqref{eq:oscillatory_integral_split}.  
Define the tail integral
\begin{equation}
F(t) \;\equiv\; \int_{t}^{\infty} f(\tau)\, d\tau,
\qquad\Rightarrow\qquad
F(t_0) = \int_{t_0}^{\infty} f(t)\, dt .
\end{equation}
Then, for any real frequency $\Omega_0$,
\begin{align}
\left| \int_{t_0}^{\infty} f(t)\, e^{i\Omega_0 t}\, dt \right|
&\le \int_{t_0}^{\infty} |f(t)|\, |e^{i\Omega_0 t}|\, dt \notag\\[4pt]
&= \int_{t_0}^{\infty} |f(t)|\, dt \notag\\[4pt]
&= \left| \int_{t_0}^{\infty} f(t)\, dt \right| \notag\\[4pt]
&= |F(t_0)|. \label{eq:direct_bound}
\end{align}
In the third line we used the fact that $f(t)$ is monotonically decreasing, so the absolute value may be taken outside the integral.

This bound is independent of $\Omega_0$ and therefore remains effective in the low-frequency regime, where oscillatory cancellation is weak and the oscillatory-phase bound of Appendix~\ref{app:OPB_bound} does not provide useful suppression.

% ====================================================================================================
\section{Supplementary Analyses}
\label{app:supplementary_analyses}
\YS{
This appendix provides supplementary convergence analyses and parameter-selection details that support the implementation of the IRM. 
These figures complement the discussion in Sec.~\ref{sec:IRM_pf} and Sec.~\ref{sec:IRM_impf_Chen_Trajectory} but are not required for understanding the method itself. 
They are included here for completeness and to document the numerical procedure used throughout the paper.
}

% ----------------------------------------------------------------------------------------------------
\subsubsection{Frequency dependence of the IRM bound}
\begin{figure*}[!htbp]
    \centering
    \includegraphics[width=0.49\linewidth]{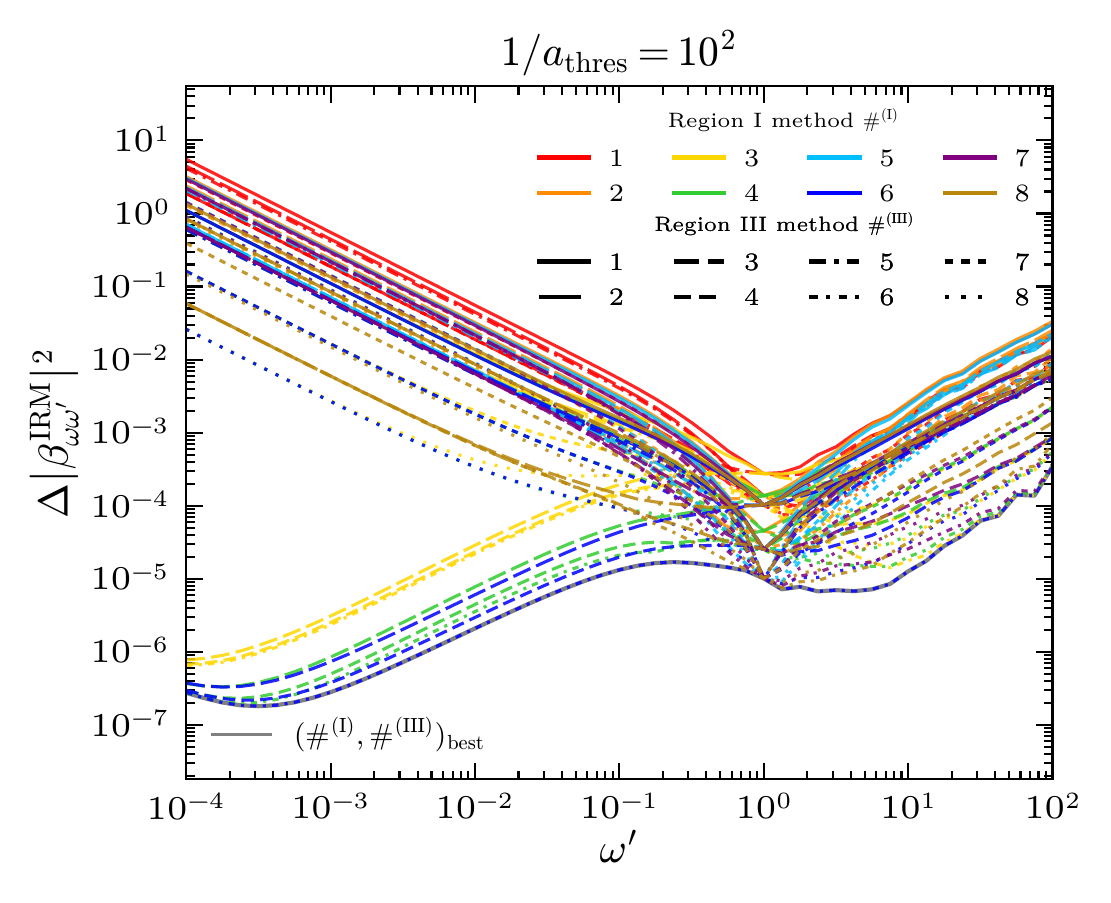}
    \includegraphics[width=0.49\linewidth]{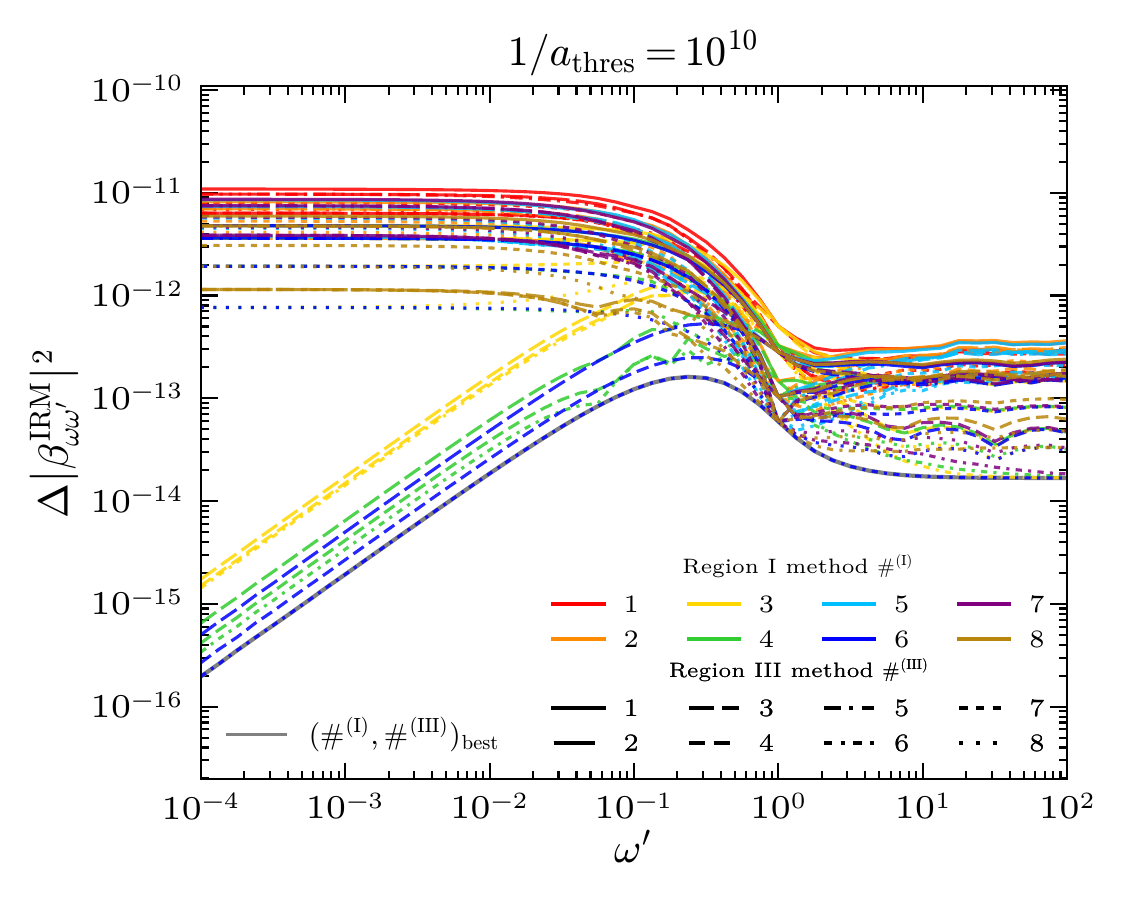}
    \caption{
    Frequency dependence of $\Delta|\beta^{\mathrm{IRM}}_{\omega\omega'}|^{2}$ for all $8\times8$ Region~I/III bounding combinations and two representative values of $1/a_{\mathrm{thres}}$.
    Colored and styled curves correspond to different bounding strategies, and the gray curve marks the optimal combination in each panel.
    }
    \label{fig:logex_Dbeta2_wprime_ij}
\end{figure*}
The line plots Fig.~\ref{fig:logex_Dbeta2_wprime_ij} highlight the frequency structure of $\Delta|\beta^{\mathrm{IRM}}_{\omega\omega'}|^{2}$ for fixed values of $1/a_{\mathrm{thres}}$.
Each panel displays all $64$ bounding combinations, with the optimal strategy shown in gray.
Distinct behavior across different frequency regimes is visible, reflecting the nontrivial coupling between frequencies and dynamical deviation parameters.
In the infrared ($\omega'\to 0$), the curves separate into two characteristic branches: for large $a_{\mathrm{thres}}$, one branch grows as $\omega'\to 0$, whereas for small $a_{\mathrm{thres}}$, the growth transitions into a plateau.
In the ultraviolet, the curves approach an approximately $\omega'$-independent floor for sufficiently small $a_{\mathrm{thres}}$.
The turnover between the infrared and ultraviolet regimes occurs around $\omega' \sim \omega = 1$, where the bounding terms involving $\omega_- = \omega - \omega'$ approach zero, leading to a local minimum in the squared-magnitude bound.

% ----------------------------------------------------------------------------------------------------
\subsubsection{Optimization of the matching strategy}
\begin{figure*}[!htbp]
    \centering
    \includegraphics[width=0.49\linewidth]{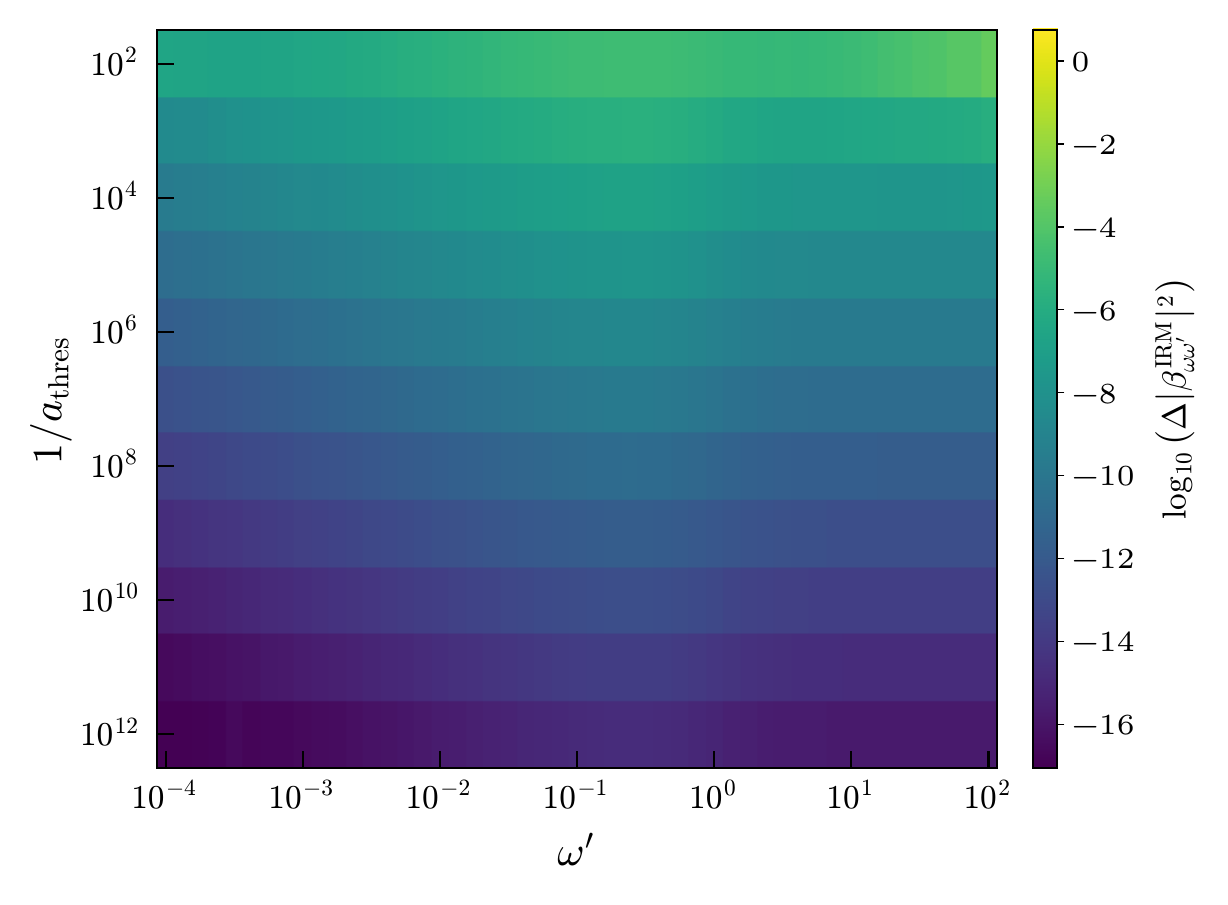}
    \includegraphics[width=0.49\linewidth]{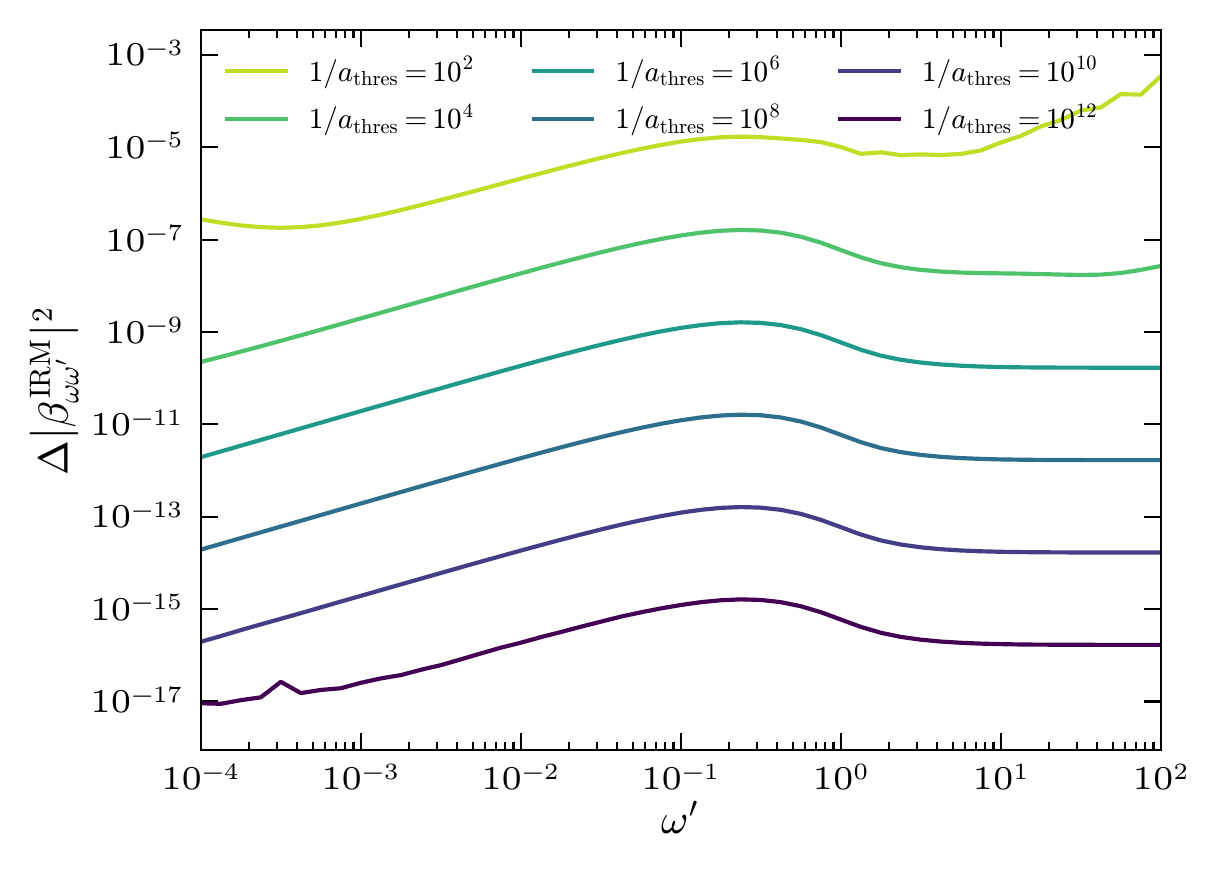}
    \caption{
    \textbf{Left:} Minimal $\Delta|\beta^{\mathrm{IRM}}_{\omega\omega'}|^{2}$ over all $8\times8$ Region~I/III bounding combinations, shown as a function of $(\omega',1/a_{\mathrm{thres}})$.
    \textbf{Right:} Frequency dependence of $\Delta|\beta^{\mathrm{IRM}}_{\omega\omega'}|^{2}$ for representative values of $1/a_{\mathrm{thres}}$.
    Increasing $1/a_{\mathrm{thres}}$ systematically suppresses the IRM deviation.}
    \label{fig:logex_Dbeta2_wprime_athres_bestij}
\end{figure*}
Fig.~\ref{fig:logex_Dbeta2_wprime_athres_bestij} summarizes the optimal $\Delta |\beta^{\mathrm{IRM}}_{\omega\omega'}|^{2}$ obtained after scanning all $8\times8$ bounding combinations \YS{and motivates the parameter choices discussed in Sec.~\ref{sec:IRM_pf}.}
The left panel shows, for each $(\omega',1/a_{\mathrm{thres}})$, the minimal squared-magnitude bound among the 64 possibilities, while the right panel displays representative frequency slices.
Two systematic features become quantitatively clear in this figure.
First, the plot now reveals the precise magnitude scaling of the squared-magnitude bound: each time $1/a_{\mathrm{thres}}$ increases by one order of magnitude, the bound $\Delta |\beta^{\mathrm{IRM}}_{\omega\omega'}|^{2}$ is likewise suppressed by roughly one order of magnitude.
This confirms that for trajectories with well-defined asymptotic behavior, one can systematically reduce the bound by pushing the matching points deeper into the asymptotic region.
Second, for fixed $1/a_{\mathrm{thres}}$, the minimal deviation shows a smooth frequency profile without divergences in either the infrared or ultraviolet.
Compared with the method-by-method plots, the optimized bound displays reduced sensitivity to frequency, indicating that adaptive bound selection yields stable accuracy across the full parameter range.

% ----------------------------------------------------------------------------------------------------
\subsubsection{Convergence and parameter selection}

\begin{figure*}[!htbp]
    \includegraphics[width=0.49\linewidth]{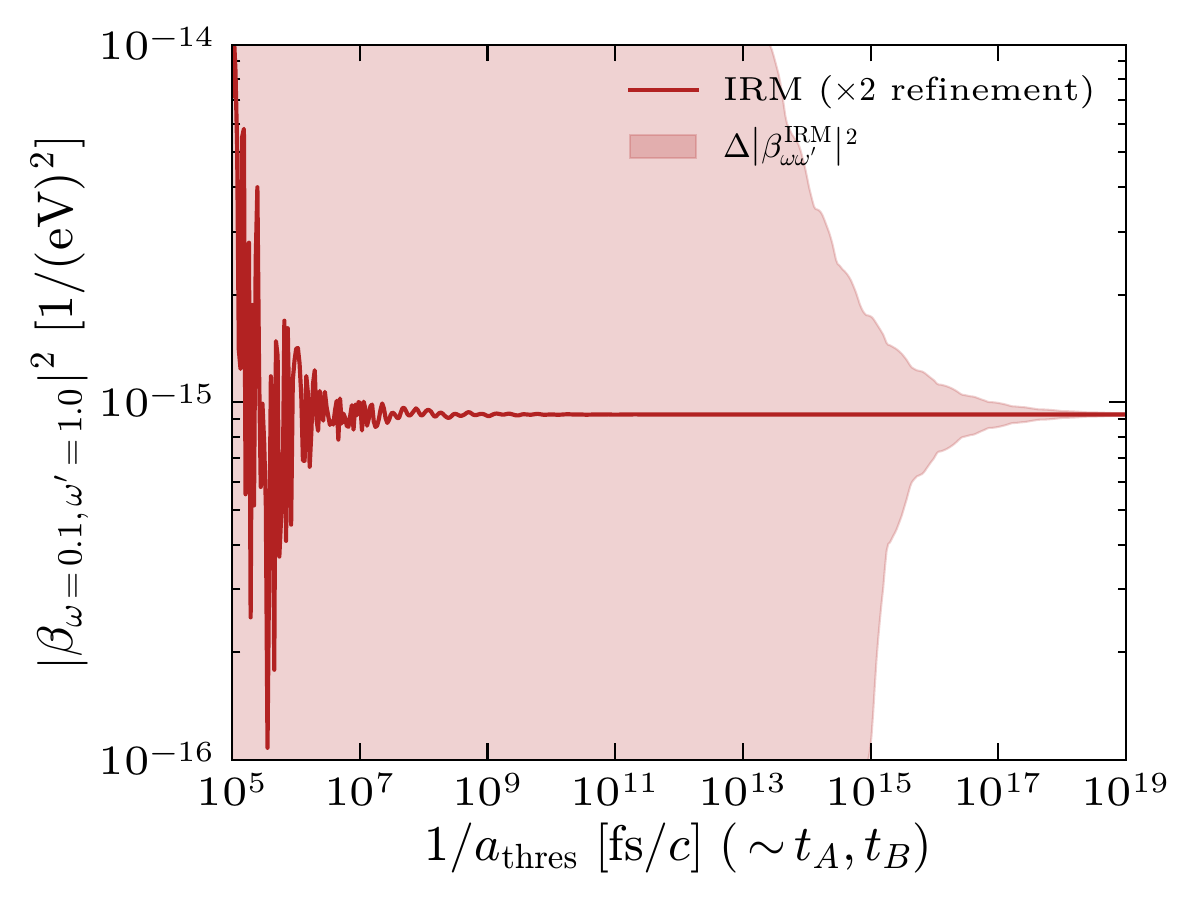}
    \includegraphics[width=0.49\linewidth]{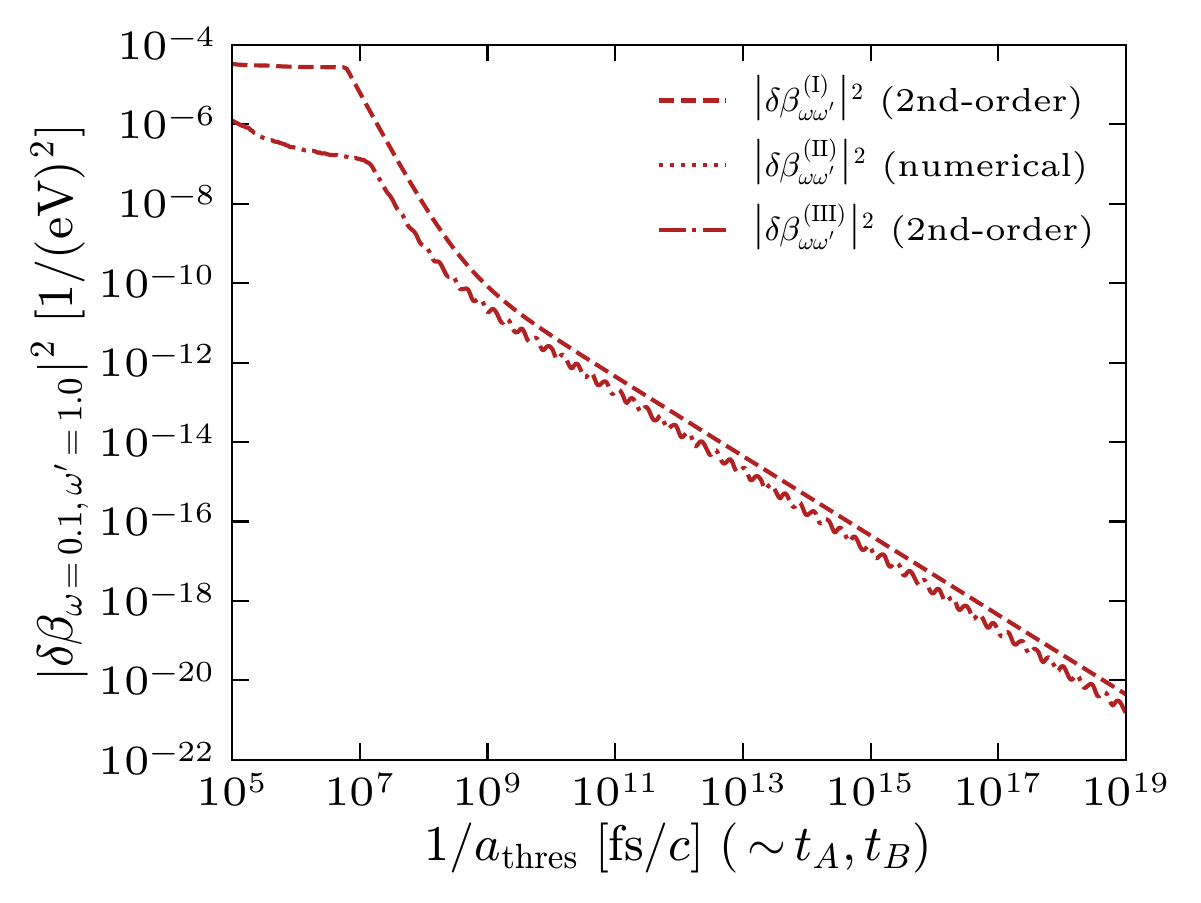}
    \includegraphics[width=0.49\linewidth]{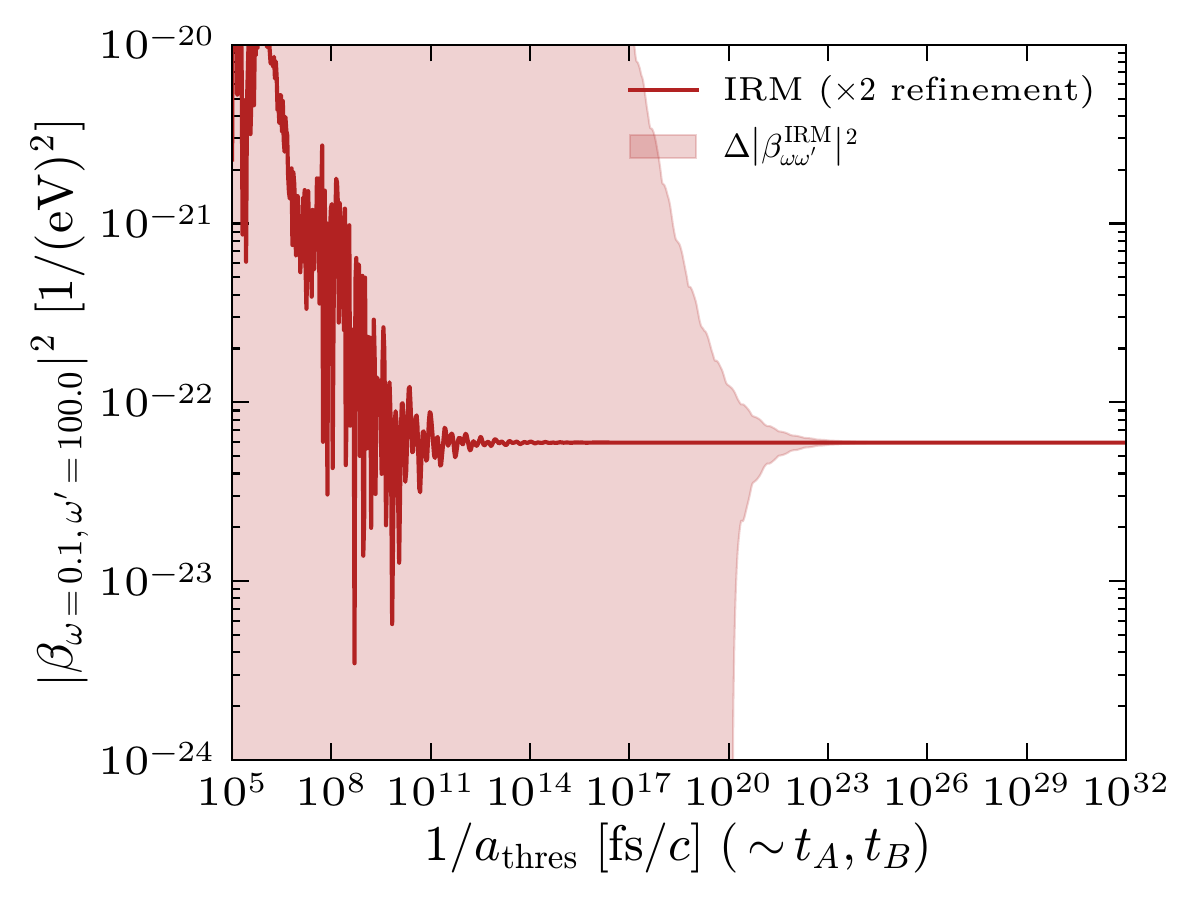}
    \includegraphics[width=0.49\linewidth]{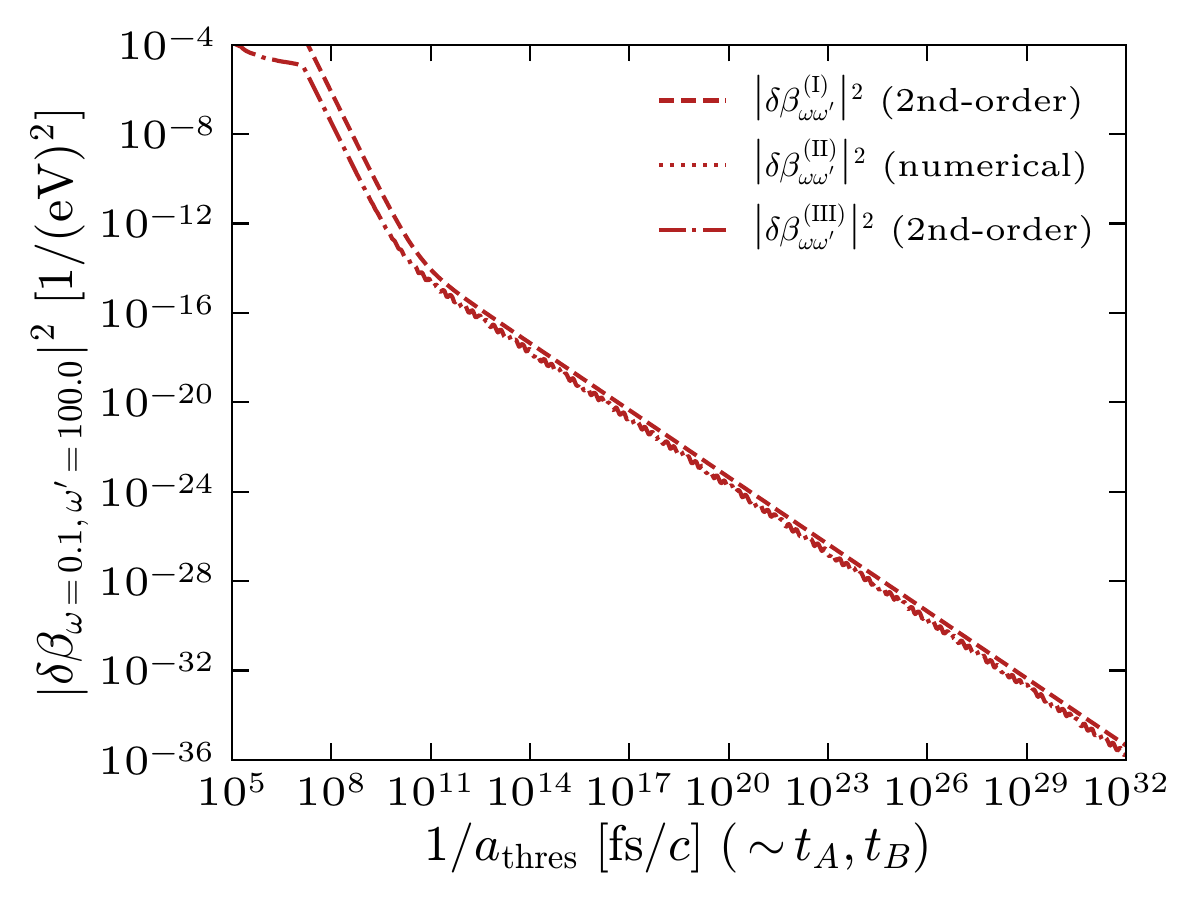}
    \caption{
    Convergence of the IRM for the Chen--Mourou trajectory with \YS{$\alpha=10^{-4}\ \rm{eV}$,} $(\omega,\omega')=(0.1~\mathrm{eV},\,1~\mathrm{eV})$ (top row), and $(\omega,\omega')=(0.1~\mathrm{eV},\,100~\mathrm{eV})$ (bottom row).  
    \textbf{Left panels:} IRM evaluations of $|\beta^{\mathrm{IRM}}_{\omega\omega'}|^{2}$ as a function of $1/a_{\mathrm{thres}}$.
    \textbf{Right panels:} Comparison of the IRM-induced deviation and the estimated numerical calculation error.
    Here $\delta \beta^{(\mathrm{II})}_{\omega\omega'}$ accounts only for the floating-point precision error and does not include the interval refinement error.
    }
    \label{fig:anabhel_convergence_w'=(1,100)_combined}
\end{figure*}
\YS{Fig.~\ref{fig:anabhel_convergence_w'=(1,100)_combined} provides additional convergence tests used to determine the matching parameter adopted in the main text.}
The left panels show the convergence of the IRM result as a function of $1/a_{\rm thres}$, while the right panels compare the estimated IRM-induced deviation from Regions~I and III with the floating-point uncertainty from the numerical integration over Region~II.
Throughout the parameter range considered, the floating-point uncertainty remains well below the IRM-induced deviation.
These convergence curves illustrate how far the asymptotic matching boundaries must be pushed to achieve a desired numerical accuracy.
For $\omega'=1~\mathrm{eV}$, the total error falls below $0.1\%\times |\beta^{\rm IRM}_{\omega\omega'}|^{2}$ at $1/a_{\rm thres}\sim10^{17}$, whereas for $\omega'=100~\mathrm{eV}$ the same accuracy is reached only after $1/a_{\rm thres}\sim10^{23}$.
\YS{Accordingly, the following conservative choices are adopted throughout this work:
\begin{align}
& 1/a_{\rm thres}=10^{18},
\qquad (\omega'<1~\mathrm{eV}), \notag\\
& 1/a_{\rm thres}=10^{32},
\qquad (1~\mathrm{eV}\le\omega'\le100~\mathrm{eV}). \notag
\end{align}
At these values, the analytic deviations associated with Regions~I and III are already well below the magnitude of the IRM result, while the floating-point uncertainty remains negligible. 
These supplementary tests therefore justify the parameter choices used in all subsequent calculations.
}
% ====================================================================================================
\clearpage
\onecolumngrid
\section{Summary of the IRM Bounds}
\label{app:Bound_Summary}
In this appendix, we summarize all the resulting bounds discussed in the main text.
The bounding strategies and corresponding bounds are listed in the following tables: \\
Table~\ref{tb:pf_delta_beta_III_bounds}: \textbf{perfectly} reflecting mirror with \textbf{late-time} \textbf{asymptotic-null} or \textbf{asymptotic-inertial} behavior, \\
Table~\ref{tb:pf_delta_beta_I_bounds}: \textbf{perfectly} reflecting mirror with \textbf{early-time} \textbf{asymptotic-null} or \textbf{asymptotic-inertial} behavior, \\
Table~\ref{tb:impf_delta_beta_III_bounds_null}: \textbf{imperfectly} reflecting mirror with \textbf{late-time} \textbf{asymptotic-null} behavior, \\  
Table~\ref{tb:impf_delta_beta_I_bounds_null}: \textbf{imperfectly} reflecting mirror with \textbf{early-time} \textbf{asymptotic-null} behavior, \\  
Table~\ref{tb:impf_delta_beta_I_bounds_static}: \textbf{imperfectly} reflecting mirror with \textbf{early-time} \textbf{asymptotic-static} behavior. \\
Throughout these tables, IBP denotes integration by parts, OPB the oscillatory-phase bound, and DMB the direct-magnitude bound.
Each row represents a distinct bounding strategy involving different numbers of IBP operations and combinations of the OPB and DMB.

% ----------------------------------------------------------------------------------------------------
% Pf, III, Asymptotic-inertial
% \clearpage
\begin{table}[h]
\setcellgapes{2pt}\makegapedcells
\caption{\label{tb:pf_delta_beta_III_bounds}
Summary of the bounding strategies and the corresponding bounds for $\delta\beta^{(\mathrm{III})}_{\omega\omega'}$ in Eq.~\eqref{eq:pf_delta_beta_III_1st_order} for a perfectly reflecting mirror following a late-time asymptotic-null or asymptotic-inertial trajectory.
}
\resizebox{\textwidth}{!}{%
\begin{tabular}{c|c|c|c}
\Xhline{1.25pt}
\textbf{$\#^{(\mathrm{III})}$} &
\textbf{Strategy} &
\textbf{$4\pi \sqrt{\omega \omega'}\delta \beta_{\omega \omega'}^{(\mathrm{III}),\mathrm{com}}$} &
\textbf{$4\pi \sqrt{\omega \omega'}\left| \delta \beta_{\omega \omega'}^{(\mathrm{III}),\mathrm{abs}} \right|$} \\
\Xhline{1.25pt}
1 &
\makecell[l]{0 IBP \\ OPB $\to \delta z^{(\mathrm{III})}, \delta\dot z^{(\mathrm{III})}$} &
\makecell[l]{%
$\displaystyle \left( i\,\frac{\Omega^{(\mathrm{III})}_{1}}{\Omega^{(\mathrm{III})}_{2}} \right) e^{-i \omega_- C_B} e^{i \Omega^{(\mathrm{III})}_{2} t_B} - {4\pi\sqrt{\omega \omega'}}\beta^{(\mathrm{III})}_{\omega \omega',i}$} &
\makecell[l]{$\displaystyle  \left| \frac{8\omega_- \Omega^{(\mathrm{III})}_{1} \delta z_B}{\Omega^{(\mathrm{III})}_{2}}\right|
+ \left| \frac{8\omega_+ \delta\dot z_B}{\Omega^{(\mathrm{III})}_{2}} \right|$} \\
\hline
2 &
\makecell[l]{0 IBP \\ OPB $\to \delta z^{(\mathrm{III})}$ \\ DMB $\to \delta\dot z^{(\mathrm{III})}$} & 
\makecell[l]{same as row 1} &
\makecell[l]{$\displaystyle  \left| \frac{8\omega_- \Omega^{(\mathrm{III})}_{1} \delta z_B}{\Omega^{(\mathrm{III})}_{2}}\right|
+ \left| \omega_+ \delta z_B \right|$ }\\
\Xhline{1.25pt}
3 &
\makecell[l]{1 IBP $\to \delta z^{(\mathrm{III})}$ \\ OPB $\to \delta\dot z^{(\mathrm{III})}$} & 
\makecell[l]{%
$\displaystyle \left (i\,\frac{\Omega^{(\mathrm{III})}_{1}}{\Omega^{(\mathrm{III})}_{2}} -\dfrac{\omega_- \Omega^{(\mathrm{III})}_1 \delta z_B}{\Omega^{(\mathrm{III})}_2} \right) e^{-i \omega_- C_B} e^{i \Omega^{(\mathrm{III})}_{2} t_B}- {4\pi\sqrt{\omega \omega'}}\beta^{(\mathrm{III})}_{\omega \omega',i}$} & 
\makecell[l]{$\displaystyle  \left| \frac{8 \left( \omega_+^2 - \omega_-^2\right) \delta \dot{z}_B}{\left( \Omega^{(\mathrm{III})}_{2}\right)^2}\right|$} \\
\hline
4 &
\makecell[l]{1 IBP $\to \delta z^{(\mathrm{III})}$ \\ DMB $\to \delta\dot z^{(\mathrm{III})}$} & 
\makecell[l]{same as row 3} &
\makecell[l]{$\displaystyle  \left| \frac{\left( \omega_+^2 - \omega_-^2\right) \delta z_B}{\Omega^{(\mathrm{III})}_{2}}\right|$} \\
\hline
5 &
\makecell[l]{1 IBP $\to \delta \dot{z}^{(\mathrm{III})}$ \\ OPB $\to \delta z^{(\mathrm{III})}$ \\ DMB $\to \delta\ddot z^{(\mathrm{III})}$} & 
\makecell[l]{$\displaystyle \left (i\,\frac{\Omega^{(\mathrm{III})}_{1}}{\Omega^{(\mathrm{III})}_{2}} -i\dfrac{\omega_+ \delta \dot{z}_B}{\Omega^{(\mathrm{III})}_2} \right) e^{-i \omega_- C_B}e^{i \Omega^{(\mathrm{III})}_{2} t_B} - {4\pi\sqrt{\omega \omega'}}\beta^{(\mathrm{III})}_{\omega \omega',i}$ }& 
\makecell[l]{$\displaystyle  \left| \frac{8\omega_- \Omega^{(\mathrm{III})}_{1} \delta z_B}{\Omega^{(\mathrm{III})}_{2}}\right|
+ \left| \frac{\omega_+ \delta\dot z_B}{\Omega^{(\mathrm{III})}_{2}} \right|$} \\
\Xhline{1.25pt}
6 &
\makecell[l]{2 IBP $\to \delta z^{(\mathrm{III})}, \delta \dot{z}^{(\mathrm{III})}$ \\ DMB $\to \delta\ddot z^{(\mathrm{III})}$} & 
\makecell[l]{$\displaystyle \left (i\,\frac{\Omega^{(\mathrm{III})}_{1}}{\Omega^{(\mathrm{III})}_{2}} -\dfrac{\omega_- \Omega^{(\mathrm{III})}_1 \delta z_B}{\Omega^{(\mathrm{III})}_2} - i\dfrac{\left( \omega_+^2 - \omega_-^2 \right )\delta \dot{z}_B }{\left( \Omega^{(\mathrm{III})}_2 \right) ^2}\right) e^{-i \omega_- C_B}e^{i \Omega^{(\mathrm{III})}_{2} t_B} - {4\pi\sqrt{\omega \omega'}} \beta^{(\mathrm{III})}_{\omega \omega',i} $ } & 
\makecell[l]{$\displaystyle  \left| \frac{\left( \omega_+^2 - \omega_-^2\right) \delta \dot{z}_B}{\left( \Omega^{(\mathrm{III})}_{2} \right )^2}\right|$} \\
\hline
7 &
\makecell[l]{2 IBP $\to \delta \dot{z}^{(\mathrm{III})}, \delta z^{(\mathrm{III})}$ \\ OPB $\to \delta \dot{z}^{(\mathrm{III})}$ \\ DMB $\to \delta\ddot z^{(\mathrm{III})}$} & 
\makecell[l]{%
\centering
$\displaystyle \left (i\,\frac{\Omega^{(\mathrm{III})}_{1}}{\Omega^{(\mathrm{III})}_{2}} -\dfrac{\omega_- \Omega^{(\mathrm{III})}_1 \delta z_B}{\Omega^{(\mathrm{III})}_2} - i\dfrac{\omega_+\delta \dot{z}_B }{ \Omega^{(\mathrm{III})}_2 }\right) e^{-i \omega_- C_B}e^{i \Omega^{(\mathrm{III})}_{2} t_B} - {4\pi\sqrt{\omega \omega'}} \beta^{(\mathrm{III})}_{\omega \omega',i} $ } & 
\makecell[l]{$\displaystyle \left| \frac{8\omega_- \Omega^{(\mathrm{III})}_{1} \delta \dot{z}_B}{\left( \Omega^{(\mathrm{III})}_{2}\right)^2}\right| + \left| \dfrac{\omega_+ \delta \dot{z}_B}{\Omega^{(\mathrm{III})}_2} \right|$} \\
\hline
8 &
\makecell[l]{2 IBP $\to \delta \dot{z}^{(\mathrm{III})}, \delta z^{(\mathrm{III})}$ \\ DMB $\to \delta \dot{z}^{(\mathrm{III})}, \delta\ddot z^{(\mathrm{III})}$} & 
\makecell[l]{same as row 7} & 
\makecell[l]{$\displaystyle \left| \frac{\omega_- \Omega^{(\mathrm{III})}_{1} \delta z_B}{\Omega^{(\mathrm{III})}_{2}}\right| + \left| \dfrac{\omega_+ \delta \dot{z}_B}{\Omega^{(\mathrm{III})}_2} \right|$} \\
\Xhline{1.25pt}
\end{tabular}
}
\end{table}
% ----------------------------------------------------------------------------------------------------
% Pf, I, Asymptotic-inertial
% \clearpage
\begin{table}[t]
\setcellgapes{2pt}\makegapedcells
\caption{\label{tb:pf_delta_beta_I_bounds}
Summary of the bounding strategies and the corresponding bounds for $\delta\beta^{(\mathrm{I})}_{\omega\omega'}$ in Eq.~\eqref{eq:pf_delta_beta_I_1st_order} for a perfectly reflecting mirror following an early-time asymptotic-null or asymptotic-inertial trajectory.
}
\resizebox{\textwidth}{!}{%
\begin{tabular}{c|c|c|c}
\Xhline{1.25pt}
\textbf{$\#^{(\mathrm{I})}$} &
\textbf{Strategy} &
\textbf{$4\pi \sqrt{\omega \omega'}\delta \beta_{\omega \omega'}^{(\mathrm{I}),\mathrm{com}}$} &
\textbf{$4\pi \sqrt{\omega \omega'}\left| \delta \beta_{\omega \omega'}^{(\mathrm{I}),\mathrm{abs}} \right|$} \\
\Xhline{1.25pt}
1 &
\makecell[l]{0 IBP \\ OPB $\to \delta z^{(\mathrm{I})}, \delta\dot z^{(\mathrm{I})}$} & 
\makecell[l]{%
$\displaystyle - \left ( i\,\frac{\Omega^{(\mathrm{I})}_{1}}{\Omega^{(\mathrm{I})}_{2}}\right) e^{-i \omega_- C_A} e^{i \Omega^{(\mathrm{I})}_{2} t_A} - {4\pi\sqrt{\omega \omega'}}\beta^{(\mathrm{I})}_{\omega \omega',i}$ } & 
\makecell[l]{$\displaystyle  \left| \frac{8\omega_- \Omega^{(\mathrm{I})}_{1} \delta z_A}{\Omega^{(\mathrm{I})}_{2}}\right|
+ \left| \frac{8\omega_+ \delta\dot z_A}{\Omega^{(\mathrm{I})}_{2}} \right|$ } \\
\hline
2 &
\makecell[l]{0 IBP \\ OPB $\to \delta z^{(\mathrm{I})}$ \\ DMB $\to \delta\dot z^{(\mathrm{I})}$} & 
\makecell[l]{same as row 1} & 
\makecell[l]{$\displaystyle  \left| \frac{8\omega_- \Omega^{(\mathrm{I})}_{1} \delta z_A}{\Omega^{(\mathrm{I})}_{2}}\right|
+ \left| \omega_+ \delta z_A \right|$} \\
\Xhline{1.25pt}
3 &
\makecell[l]{1 IBP $\to \delta z^{(\mathrm{I})}$ \\ OPB $\to \delta\dot z^{(\mathrm{I})}$} & 
\makecell[l]{%
$\displaystyle - \left (i\,\frac{\Omega^{(\mathrm{I})}_{1}}{\Omega^{(\mathrm{I})}_{2}} -\dfrac{\omega_- \Omega^{(\mathrm{I})}_1 \delta z_A}{\Omega^{(\mathrm{I})}_2} \right) e^{-i \omega_- C_A} e^{i \Omega^{(\mathrm{I})}_{2} t_A} - {4\pi\sqrt{\omega \omega'}} \beta^{(\mathrm{I})}_{\omega \omega',i}$ } & 
\makecell[l]{$\displaystyle  \left| \frac{8 \left( \omega_+^2 - \omega_-^2\right) \delta \dot{z}_A}{\left( \Omega^{(\mathrm{I})}_{2}\right)^2}\right|$} \\
\hline
4 &
\makecell[l]{1 IBP $\to \delta z^{(\mathrm{I})}$ \\ DMB $\to \delta\dot z^{(\mathrm{I})}$} & 
\makecell[l]{same as row 3} & 
\makecell[l]{$\displaystyle  \left| \frac{\left( \omega_+^2 - \omega_-^2\right) \delta z_A}{\Omega^{(\mathrm{I})}_{2}}\right|$} \\
\hline
5 &
\makecell[l]{1 IBP $\to \delta \dot{z}^{(\mathrm{I})}$ \\ OPB $\to \delta z^{(\mathrm{I})}$ \\ DMB $\to \delta\ddot z^{(\mathrm{I})}$} & 
\makecell[l]{$\displaystyle - \left (i\,\frac{\Omega^{(\mathrm{I})}_{1}}{\Omega^{(\mathrm{I})}_{2}} -i\dfrac{\omega_+ \delta \dot{z}_A}{\Omega^{(\mathrm{I})}_2} \right) e^{-i \omega_- C_A} e^{i \Omega^{(\mathrm{I})}_{2} t_A} - {4\pi\sqrt{\omega \omega'}} \beta^{(\mathrm{I})}_{\omega \omega',i}$} & 
\makecell[l]{$\displaystyle  \left| \frac{8\omega_- \Omega^{(\mathrm{I})}_{1} \delta z_A}{\Omega^{(\mathrm{I})}_{2}}\right|
+ \left| \frac{\omega_+ \delta\dot z_A}{\Omega^{(\mathrm{I})}_{2}} \right|$} \\
\Xhline{1.25pt}
6 &
\makecell[l]{2 IBP $\to \delta z^{(\mathrm{I})}, \delta \dot{z}^{(\mathrm{I})}$ \\ DMB $\to \delta\ddot z^{(\mathrm{I})}$} & 
\makecell[l]{$\displaystyle - \left (i\,\frac{\Omega^{(\mathrm{I})}_{1}}{\Omega^{(\mathrm{I})}_{2}} -\dfrac{\omega_- \Omega^{(\mathrm{I})}_1 \delta z_A}{\Omega^{(\mathrm{I})}_2} - i\dfrac{\left( \omega_+^2 - \omega_-^2 \right )\delta \dot{z}_A }{\left( \Omega^{(\mathrm{I})}_2 \right) ^2}\right) e^{-i \omega_- C_A} e^{i \Omega^{(\mathrm{I})}_{2} t_A} - {4\pi\sqrt{\omega \omega'}} \beta^{(\mathrm{I})}_{\omega \omega',i} $ } & 
\makecell[l]{$\displaystyle  \left| \frac{\left( \omega_+^2 - \omega_-^2\right) \delta \dot{z}_A}{\left( \Omega^{(\mathrm{I})}_{2} \right )^2}\right|$} \\
\hline
7 &
\makecell[l]{2 IBP $\to \delta \dot{z}^{(\mathrm{I})}, \delta z^{(\mathrm{I})}$ \\ OPB $\to \delta \dot{z}^{(\mathrm{I})}$ \\ DMB $\to \delta\ddot z^{(\mathrm{I})}$} & 
\makecell[l]{%
$\displaystyle - \left (i\,\frac{\Omega^{(\mathrm{I})}_{1}}{\Omega^{(\mathrm{I})}_{2}} -\dfrac{\omega_- \Omega^{(\mathrm{I})}_1 \delta z_A}{\Omega^{(\mathrm{I})}_2} - i\dfrac{\omega_+\delta \dot{z}_A }{ \Omega^{(\mathrm{I})}_2 }\right) e^{-i \omega_- C_A} e^{i \Omega^{(\mathrm{I})}_{2} t_A} - {4\pi\sqrt{\omega \omega'}} \beta^{(\mathrm{I})}_{\omega \omega',i} $ } & 
\makecell[l]{$\displaystyle \left| \frac{8\omega_- \Omega^{(\mathrm{I})}_{1} \delta \dot{z}_A}{\left( \Omega^{(\mathrm{I})}_{2}\right)^2}\right| + \left| \dfrac{\omega_+ \delta \dot{z}_A}{\Omega^{(\mathrm{I})}_2} \right|$} \\
\hline
8 &
\makecell[l]{2 IBP $\to \delta \dot{z}^{(\mathrm{I})}, \delta z^{(\mathrm{I})}$ \\ DMB $\to \delta \dot{z}^{(\mathrm{I})}, \delta\ddot z^{(\mathrm{I})}$} & 
\makecell[l]{same as row 7} & 
\makecell[l]{$\displaystyle \left| \frac{\omega_- \Omega^{(\mathrm{I})}_{1} \delta z_A}{\Omega^{(\mathrm{I})}_{2}}\right| + \left| \dfrac{\omega_+ \delta \dot{z}_A}{\Omega^{(\mathrm{I})}_2} \right|$} \\
\Xhline{1.25pt}
\end{tabular}
}
\end{table}
% ----------------------------------------------------------------------------------------------------
% Impf, III, Asymptotic-null
% \clearpage
\begin{table}[h]
\setcellgapes{2pt}\makegapedcells
\caption{\label{tb:impf_delta_beta_III_bounds_null}
Summary of the bounding strategies and the corresponding bounds for $\delta\beta^{(\mathrm{III})}_{\omega\omega'}$ in Eq.~\eqref{eq:impf_delta_beta_III_1st_order} for an imperfectly reflecting mirror following a late-time asymptotic-null trajectory.
}
\resizebox{\textwidth}{!}{%
\begin{tabular}{c|c|c|c}
\Xhline{1.25pt}
\textbf{$\#^{(\mathrm{III})}$} &
\textbf{Strategy} &
\textbf{$\dfrac{4\pi\sqrt{\omega \omega'}}{\alpha} \delta \beta_{\omega \omega'}^{(\mathrm{III}),\mathrm{com}}$} &
\textbf{$\dfrac{4\pi\sqrt{\omega \omega'}}{\alpha} \left| \delta \beta_{\omega \omega'}^{(\mathrm{III}),\mathrm{abs}} \right|$} \\
\Xhline{1.25pt}
1 &
\makecell[l]{0 IBP \\ OPB $\to \left(-\delta \dot{z}^{(\mathrm{III})}\right)^{1/2}, \left(-\delta \dot{z}^{(\mathrm{III})}\right)^{3/2}, \delta z^{(\mathrm{III})}\left(-\delta \dot{z}^{(\mathrm{III})}\right)^{1/2}$} & 
\makecell[l]{$\displaystyle -\dfrac{4\pi\sqrt{\omega \omega'}}{\alpha} \beta^{(\mathrm{III})}_{\omega \omega',i} $ } & 
\makecell[l]{ $\displaystyle  \left| \frac{8\sqrt{2}\left(-\delta \dot{z}_B\right)^{1/2}}{\Omega^{(\mathrm{III})}_{2}} \right|
+ \left| \frac{2\sqrt{2}\left(-\delta \dot{z}_B\right)^{3/2}}{\Omega^{(\mathrm{III})}_{2}} \right|
+ \left| \frac{8\sqrt{2}\omega_-\delta z_B \left(-\delta \dot{z}_B\right)^{1/2}}{\Omega^{(\mathrm{III})}_{2}} \right|$} \\
\Xhline{1.25pt}
2 &
\makecell[l]{1 IBP $\to \left( -\delta \dot{z}^{\mathrm{(III)}}\right)^{3/2}$ \\ OPB $\to \left(-\delta \dot{z}^{(\mathrm{III})}\right)^{1/2}, \delta z^{(\mathrm{III})}\left(-\delta \dot{z}^{(\mathrm{III})}\right)^{1/2}$ \\ DMB $\to \left(-\delta \dot{z}^{(\mathrm{III})}\right)^{1/2} \left(-\delta \ddot{z}^{(\mathrm{III})} \right)$} & 
\makecell[l]{$\displaystyle \left( -\,\frac{\sqrt{2}\left(-\delta \dot{z}_B\right)^{3/2} }{4\Omega^{\mathrm{(III)}}_{2}} \right)  e^{-i \omega_- C_B} e^{i \Omega^{\mathrm{(III)}}_{2} t_B} - \dfrac{4\pi\sqrt{\omega \omega'}}{\alpha}\beta^{\mathrm{(III)}}_{\omega \omega',i} $ } & 
\makecell[l]{ $\displaystyle  \left| \frac{8\sqrt{2}\left(-\delta \dot{z}_B\right)^{1/2}}{\Omega^{(\mathrm{III})}_{2}} \right|
+ \left| \frac{\sqrt{2}\left(-\delta \dot{z}_B\right)^{3/2}}{4\Omega^{(\mathrm{III})}_{2}} \right|
+ \left| \frac{8\sqrt{2}\omega_-\delta z_B \left(-\delta \dot{z}_B\right)^{1/2}}{\Omega^{(\mathrm{III})}_{2}} \right|$} \\
\Xhline{1.25pt}
\end{tabular}
}
\end{table}
% ----------------------------------------------------------------------------------------------------
% Impf, I, Asymptotic-null
% \clearpage
\begin{table}[h]
\setcellgapes{2pt}\makegapedcells
\caption{\label{tb:impf_delta_beta_I_bounds_null}
Summary of the bounding strategies and the corresponding bounds for $\delta\beta^{(\mathrm{I})}_{\omega\omega'}$ for an imperfectly reflecting mirror following an early-time asymptotic-null trajectory.
}
\resizebox{\textwidth}{!}{%
\begin{tabular}{c|c|c|c}
\Xhline{1.25pt}
\textbf{$\#^{(\mathrm{I})}$} &
\textbf{Strategy} &
\textbf{$\dfrac{4\pi\sqrt{\omega \omega'}}{\alpha} \delta \beta_{\omega \omega'}^{(\mathrm{I}),\mathrm{com}}$} &
\textbf{$\dfrac{4\pi\sqrt{\omega \omega'}}{\alpha} \left| \delta \beta_{\omega \omega'}^{(\mathrm{I}),\mathrm{abs}} \right|$} \\
\Xhline{1.25pt}
1 &
\makecell[l]{0 IBP \\ OPB $\to \left(-\delta \dot{z}^{(\mathrm{I})}\right)^{1/2}, \left(-\delta \dot{z}^{(\mathrm{I})}\right)^{3/2}, \delta z^{(\mathrm{I})}\left(-\delta \dot{z}^{(\mathrm{I})}\right)^{1/2}$} & 
\makecell[l]{$\displaystyle -\dfrac{4\pi\sqrt{\omega \omega'}}{\alpha} \beta^{(\mathrm{I})}_{\omega \omega',i} $ } & 
\makecell[l]{ $\displaystyle  \left| \frac{8\sqrt{2}\left(-\delta \dot{z}_A\right)^{1/2}}{\Omega^{(\mathrm{I})}_{2}} \right|
+ \left| \frac{2\sqrt{2}\left(-\delta \dot{z}_A\right)^{3/2}}{\Omega^{(\mathrm{I})}_{2}} \right|
+ \left| \frac{8\sqrt{2}\omega_-\delta z_A \left(-\delta \dot{z}_A\right)^{1/2}}{\Omega^{(\mathrm{I})}_{2}} \right|$} \\
\Xhline{1.25pt}
2 &
\makecell[l]{1 IBP $\to \left( -\delta \dot{z}^{\mathrm{(I)}}\right)^{3/2}$ \\ OPB $\to \left(-\delta \dot{z}^{(\mathrm{I})}\right)^{1/2}, \delta z^{(\mathrm{I})}\left(-\delta \dot{z}^{(\mathrm{I})}\right)^{1/2}$ \\ DMB $\to \left(-\delta \dot{z}^{(\mathrm{I})}\right)^{1/2} \left(-\delta \ddot{z}^{(\mathrm{I})} \right)$} & 
\makecell[l]{$\displaystyle \left( \,\frac{\sqrt{2}\left(-\delta \dot{z}_A\right)^{3/2}}{4\Omega^{\mathrm{(I)}}_{2}} \right)  e^{-i \omega_- C_A} e^{i \Omega^{\mathrm{(I)}}_{2} t_A} - \dfrac{4\pi\sqrt{\omega \omega'}}{\alpha}\beta^{\mathrm{(I)}}_{\omega \omega',i} $ } & 
\makecell[l]{ $\displaystyle  \left| \frac{8\sqrt{2}\left(-\delta \dot{z}_A\right)^{1/2}}{\Omega^{(\mathrm{I})}_{2}} \right|
+ \left| \frac{\sqrt{2}\left(-\delta \dot{z}_A\right)^{3/2}}{4\Omega^{(\mathrm{I})}_{2}} \right|
+ \left| \frac{8\sqrt{2}\omega_-\delta z_A \left(-\delta \dot{z}_A\right)^{1/2}}{\Omega^{(\mathrm{I})}_{2}} \right|$} \\
\Xhline{1.25pt}
\end{tabular}
}
\end{table}
% ----------------------------------------------------------------------------------------------------
% Impf, I, Asymptotic-static
% \clearpage
\begin{table}[t]
\setcellgapes{2pt}\makegapedcells
\caption{\label{tb:impf_delta_beta_I_bounds_static}
Summary of the bounding strategies and the corresponding bounds for $\delta\beta^{(\mathrm{I})}_{\omega\omega'}$ in Eq.~\eqref{eq:impf_delta_beta_I_1st_order_static} for an imperfectly reflecting mirror following an early-time asymptotic-static trajectory.
}
\resizebox{\textwidth}{!}{%
\begin{tabular}{c|c|c|c}
\Xhline{1.25pt}
\textbf{$\#^{\mathrm{(I)}}$} &
\textbf{Strategy} &
\textbf{$\dfrac{4\pi\sqrt{\omega \omega'}}{\alpha} \delta \beta_{\omega \omega'}^{\mathrm{(I)},\mathrm{com}}$} &
\textbf{$\dfrac{4\pi\sqrt{\omega \omega'}}{\alpha} \left| \delta \beta_{\omega \omega'}^{\mathrm{(I)},\mathrm{abs}} \right|$} \\
\Xhline{1.25pt}
1 &
\makecell[l]{0 IBP \\ OPB $\to \delta z^{\mathrm{(I)}}, \left(\delta z^{\mathrm{(I)}}\right)^2, \left(\delta \dot{z}^{\mathrm{(I)}}\right)^2$} & 
\makecell[l]{$\displaystyle \left( -\,\frac{1}{\Omega^{\mathrm{(I)}}_{2}} \right)  e^{-i \omega_- C_A} e^{i \Omega^{\mathrm{(I)}}_{2} t_A} - \dfrac{4\pi\sqrt{\omega \omega'}}{\alpha}\beta^{\mathrm{(I)}}_{\omega \omega',i} $ } & 
\makecell[l]{$\displaystyle \left| \frac{8\omega_- \delta z_A}{\Omega^{\mathrm{(I)}}_{2}}\right|
+ \left| \frac{4\omega_-^2 \delta z_A^2}{\Omega^{\mathrm{(I)}}_{2}} \right|
+ \left| \frac{4\delta \dot{z}_A^2}{\Omega^{\mathrm{(I)}}_{2}} \right| $ }\\
\Xhline{1.25pt}
2 &
\makecell[l]{1 IBP $\to \delta z^{\mathrm{(I)}}$ \\ OPB $\to \delta \dot{z}^{\mathrm{(I)}}, \left(\delta z^{\mathrm{(I)}}\right)^2, \left(\delta \dot{z}^{\mathrm{(I)}}\right)^2$} & 
\makecell[l]{%
\centering
$\displaystyle 
\left ( -\,\frac{1}{\Omega^{\mathrm{(I)}}_{2}} -i \dfrac{\omega_- \delta z_A}{\Omega^{\mathrm{(I)}}_2} \right) e^{-i \omega_- C_A} e^{i \Omega^{\mathrm{(I)}}_{2} t_A} - \dfrac{4\pi\sqrt{\omega \omega'}}{\alpha}\beta^{\mathrm{(I)}}_{\omega \omega',i}$} & 
\makecell[l]{$\displaystyle  \left| \frac{8\omega_- \delta \dot{z}_A}{\left(\Omega^{\mathrm{(I)}}_{2} \right)^2}\right|
+ \left| \frac{4\omega_-^2 \delta z_A^2}{\Omega^{\mathrm{(I)}}_{2}} \right|
+ \left| \frac{4\delta \dot{z}_A^2}{\Omega^{\mathrm{(I)}}_{2}} \right| $} \\
\hline
3 &
\makecell[l]{1 IBP $\to \delta z^{\mathrm{(I)}}$ \\ OPB $\to \left(\delta z^{\mathrm{(I)}}\right)^2, \left(\delta \dot{z}^{\mathrm{(I)}}\right)^2$ \\ DMB $\to \delta\dot z^{\mathrm{(I)}}$} & 
\makecell[l]{same as row 2} & 
\makecell[l]{$\displaystyle  \left| \frac{\omega_- \delta z_A}{\Omega^{\mathrm{(I)}}_{2}}\right|
+ \left| \frac{4\omega_-^2 \delta z_A^2}{\Omega^{\mathrm{(I)}}_{2}} \right|
+ \left| \frac{4\delta \dot{z}_A^2}{\Omega^{\mathrm{(I)}}_{2}} \right| $ }\\
\hline
4 &
\makecell[l]{1 IBP $\to \left( \delta z^{\mathrm{(I)}}\right)^2$ \\ OPB $\to \delta z^{\mathrm{(I)}}, \delta z^{\mathrm{(I)}}\delta \dot{z}^{\mathrm{(I)}}, \left(\delta \dot{z}^{\mathrm{(I)}}\right)^2$} & 
\makecell[l]{%
\centering
$\displaystyle \left ( -\,\frac{1}{\Omega^{\mathrm{(I)}}_{2}} +\dfrac{\omega_-^2 \delta z_A^2}{2\Omega^{\mathrm{(I)}}_2} \right) e^{-i \omega_- C_A} e^{i \Omega^{\mathrm{(I)}}_{2} t_A} - \dfrac{4\pi\sqrt{\omega \omega'}}{\alpha}\beta^{\mathrm{(I)}}_{\omega \omega',i}$} & 
\makecell[l]{$\displaystyle  \left| \frac{8\omega_- \delta z_A}{\Omega^{\mathrm{(I)}}_{2}}\right|
+ \left| \frac{8\omega_-^2 \delta z_A \delta \dot{z}_A}{\left( \Omega^{\mathrm{(I)}}_{2} \right)^2} \right|
+ \left| \frac{4\delta \dot{z}_A^2}{\Omega^{\mathrm{(I)}}_{2}} \right| $} \\
\hline
5 &
\makecell[l]{1 IBP $\to \left( \delta z^{\mathrm{(I)}}\right)^2$ \\ OPB $\to \delta z^{\mathrm{(I)}}, \left(\delta \dot{z}^{\mathrm{(I)}}\right)^2$ \\ DMB $\to \delta z^{\mathrm{(I)}}\delta \dot{z}^{\mathrm{(I)}}$} & 
\makecell[l]{same as row 4} & 
\makecell[l]{$\displaystyle  \left| \frac{8\omega_- \delta z_A}{\Omega^{\mathrm{(I)}}_{2}}\right|
+ \left| \frac{\omega_-^2 \delta z_A^2}{2 \Omega^{\mathrm{(I)}}_{2}} \right|
+ \left| \frac{4\delta \dot{z}_A^2}{\Omega^{\mathrm{(I)}}_{2}} \right| $} \\
\hline
6 &
\makecell[l]{1 IBP $\to \left( \delta \dot{z}^{\mathrm{(I)}}\right)^2$ \\ OPB $\to \delta z^{\mathrm{(I)}}, \left(\delta z^{\mathrm{(I)}}\right)^2$ \\ DMB $\to \delta \dot{z}^{\mathrm{(I)}}\delta \ddot{z}^{\mathrm{(I)}}$} & 
\makecell[l]{$\displaystyle \left ( -\,\frac{1}{\Omega^{\mathrm{(I)}}_{2}} + \dfrac{\delta \dot{z}_A^2}{2\Omega^{\mathrm{(I)}}_2} \right) e^{-i \omega_- C_A} e^{i \Omega^{\mathrm{(I)}}_{2} t_A} - \dfrac{4\pi\sqrt{\omega \omega'}}{\alpha}\beta^{\mathrm{(I)}}_{\omega \omega',i}$} & 
\makecell[l]{$\displaystyle  \left| \frac{8\omega_- \delta z_A}{\Omega^{\mathrm{(I)}}_{2}}\right|
+ \left| \frac{4\omega_-^2 \delta z_A^2}{\Omega^{\mathrm{(I)}}_{2}} \right|
+ \left| \frac{\delta \dot{z}_A^2}{2\Omega^{\mathrm{(I)}}_{2}} \right| $ }\\
\Xhline{1.25pt}
7 &
\makecell[l]{2 IBP $\to \delta z^{\mathrm{(I)}}, \delta \dot{z}^{\mathrm{(I)}}$ \\ OPB $\to \left(\delta z^{\mathrm{(I)}}\right)^2,\left( \delta \dot{z}^{\mathrm{(I)}}\right)^2$ \\ DMB $\to \delta\ddot z^{\mathrm{(I)}}$} &
\makecell[l]{$\displaystyle 
\left ( -\,\frac{1}{\Omega^{\mathrm{(I)}}_{2}} -i\dfrac{\omega_- \delta z_A}{\Omega^{\mathrm{(I)}}_2} + \dfrac{\omega_- \delta \dot{z}_A}{\left(\Omega^{\mathrm{(I)}}_2\right)^2} \right) e^{-i \omega_- C_A} e^{i \Omega^{\mathrm{(I)}}_{2} t_A } - \dfrac{4\pi\sqrt{\omega \omega'}}{\alpha}\beta^{\mathrm{(I)}}_{\omega \omega',i}$ } & 
\makecell[l]{$\displaystyle  \left| \frac{\omega_- \delta \dot{z}_A}{\left(\Omega^{\mathrm{(I)}}_{2} \right)^2}\right|
+ \left| \frac{4\omega_-^2 \delta z_A^2}{\Omega^{\mathrm{(I)}}_{2}} \right|
+ \left| \frac{4\delta \dot{z}_A^2}{\Omega^{\mathrm{(I)}}_{2}} \right| $ } \\
\hline
8 &
\makecell[l]{2 IBP $\to \delta z^{\mathrm{(I)}}, \left( \delta z^{\mathrm{(I)}} \right)^2$ \\ OPB $\to \delta \dot{z}^{\mathrm{(I)}}, \delta z^{\mathrm{(I)}} \delta \dot{z}^{\mathrm{(I)}}, \left( \delta \dot{z}^{\mathrm{(I)}} \right)^2 $} & 
\makecell[l]{%
\centering
$\displaystyle 
\left ( -\,\frac{1}{\Omega^{\mathrm{(I)}}_{2}} -i\dfrac{\omega_- \delta z_A}{\Omega^{\mathrm{(I)}}_2} + \dfrac{\omega_-^2 \delta z_A^2}{2\Omega^{\mathrm{(I)}}_2} \right) e^{-i \omega_- C_A} e^{i \Omega^{\mathrm{(I)}}_{2} t_A} - \dfrac{4\pi\sqrt{\omega \omega'}}{\alpha}\beta^{\mathrm{(I)}}_{\omega \omega',i}$ } & 
\makecell[l]{$\displaystyle  \left| \frac{8\omega_- \delta \dot{z}_A}{\left( \Omega^{\mathrm{(I)}}_{2} \right)^2}\right|
+ \left| \frac{8\omega_-^2 \delta z_A \delta \dot{z}_A}{\left( \Omega^{\mathrm{(I)}}_{2} \right)^2} \right|
+ \left| \frac{4\delta \dot{z}_A^2}{\Omega^{\mathrm{(I)}}_{2}} \right| $} \\
\hline
9 &
\makecell[l]{2 IBP $\to \delta z^{\mathrm{(I)}}, \left( \delta z^{\mathrm{(I)}} \right)^2$ \\ OPB $\to \delta \dot{z}^{\mathrm{(I)}}, \left( \delta \dot{z}^{\mathrm{(I)}} \right)^2 $ \\ DMB $\to \delta z^{\mathrm{(I)}} \delta \dot{z}^{\mathrm{(I)}}$} & 
\makecell[l]{same as row 8} & 
\makecell[l]{$\displaystyle  \left| \frac{8\omega_- \delta \dot{z}_A}{\left( \Omega^{\mathrm{(I)}}_{2} \right)^2}\right|
+ \left| \frac{\omega_-^2 \delta z_A^2}{2\Omega^{\mathrm{(I)}}_{2}} \right|
+ \left| \frac{4\delta \dot{z}_A^2}{\Omega^{\mathrm{(I)}}_{2}} \right| $ } \\
\hline
10 &
\makecell[l]{2 IBP $\to \delta z^{\mathrm{(I)}}, \left( \delta z^{\mathrm{(I)}} \right)^2$ \\ OPB $\to \delta z^{\mathrm{(I)}} \delta \dot{z}^{\mathrm{(I)}}, \left( \delta \dot{z}^{\mathrm{(I)}} \right)^2 $ \\ DMB $\to \delta \dot{z}^{\mathrm{(I)}}$} & 
\makecell[l]{same as row 8} & 
\makecell[l]{$\displaystyle  \left| \frac{\omega_- \delta z_A}{\Omega^{\mathrm{(I)}}_{2}}\right|
+ \left| \frac{8\omega_-^2 \delta z_A \delta \dot{z}_A}{\left( \Omega^{\mathrm{(I)}}_{2} \right)^2} \right|
+ \left| \frac{4\delta \dot{z}_A^2}{\Omega^{\mathrm{(I)}}_{2}} \right| $} \\
\hline
11 &
\makecell[l]{2 IBP $\to \delta z^{\mathrm{(I)}}, \left( \delta z^{\mathrm{(I)}} \right)^2$ \\ OPB $\to \left( \delta \dot{z}^{\mathrm{(I)}} \right)^2 $ \\ DMB $\to \delta \dot{z}^{\mathrm{(I)}}, \delta z^{\mathrm{(I)}} \delta \dot{z}^{\mathrm{(I)}}$} & 
\makecell[l]{same as row 8} & 
\makecell[l]{$\displaystyle  \left| \frac{\omega_- \delta z_A}{\Omega^{\mathrm{(I)}}_{2}}\right|
+ \left| \frac{\omega_-^2 \delta z_A^2}{2\Omega^{\mathrm{(I)}}_{2}} \right|
+ \left| \frac{4\delta \dot{z}_A^2}{\Omega^{\mathrm{(I)}}_{2}} \right| $} \\
\hline
12 &
\makecell[l]{2 IBP $\to \delta z^{\mathrm{(I)}}, \left( \delta \dot{z}^{\mathrm{(I)}} \right)^2$ \\ OPB $\to \delta \dot{z}^{\mathrm{(I)}}, \left( \delta z^{\mathrm{(I)}} \right)^2$ \\ DMB $\to \delta \dot{z}^{\mathrm{(I)}} \delta \ddot{z}^{\mathrm{(I)}}$} & 
\makecell[l]{%
\centering
$\displaystyle 
\left ( -\,\frac{1}{\Omega^{\mathrm{(I)}}_{2}} -i\dfrac{\omega_- \delta z_A}{\Omega^{\mathrm{(I)}}_2} + \dfrac{\delta \dot{z}_A^2}{2\Omega^{\mathrm{(I)}}_2} \right) e^{-i \omega_- C_A} e^{i \Omega^{\mathrm{(I)}}_{2} t_A} - \dfrac{4\pi\sqrt{\omega \omega'}}{\alpha}\beta^{\mathrm{(I)}}_{\omega \omega',i}$ } & 
\makecell[l]{$\displaystyle  \left| \frac{8\omega_- \delta \dot{z}_A}{\left( \Omega^{\mathrm{(I)}}_{2} \right)^2}\right|
+ \left| \frac{4\omega_-^2 \delta z_A^2}{\Omega^{\mathrm{(I)}}_{2}} \right|
+ \left| \frac{\delta \dot{z}_A^2}{2\Omega^{\mathrm{(I)}}_{2}} \right| $} \\
\hline
13 &
\makecell[l]{2 IBP $\to \delta z^{\mathrm{(I)}}, \left( \delta \dot{z}^{\mathrm{(I)}} \right)^2$ \\ OPB $\to \left( \delta z^{\mathrm{(I)}} \right)^2$ \\ DMB $\to \delta \dot{z}^{\mathrm{(I)}}, \delta \dot{z}^{\mathrm{(I)}} \delta \ddot{z}^{\mathrm{(I)}}$} & 
\makecell[l]{same as row 12} & 
\makecell[l]{$\displaystyle  \left| \frac{\omega_- \delta z_A}{\Omega^{\mathrm{(I)}}_{2}}\right|
+ \left| \frac{4\omega_-^2 \delta z_A^2}{\Omega^{\mathrm{(I)}}_{2}} \right|
+ \left| \frac{\delta \dot{z}_A^2}{2\Omega^{\mathrm{(I)}}_{2}} \right| $} \\
\hline
14 &
\makecell[l]{2 IBP $\to \left( \delta z^{\mathrm{(I)}} \right)^2, \left( \delta \dot{z}^{\mathrm{(I)}} \right)^2$ \\ OPB $\to \delta z^{\mathrm{(I)}}, \delta z^{\mathrm{(I)}} \delta \dot{z}^{\mathrm{(I)}}$ \\ DMB $ \to \delta \dot{z}^{\mathrm{(I)}} \delta \ddot{z}^{\mathrm{(I)}}$} & 
\makecell[l]{%
\centering
$\displaystyle 
\left ( -\,\frac{1}{\Omega^{\mathrm{(I)}}_{2}} +\dfrac{\omega_-^2 \delta z_A^2}{2\Omega^{\mathrm{(I)}}_2} + \dfrac{\delta \dot{z}_A^2}{2\Omega^{\mathrm{(I)}}_2} \right) e^{-i \omega_- C_A} e^{i \Omega^{\mathrm{(I)}}_{2} t_A} - \dfrac{4\pi\sqrt{\omega \omega'}}{\alpha}\beta^{\mathrm{(I)}}_{\omega \omega',i}$} & 
\makecell[l]{$\displaystyle  \left| \frac{8\omega_- \delta z_A}{\Omega^{\mathrm{(I)}}_{2}}\right|
+ \left| \frac{8\omega_-^2 \delta z_A \delta \dot{z}_A}{\left( \Omega^{\mathrm{(I)}}_{2} \right)^2} \right|
+ \left| \frac{\delta \dot{z}_A^2}{2\Omega^{\mathrm{(I)}}_{2}} \right| $ } \\
\hline
15 &
\makecell[l]{2 IBP $\to \left( \delta z^{\mathrm{(I)}} \right)^2, \left( \delta \dot{z}^{\mathrm{(I)}} \right)^2$ \\ OPB $\to \delta z^{\mathrm{(I)}}$ \\ DMB $\to \delta z^{\mathrm{(I)}} \delta \dot{z}^{\mathrm{(I)}}, \delta \dot{z}^{\mathrm{(I)}} \delta \ddot{z}^{\mathrm{(I)}}$} & 
\makecell[l]{same as row 14} & 
\makecell[l]{$\displaystyle  \left| \frac{8\omega_- \delta z_A}{\Omega^{\mathrm{(I)}}_{2}}\right|
+ \left| \frac{\omega_-^2 \delta z_A^2}{2\Omega^{\mathrm{(I)}}_{2}} \right|
+ \left| \frac{\delta \dot{z}_A^2}{2\Omega^{\mathrm{(I)}}_{2}} \right| $} \\
\Xhline{1.25pt}
\end{tabular}
}
\end{table}
\clearpage
\begin{table}[t]
\addtocounter{table}{-1}
\setcellgapes{2pt}\makegapedcells
\caption{(continued)}
\resizebox{\textwidth}{!}{%
\begin{tabular}{c|c|c|c}
\Xhline{1.25pt}
\textbf{$\#^{\mathrm{(I)}}$} &
\textbf{Strategy} &
\textbf{$\dfrac{4\pi\sqrt{\omega \omega'}}{\alpha} \delta \beta_{\omega \omega'}^{\mathrm{(I)},\mathrm{com}}$} &
\textbf{$\dfrac{4\pi\sqrt{\omega \omega'}}{\alpha} \left| \delta \beta_{\omega \omega'}^{\mathrm{(I)},\mathrm{abs}} \right|$} \\
\Xhline{1.25pt}
16 &
\makecell[l]{3 IBP $\to \delta z^{\mathrm{(I)}}, \delta \dot{z}^{\mathrm{(I)}}, \left( \delta z^{\mathrm{(I)}} \right)^2$ \\ OPB $\to \delta z^{\mathrm{(I)}}\delta \dot{z}^{\mathrm{(I)}},\left( \delta \dot{z}^{\mathrm{(I)}}\right)^2$ \\ DMB $\to \delta\ddot z^{\mathrm{(I)}}$} & 
\makecell[l]{%
\centering
$\displaystyle \left ( -\,\frac{1}{\Omega^{\mathrm{(I)}}_{2}} -i\dfrac{\omega_- \delta z_A}{\Omega^{\mathrm{(I)}}_2} + \dfrac{\omega_- \delta \dot{z}_A}{\left(\Omega^{\mathrm{(I)}}_2\right)^2} + \dfrac{\omega_-^2 \delta z_A^2}{2\Omega^{\mathrm{(I)}}_2}\right) e^{-i \omega_- C_A} e^{i \Omega^{\mathrm{(I)}}_{2} t_A}  - \dfrac{4\pi\sqrt{\omega \omega'}}{\alpha}\beta^{\mathrm{(I)}}_{\omega \omega',i} $ } & 
\makecell[l]{$\displaystyle  \left| \frac{\omega_- \delta \dot{z}_A}{\left( \Omega^{\mathrm{(I)}}_{2} \right)^2}\right|
+ \left| \frac{8\omega_-^2 \delta z_A \delta \dot{z}_A}{\left( \Omega^{\mathrm{(I)}}_{2} \right)^2} \right|
+ \left| \frac{4\delta \dot{z}_A^2}{\Omega^{\mathrm{(I)}}_{2}} \right| $ } \\
\hline
17 &
\makecell[l]{3 IBP $\to \delta z^{\mathrm{(I)}}, \delta \dot{z}^{\mathrm{(I)}}, \left( \delta z^{\mathrm{(I)}} \right)^2$ \\ OPB $\to \left( \delta \dot{z}^{\mathrm{(I)}}\right)^2$ \\ DMB $\to \delta z^{\mathrm{(I)}}\delta \dot{z}^{\mathrm{(I)}}, \delta\ddot z^{\mathrm{(I)}}$} & 
\makecell[l]{same as row 16} & 
\makecell[l]{$\displaystyle  \left| \frac{\omega_- \delta \dot{z}_A}{\left( \Omega^{\mathrm{(I)}}_{2} \right)^2}\right|
+ \left| \frac{8\omega_-^2 \delta z_A \delta \dot{z}_A}{\left( \Omega^{\mathrm{(I)}}_{2} \right)^2} \right|
+ \left| \frac{4\delta \dot{z}_A^2}{\Omega^{\mathrm{(I)}}_{2}} \right| $ } \\
\hline
18 &
\makecell[l]{3 IBP $\to \delta z^{\mathrm{(I)}}, \delta \dot{z}^{\mathrm{(I)}}, \left( \delta \dot{z}^{\mathrm{(I)}} \right)^2$ \\ OPB $\to \left( \delta z^{\mathrm{(I)}} \right)^2 $ \\ DMB $\to \delta\ddot z^{\mathrm{(I)}}, \delta \dot{z}^{\mathrm{(I)}} \delta \ddot{z}^{\mathrm{(I)}}$} & 
\makecell[l]{$\displaystyle 
\left ( -\,\frac{1}{\Omega^{\mathrm{(I)}}_{2}} -i\dfrac{\omega_- \delta z_A}{\Omega^{\mathrm{(I)}}_2} + \dfrac{\omega_- \delta \dot{z}_A}{\left(\Omega^{\mathrm{(I)}}_2\right)^2} + \dfrac{\delta \dot{z}_A^2}{2\Omega^{\mathrm{(I)}}_2}\right) e^{-i \omega_- C_A} e^{i \Omega^{\mathrm{(I)}}_{2} t_A}  - \dfrac{4\pi\sqrt{\omega \omega'}}{\alpha}\beta^{\mathrm{(I)}}_{\omega \omega',i}$ } & 
\makecell[l]{$\displaystyle  \left| \frac{\omega_- \delta \dot{z}_A}{\left( \Omega^{\mathrm{(I)}}_{2} \right)^2}\right|
+ \left| \frac{4\omega_-^2 \delta z_A^2}{\Omega^{\mathrm{(I)}}_{2}} \right|
+ \left| \frac{\delta \dot{z}_A^2}{2\Omega^{\mathrm{(I)}}_{2}} \right| $ }\\
\Xhline{1.25pt}
19 &
\makecell[l]{3 IBP $\to \delta z^{\mathrm{(I)}}, \left( \delta z^{\mathrm{(I)}} \right)^2, \left( \delta \dot{z}^{\mathrm{(I)}} \right)^2$ \\ OPB $\to \delta \dot{z}^{\mathrm{(I)}}, \delta z^{\mathrm{(I)}} \delta \dot{z}^{\mathrm{(I)}}$ \\ DMB $\to \delta \dot{z}^{\mathrm{(I)}} \delta \ddot{z}^{\mathrm{(I)}}$} & 
\makecell[l]{%
\centering
$\displaystyle 
\left ( -\,\frac{1}{\Omega^{\mathrm{(I)}}_{2}} -i\dfrac{\omega_- \delta z_A}{\Omega^{\mathrm{(I)}}_2} + \dfrac{\omega_-^2 \delta z_A^2}{2\Omega^{\mathrm{(I)}}_2}+\dfrac{\delta \dot{z}_A^2}{2\Omega^{\mathrm{(I)}}_2} \right) e^{-i \omega_- C_A} e^{i \Omega^{\mathrm{(I)}}_{2} t_A}  - \dfrac{4\pi\sqrt{\omega \omega'}}{\alpha}\beta^{\mathrm{(I)}}_{\omega \omega',i}$ } & 
\makecell[l]{$\displaystyle  \left| \frac{8\omega_- \delta \dot{z}_A}{\left( \Omega^{\mathrm{(I)}}_{2} \right)^2}\right|
+ \left| \frac{8\omega_-^2 \delta z_A \delta \dot{z}_A}{\left( \Omega^{\mathrm{(I)}}_{2} \right)^2} \right|
+ \left| \frac{\delta \dot{z}_A^2}{2\Omega^{\mathrm{(I)}}_{2}} \right| $} \\
\hline
20 &
\makecell[l]{3 IBP $\to \delta z^{\mathrm{(I)}}, \left( \delta z^{\mathrm{(I)}} \right)^2, \left( \delta \dot{z}^{\mathrm{(I)}} \right)^2$ \\ OPB $\to \delta \dot{z}^{\mathrm{(I)}}$ \\ DMB $\to \delta z^{\mathrm{(I)}} \delta \dot{z}^{\mathrm{(I)}}, \delta \dot{z}^{\mathrm{(I)}} \delta \ddot{z}^{\mathrm{(I)}}$} & 
\makecell[l]{same as row 19} & 
\makecell[l]{$\displaystyle  \left| \frac{8\omega_- \delta \dot{z}_A}{\left( \Omega^{\mathrm{(I)}}_{2} \right)^2}\right|
+ \left| \frac{\omega_-^2 \delta z_A^2}{2\Omega^{\mathrm{(I)}}_{2}} \right|
+ \left| \frac{\delta \dot{z}_A^2}{2\Omega^{\mathrm{(I)}}_{2}} \right| $} \\
\hline
21 &
\makecell[l]{3 IBP $\to \delta z^{\mathrm{(I)}}, \left( \delta z^{\mathrm{(I)}} \right)^2, \left( \delta \dot{z}^{\mathrm{(I)}} \right)^2$ \\ OPB $\to \delta z^{\mathrm{(I)}} \delta \dot{z}^{\mathrm{(I)}}$ \\ DMB $\to \delta \dot{z}^{\mathrm{(I)}}, \delta \dot{z}^{\mathrm{(I)}} \delta \ddot{z}^{\mathrm{(I)}}$} & 
\makecell[l]{same as row 19} &
\makecell[l]{$\displaystyle  \left| \frac{\omega_- \delta z_A}{\Omega^{\mathrm{(I)}}_{2}}\right|
+ \left| \frac{8\omega_-^2 \delta z_A \delta \dot{z}_A}{\left( \Omega^{\mathrm{(I)}}_{2} \right)^2} \right|
+ \left| \frac{\delta \dot{z}_A^2}{2\Omega^{\mathrm{(I)}}_{2}} \right| $} \\
\hline
22 &
\makecell[l]{3 IBP $\to \delta z^{\mathrm{(I)}}, \left( \delta z^{\mathrm{(I)}} \right)^2, \left( \delta \dot{z}^{\mathrm{(I)}} \right)^2$ \\ DMB $\to \delta \dot{z}^{\mathrm{(I)}}, \delta z^{\mathrm{(I)}} \delta \dot{z}^{\mathrm{(I)}}, \delta \dot{z}^{\mathrm{(I)}} \delta \ddot{z}^{\mathrm{(I)}}$} &
\makecell[l]{same as row 19} &
\makecell[l]{$\displaystyle  \left| \frac{\omega_- \delta z_A}{\Omega^{\mathrm{(I)}}_{2}}\right|
+ \left| \frac{\omega_-^2 \delta z_A^2}{2\Omega^{\mathrm{(I)}}_{2}} \right|
+ \left| \frac{\delta \dot{z}_A^2}{2\Omega^{\mathrm{(I)}}_{2}} \right| $} \\
\Xhline{1.25pt}
23 &
\makecell[l]{4 IBP $\to \delta z^{\mathrm{(I)}}, \delta \dot{z}^{\mathrm{(I)}}, \left( \delta z^{\mathrm{(I)}} \right)^2, \left( \delta \dot{z}^{\mathrm{(I)}} \right)^2$ \\ OPB $\to \delta z^{\mathrm{(I)}}\delta \dot{z}^{\mathrm{(I)}}$ \\ DMB $\to \delta\ddot z^{\mathrm{(I)}}, \delta \dot{z}^{\mathrm{(I)}}\delta\ddot z^{\mathrm{(I)}}$} & 
\makecell[l]{%
$\displaystyle 
\left ( -\,\frac{1}{\Omega^{\mathrm{(I)}}_{2}} -i\dfrac{\omega_- \delta z_A}{\Omega^{\mathrm{(I)}}_2} + \dfrac{\omega_- \delta \dot{z}_A}{\left(\Omega^{\mathrm{(I)}}_2\right)^2} + \dfrac{\omega_-^2 \delta z_A^2}{2\Omega^{\mathrm{(I)}}_2} + \dfrac{\delta \dot{z}_A^2}{2\Omega^{\mathrm{(I)}}_2} \right) e^{-i \omega_- C_A} e^{i \Omega^{\mathrm{(I)}}_{2} t_A} - \dfrac{4\pi\sqrt{\omega \omega'}}{\alpha}\beta^{\mathrm{(I)}}_{\omega \omega',i}$ } & 
\makecell[l]{$\displaystyle  \left| \frac{\omega_- \delta \dot{z}_A}{\left( \Omega^{\mathrm{(I)}}_{2} \right)^2}\right|
+ \left| \frac{8\omega_-^2 \delta z_A \delta \dot{z}_A}{\left( \Omega^{\mathrm{(I)}}_{2} \right)^2} \right|
+ \left| \frac{\delta \dot{z}_A^2}{2\Omega^{\mathrm{(I)}}_{2}} \right| $ } \\
\hline
24 &
\makecell[l]{4 IBP $\to \delta z^{\mathrm{(I)}}, \delta \dot{z}^{\mathrm{(I)}}, \left( \delta z^{\mathrm{(I)}} \right)^2, \left( \delta \dot{z}^{\mathrm{(I)}} \right)^2$ \\ DMB $\to \delta\ddot z^{\mathrm{(I)}}, \delta z^{\mathrm{(I)}}\delta \dot{z}^{\mathrm{(I)}}, \delta \dot{z}^{\mathrm{(I)}}\delta\ddot z^{\mathrm{(I)}}$} &
\makecell[l]{same as row 23} &
\makecell[l]{$\displaystyle  \left| \frac{\omega_- \delta \dot{z}_A}{\left( \Omega^{\mathrm{(I)}}_{2} \right)^2}\right|
+ \left| \frac{\omega_-^2 \delta z_A^2}{2\Omega^{\mathrm{(I)}}_{2}} \right|
+ \left| \frac{\delta \dot{z}_A^2}{2\Omega^{\mathrm{(I)}}_{2}} \right| $} \\
\Xhline{1.25pt}
\end{tabular}
}
\end{table}

% ====================================================================================================
\twocolumngrid
\bibliography{apssamp}% Produces the bibliography via BibTeX.
\end{document}